\documentclass[aps,twocolumn,10pt,prc,floatfix,showpacs,preprintnumbers,amsmath,amssymb,nofootinbib,superscriptaddress]{revtex4-1}

\usepackage{graphicx}%
\usepackage{dcolumn}%
\usepackage{bm}%
\usepackage[mathlines]{lineno}%
\usepackage[dvipsnames]{xcolor}
\usepackage{multirow}
\usepackage{diagbox}
\usepackage{comment}

\usepackage[utf8]{inputenc}
\usepackage[T1]{fontenc}
\usepackage{etoolbox}
\usepackage{mathtools,slashed}
\usepackage{cases}
\usepackage{lipsum}
\usepackage{placeins}

\usepackage{caption}
\usepackage{subcaption}
\usepackage{chngcntr} %
\usepackage{apptools} %

\usepackage[pdfstartview=FitH,
            CJKbookmarks=true,
            bookmarksnumbered=true,
            bookmarksopen=true,
            colorlinks,
            linkcolor=blue,
            anchorcolor=blue,
            citecolor=blue,
            urlcolor=blue,
            ]{hyperref}

\draft %

\newcommand{\bv}{\mathbf}

\newcommand{\etal}{\textit{et al.}}

\newcommand{\tp}{t^{\prime}}
\newcommand{\var}[1]{\mathrm{Var}\Bigl(#1\Bigr)}
\newcommand{\cov}[1]{\mathrm{Cov}\Bigl(#1\Bigr)}
\newcommand{\vars}[1]{\mathrm{Var}\left(#1\right)}
\newcommand{\covs}[1]{\mathrm{Cov}\left(#1\right)}

\newcommand{\Djt}{\Delta_j(t)}
\newcommand{\Djtp}{\Delta_j(\tp)}
\newcommand{\Dit}{\Delta_i(t)}
\newcommand{\Ditp}{\Delta_i(\tp)}

\newcommand{\geom}[2]{\dfrac{1-#1^{2#2}}{1-#1^2}}
\newcommand{\sumgeom}[1]{\geom{\alpha}{#1}+\geom{\beta}{#1}}
\newcommand{\bgeom}[1]{\geom{\beta}{#1}}

\newcommand{\textgeom}[2]{\frac{1-#1^{2#2}}{1-#1^2}}
\newcommand{\textbgeom}[1]{\textgeom{\beta}{#1}}

\newcommand{\lrpars}[1]{\left(#1 \right)}
\newcommand{\lnp}[1]{\ln \lrpars{#1}}

\newcommand{\logdet}[3]{\dfrac12 \lnp{\dfrac{#1 \, #2}{#3}}}

\newcommand\wmsphi{\ensuremath{\Phi_{\mathrm{WMS}}}}
\newcommand\phiR{\ensuremath{\Phi_{\mathrm{R}}}}

\DeclareMathAlphabet{\mathbbb}{U}{bbold}{m}{n}
\DeclareRobustCommand{\rchi}{{\mathpalette\irchi\relax}}
\newcommand{\irchi}[2]{\raisebox{\depth}{$#1\chi$}} %
\newcommand\underrel[3][]{\mathrel{\mathop{#3}\limits_{%
      \ifx c#1\relax\mathclap{#2}\else#2\fi}}}

\newcommand\alphaeq{\stackrel{\mathclap{\normalfont\mbox{\small{a}}}}{=}}
\newcommand\betaeq{\stackrel{\mathclap{\normalfont\mbox{\small{b}}}}{=}}
\newcommand\gammaeq{\stackrel{\mathclap{\normalfont\mbox{\small{c}}}}{=}}
\newcommand\deltaeq{\stackrel{\mathclap{\normalfont\mbox{\small{d}}}}{=}}

\newcommand*{\figuretitle}[1]{%
    {\centering  
    \textbf{#1}
    \par\smallskip}}

\begin{document}

\title{Simple physical systems as a reference for multivariate information dynamics}

\author{Alberto Liardi}
    \email[]{a.liardi@imperial.ac.uk}
\affiliation{Department of Computing, Imperial College London, UK}
\affiliation{Centre for Complexity Science, Imperial College London, UK}
\affiliation{Department of Mathematics, Imperial College London, UK}

\author{Madalina I Sas}
\affiliation{Department of Computing, Imperial College London, UK}
\affiliation{Centre for Complexity Science, Imperial College London, UK}

\author{George Blackburne}
\affiliation{Department of Experimental Psychology, University College London, UK}
\affiliation{Department of Computing, Imperial College London, UK}

\author{\mbox{William J Knottenbelt}}
 \affiliation{Department of Computing, Imperial College London, UK}

\author{Pedro A.M. Mediano}
\affiliation{Department of Computing, Imperial College London, UK}
\affiliation{Division of Psychology and Language Sciences, University College London, UK}

\author{Henrik Jeldtoft Jensen}
\affiliation{Department of Mathematics, Imperial College London, UK}
\affiliation{Centre for Complexity Science, Imperial College London, UK}
\affiliation{Department of Mathematical and Computing Science, Science Institute of Tokyo, Japan}

\begin{abstract}
Understanding a complex system entails capturing the non-trivial collective phenomena that arise from interactions between its different parts. Information theory is a flexible and robust framework to study such behaviours, with several measures designed to quantify and characterise the interdependencies among the system's components.
However, since these estimators rely on the statistical distributions of observed quantities, it is crucial to examine the relationships between information-theoretic measures and the system's underlying mechanistic structure.
To this end, here we present an information-theoretic analytical investigation of an elementary system of interactive random walkers subject to Gaussian noise.
Focusing on partial information decomposition, causal emergence, and integrated information, our results help us develop some intuitions on their relationship with the physical parameters of the system. 
For instance, we observe that uncoupled systems can exhibit emergent properties, in a way that we suggest may be better described as ``statistically autonomous''. 
Overall, we observe that in this simple scenario information measures align more reliably with the system's mechanistic properties when calculated at the level of microscopic components, rather than their coarse-grained counterparts, and over timescales comparable with the system's intrinsic dynamics.  
Moreover, we show that approaches that separate the contributions of the system's dynamics and steady-state distribution (e.g.\ via causal perturbations) may help strengthen the interpretation of information-theoretic analyses.
\end{abstract}


\maketitle %

\section{Introduction} \label{sec:introduction}

Information theory has become a convenient and powerful methodology for the characterisation of multivariate dynamical systems \cite{cover1991information, mackay2003information}. The formalism initiated by Shannon \cite{shannon1948mathematical} has evolved into a sophisticated field able to reveal from observable data the fundamental organisation and information-processing properties of complex systems.

Present-day information theory is a broad and lively research subject with applications spanning many fields, such as computer science and machine learning \cite{tishby2000information, shwartz2017opening, proca2024synergistic}, neuroscience \cite{timme2018tutorial, stone2019informationtheorytutorialintroduction, luppi2024information}, biology \cite{smith2000concept, adami2004information}, ecology \cite{margalef1973information,rajpal2023quantifying} and beyond \cite{Ben-Naim2024}. 
Information theory offers a variety of metrics to assess the interdependencies and complexity of dynamical processes, with mutual information and transfer entropy being among the most popular.  
Notably, recent advances in the field now allow for more fine-grained and profound characterisation of multivariate systems \cite{williams2010nonnegative, mediano2019beyond, ince2017partial, varley2023partial, varley2024generalized}, including the interrogation of relationships between microscopic and macroscopic scales \cite{rosas_information-theoretic_2018,hoel2013quantifying,rosas2024software}.
Nonetheless, linking experimental results using these information-theoretic quantities to our intuitions about the mechanisms behind complex phenomena remains a challenging problem.

One common suspect that often leads to spurious conclusions is the so-called ``ghost of causality'' in statistical models \cite{pearl2009causality, pearl2009causal, pearl2016causal}. Specifically, while many works have investigated structural causal inference using information flow \cite{wiener1956theory, duan2013direct, eichler2013causal, martinez2024decomposing}, it has also been shown that this flow is not always faithful to the underlying causal structure~\cite{janzing2013quantifying}. 
Within complexity science, an increasingly popular view is to distinguish between ``mechanisms'' (i.e.\ the underlying structure of the system) and ``behaviours'' (i.e.\ the properties of the system's trajectories in its state space)~\cite{rosas2022disentangling}. Elucidating the relationship between the two, however, is complicated -- even more so 
when analysing real-world complex systems, where the high number of degrees of freedom and the complexity of interactions make it extremely difficult to interpret the underlying mechanical structure. 

While the field has thoroughly explored the formal properties of elementary information-theoretic quantities, less attention has been given to more recent measures -- those which we likely stand most to gain from for understanding and engineering complex systems.
To fill this gap, here we investigate how specific information-theoretic relationships are related to functional or mechanistic properties by analytically studying a simple system of interacting random walkers, for which physical intuitions and mathematical tools are available.
With this model at hand, we study how information-theoretic metrics vary when the interaction strength changes, and how this affects the information dynamics of the system. 
Importantly, we show that in general information-theoretic behaviours differ from the underlying mechanisms,
and outline some conditions that may help strengthen the interpretation of such measures applied to empirical data.

The rest of this paper is organised as follows: in Sec.~\ref{sec:warnings} we provide a simple and known example of a non-intuitive property of information-theoretic quantities, which justifies this study. Then, we present the one-dimensional model of interacting random walkers that will be our main case study (Sec.~\ref{sec:model}). In Sec.~\ref{sec:results}, we report the analytical and numerical results for various information-theoretic metrics, and we discuss their implications for the study of real data in Sec.~\ref{sec:discussion}. 
In the supplementary material, we report the detailed calculations of all quantities considered, additional investigations of their behaviours, and a brief comparative study of theoretical and numerical results.

\section{A simple example} \label{sec:warnings}

To illustrate how interpreting the behaviour of statistical quantities can be subtle, %
let us focus on the behaviour of mutual information (MI) in a stochastic system. 
Given two random variables $X,Y$, the mutual information between them assesses the amount of information that we gain about $X$ when $Y$ is observed. In other words, it measures the knowledge that $Y$ brings about $X$. Denoting Shannon's entropy by $H$, the MI between $X$ and $Y$ can be written as
\begin{equation}
    I(X;Y) = H(Y) - H(Y|X) \,.
\end{equation}
For further discussion and mathematical details, we refer to App.~\ref{app:MI}.

An interesting property of MI is that it can vanish in two different scenarios:
\begin{enumerate}
    \item if $X$ and $Y$ are independent variables, and thus ${H(Y) = H(Y|X) > 0}$,
    \item if $Y$ is a deterministic function of $X$, and thus ${H(Y) = H(Y|X) = 0}$.
\end{enumerate}
Therefore, MI depends on properties of both the channel (i.e.\ the relation between $X$ and $Y$) and the input distribution (in this case $Y$), and hence a change in MI can't be easily attributed to either. This may complicate the interpretation of the behaviour of the mutual information.

\begin{figure}[]
\centering
\hspace*{-0.5cm}
\includegraphics[width=1\columnwidth]{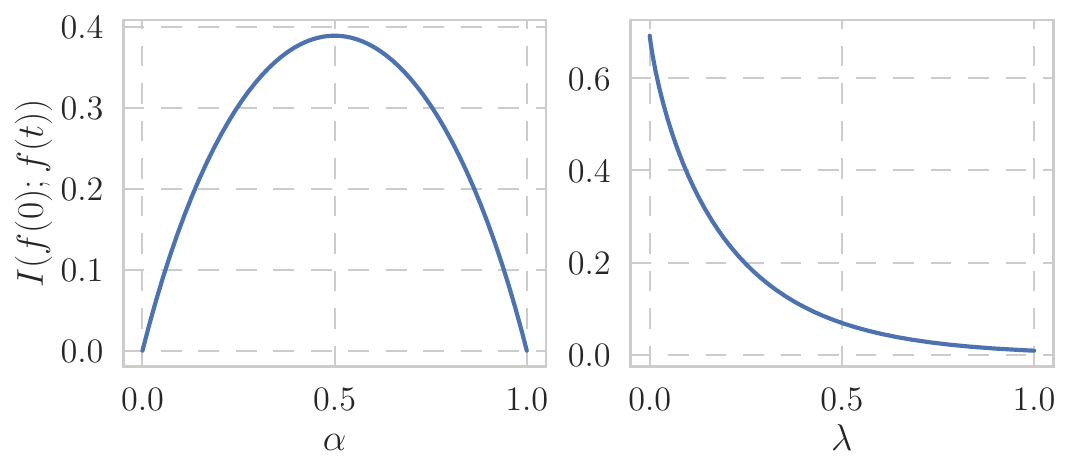}
    \caption{\textbf{Mutual information changes with both initial conditions (input) and temporal evolution (channel).} Mutual information $I(f(0);f(t))$ between the signal at time $t=0$ and time $t=1$ for different values of $(\alpha,\lambda)$. In the left panel, the swapping rate is fixed at $\lambda=0.1$, providing a non-monotonic MI as $\alpha$ varies. In the right panel, for fixed $\alpha=0.5$, the mutual information decreases monotonously as the swapping rate $\lambda$ induces more randomness in the system.
    See App.~\ref{app:swap_sig} for details.}
\label{fig:Swap_Signal}
\end{figure}

To better illustrate this point, consider a time-dependent signal $f(t)$ which can take the two values $+A$ and $-A$. We set the initial condition $f(0)= A$ with probability $\alpha$, and let the signal evolve in time, swapping between $\pm A$ at a constant rate $\lambda$. 
We want to compute the mutual information $I(f(0);f(t))$ between the signal at time zero and a later time $t$, i.e.\ we measure how much information the starting state provides about the one at time $t$. This can be easily achieved by working directly on the probabilities $p(f(0))$, $p(f(t))$, and $p(f(0);f(t))$. We refer to App.~\ref{app:swap_sig} for the mathematical details. 
To get a feeling of how the mutual information behaves when %
the interdependencies among channels and the property of the input distribution are changed, we let $\alpha$ vary from 0 to 1 while $\lambda$ is fixed (Fig.~\ref{fig:Swap_Signal}). Interestingly, the MI follows a non-monotonous behaviour: it is positive for $0<\alpha<1$, then vanishes for $\alpha=0,1$, since the value at $f(0)$ is assumed with probability 1. This is due to a vanishing entropy in the marginal distribution, as the system becomes fully deterministic (case 2 above). 
Similarly, for fixed $\alpha$, the increased stochasticity induced by a larger $\lambda$ drives the mutual information to zero (case 1 above).

This behaviour is of course well known in information theory and stochastic processes. Nonetheless, it is a useful reminder that a decrease in the mutual information may be either indicative of %
a weaker interdependence between variables or a more deterministic nature of the input distribution -- two very different mechanical properties of the system.

We will now elaborate on this theme by focusing on independent and coupled random walkers, both in a stationary and non-stationary case. We present analytic and numerical results for a variety of measures that are currently commonly employed to discuss higher-order properties in complex systems.

\section{Model} \label{sec:model}

Our mathematical laboratorium consists of a set of $N$ random walkers (RWs) that move in a continuous 1-dimensional space in discrete time. 
We consider an equation of motion that follows an Ornstein-Uhlenbeck (OU) process \cite{uhlenbeck1930theory}, in which the RWs are also coupled with each other through springs, forming an open-chain structure. 

Starting from the Langevin equation in continuous time, for the central walkers $i=2,3,...,N-1$ we have
\begin{equation} \label{eq:langevin}
\begin{aligned}
m \frac{d^2 x_i}{dt^2} = & -\nu \frac{dx_i}{dt} - \theta [x_i(t)-ia] + \gamma [x_{i+1}(t) - x_i(t) - a] +\\
& - \gamma [x_i(t) - x_{i-1}(t) - a] + \eta_i(t) \,,
\end{aligned}
\end{equation}
where $\nu$ is the friction coefficient, $\gamma$ the spring constant, $a$ the resting distance of the springs, $\theta$ the OU restoring constant, and $\eta_i(t)$ a stochastic Gaussian term sampled from $\mathcal{N}(0,\sigma^2)$. 
Restricting ourselves to the overdamped dynamics regime ($m\rightarrow0$) in discrete time, we have:
\begin{equation}
\begin{aligned}
x_i(t + dt) = & x_i(t) -\frac{dt}{\nu}\theta [x_i(t)-ia] + \\
& + \frac{\gamma dt}{\nu} (x_{i+1}(t) + x_{i-1}(t) - 2x_i(t)) + \frac{dt}{\nu} \eta_i(t) \, .
\end{aligned}
\end{equation}
In other words, the dynamics of a RW is characterised by the interactions with its two nearest neighbours mediated by $\gamma$, and a restoring force due to $\theta$.
Finally, we can rewrite the equations in terms of the deviations from the resting positions of the springs ${\Delta_i(t) \equiv x_i(t) - ia} $, obtaining
\begin{equation}
\begin{aligned} \label{eq:eom_ith}
\Delta_i(t+1) = & (1-\theta-2\gamma)\Delta_i(t) + \gamma (\Delta_{i+1}(t) + \Delta_{i-1}(t)) +\\
& + \eta_i(t) \, ,
\end{aligned}
\end{equation}
where for ease of notation we let $dt=\nu=1$\footnote{Setting $dt=1$ defines the system's timescale as one timestep, with arbitrary unit of time. Accordingly, references to short or long timescales should be understood relative to the system’s intrinsic dynamics, which can correspond to various actual durations across different physical processes.}. 
With a small abuse of language, throughout this work, we will refer to $\Delta_i$ as both deviations and positions of the RWs.
Focusing now on the edges $i=1,N$, similar considerations lead to the following equations of motion:
\begin{gather} 
\Delta_1(t+1) = (1-\theta-\gamma)\Delta_1(t) + \gamma \Delta_{2}(t) + \eta_1(t) \, , \notag \\ \label{eq:eom_1Nth}
\Delta_N(t+1) = (1-\theta-\gamma)\Delta_N(t) + \gamma \Delta_{N-1}(t) + \eta_N(t) \, .
\end{gather}
We remind that the OU process taken into consideration is stationary only for ${\theta\in(0,1)}$. Thus, for our analyses, we restrict to this range of the parameter.

Eqs.~\eqref{eq:eom_ith}-\eqref{eq:eom_1Nth} can be rewritten in the more compact matrix form
\begin{equation}     \label{eq:eom_RW}
    \bv{\Delta}(t+1)=\bv{M\Delta}(t)+\boldsymbol{\eta}(t) \,, 
\end{equation}
where we introduced the vectors ${\bv{\Delta}(t)=(\Delta_1(t),\ldots,\Delta_N(t))}$ and ${\boldsymbol{\eta}(t)=(\eta_1(t),\ldots,\eta_N(t))}$, the shorthand ${\alpha\equiv1 - \theta - \gamma}$ and ${\beta\equiv1 - \theta - 2\gamma}$, and the transition matrix
\begin{gather}    \label{eq:M_OU_NRW}
    \bv{M} = 
    \begin{pmatrix}
    \alpha & \gamma & 0 & \cdots & 0 & 0 \\
    \gamma & \beta & \gamma & 0 & \cdots & 0 \\
    \vdots & \vdots & \ddots & \ddots & \vdots & \vdots \\
    0 & \cdots & 0 & \gamma & \beta & \gamma \\
    0 & 0 & \cdots & 0 & \gamma & \alpha
\end{pmatrix} \, .
\end{gather}

Assuming as initial conditions ${\bv{\Delta}(0)=0}$, 
the dynamical equations for $\bv{\Delta}$ for any two time points $t, t'\,(t'>t)$ become
\begin{align}   \label{eq:delta_ev}
    \bv{\Delta}(t^{\prime}) &= \boldsymbol{M}^{t^{\prime}-t}\boldsymbol{\Delta}(t) + \boldsymbol{\rchi}_{tt^{\prime}} \,,\\
    \boldsymbol{\rchi}_{tt^{\prime}} &= 
    \begin{aligned}[t] \label{eq:chi_ev}
    & \boldsymbol{M}^{t^{\prime}-t-1}\boldsymbol{\eta}(t) + \boldsymbol{M}^{t^{\prime}-t-2}\boldsymbol{\eta}(t+1) + \ldots+ \\
    & + \boldsymbol{M}\boldsymbol{\eta}(t^{\prime}-2) + \boldsymbol{\eta}(t^{\prime}-1) \,.
    \end{aligned}
\end{align}
In Eq.~\eqref{eq:delta_ev} we can identify two distinctly different parts: ${\boldsymbol{M}^{t^{\prime}-t}\boldsymbol{\Delta}(t)}$ describes the deterministic evolution of the RWs, containing the interaction terms and restoring forces, whereas ${\boldsymbol{\rchi}_{tt^{\prime}}}$ encompasses the accumulated stochastic effects, as its role is to propagate the noise contributions added at each timestep.

Importantly, the knowledge of $\boldsymbol{\rchi}_{tt^{\prime}}$ is sufficient for calculating most information quantities, as the stochastic terms alone determine variances and covariances of the deviations. 
Moreover, we recognise that Eq.~\eqref{eq:delta_ev} expresses the evolution of the system as a Vector Autoregression (VAR) model, for which information-theoretic results are available in the literature \cite{barnett2009granger, barnett2011behaviour, barnett2012transfer}. We bridge our model with such findings and discuss their implications in Sec.~\ref{sec:discussion}. 
More details on the model can be found in App.~\ref{app:N-RWcase}.

To keep the problem analytically tractable and preserve an intuitive understanding, in this work we solve Eqs.~\eqref{eq:delta_ev}-\eqref{eq:chi_ev} under two important assumptions:  
\begin{enumerate}
    \item we use a perturbative approach expanding the equations of motion to first order in $\gamma$, meaning that the dynamics of the walkers are only slightly affected by the coupling. In practice, this accounts for $\gamma$ to be small compared to the stochastic terms $\eta_i$.
    \item we focus on the central walkers of the chain, i.e.\ all walkers except for the two endpoints. This choice ensures symmetry among the RWs, simplifying the calculations. For large systems, ignoring the edges has negligible effects.
\end{enumerate}
Hence, our results will be exact in the limit of vanishing $\gamma$ and  $N\rightarrow\infty$. 
Since assumption 1 may pose important limitations on the explanatory power of the system, we additionally consider the case of $N=2$ random walkers, which can be solved exactly for all $\gamma$ (see App. \ref{app:2-RWcase}), complementing the analysis of the $N>2$ RWs. 

To address how information-theoretic measures can inform us about macroscopic properties of the system, we also consider the dynamics of the centre of mass (c.o.m.\@) of the RWs, defined as:
\begin{equation} \label{eq:com_def_NRW}
    V(t) = \frac{1}{N}\sum_{j=1}^N \Delta_j(t) \,.
\end{equation}
For subsequent timesteps, the contributions of the internal elastic forces balance out according to Newton's third law, so that $\gamma$ vanishes from the expression of $V(\tp)$. This leads to the following equation of motion:
\begin{equation} \label{eq:eom_V}
\begin{aligned}
    V(t^{\prime}) & = \frac{1}{N}\sum_{j=1}^N \Delta_j(t^{\prime}) \\
    & \alphaeq \dfrac{(1-\theta)^{\tp-t}}{N}\sum_{j=1}^N \Delta_j(t)+\sum_{j=1}^N\sum_{\tilde{t}=t}^{{\tp}-1} \dfrac{(1-\theta)^{\tp-\tilde{t}-1}}{N} \eta_j(\tilde{t}) \\
    & = (1-\theta)^{\tp-t}V(t) + \rchi_{tt^{\prime}_V} \,,
\end{aligned}
\end{equation}
where in (a) we used Eq.~\eqref{eq:eom_ith} recursively $\tp-t$ times.
In other words, 
regardless of whether the walkers are interacting or not, their c.o.m.\ behaves as a free random walker itself, driven by a noise resulting from the cumulative effect of the noise of the individual walkers. Thus, although the c.o.m.\ is constituted by the positions of the individual walkers, one may consider it as a ``statistically autonomous''~\cite{rosas2024software, barnett2023dynamical} random walker by itself.

The simplicity of this setup allows for analytical calculations of information measures. Thus, the interplay between the interaction terms $(\theta, \gamma)$ and the stochastic drive ($\eta$), makes an ideal test bed to investigate questions concerning how information-theoretic quantities relate to mechanistic causal properties. 

In the following sections, we investigate the behaviour of some of the most used information-theoretic measures in the system of $N>2$ RWs analysed to first order in $\gamma$, and exact in $\gamma$ for the case of $N=2$ RWs.
To lighten the mathematical formalism and to foster an intuitive understanding, in the main text we present the results for the case of $\theta=0$, leaving the more general case in Apps.~\ref{app:N-RWcase}-\ref{app:2-RWcase}. 
Although setting $\theta=0$ leads to a non-stationary dynamics, with $\vars{\Djt}\to\infty$ for $t\to\infty$, we observe that the qualitative behaviour of the information measures studied remains the same in the stationary case. A discussion on the role of stationarity is included in Sec.~\ref{sec:discussion}.

\section{Results} \label{sec:results}

\subsection{Mutual Information} \label{sec:results_MI}

We begin our analysis by looking at the time-delayed mutual information (TDMI) between various elements of the system of $N>2$ RWs. 
Since its introduction by Fraser and Swinney \cite{fraser1986independent}, TDMI has become a crucial quantity in the study of the information dynamics of complex systems. 
TDMI quantifies the statistical dependence between a variable and a time-delayed version of another, reflecting how information from a past state contributes to predicting a future state. 
This metric is commonly used to reveal underlying temporal structures, finding applications in numerous fields \cite{james2011anatomy}.
In our case, given two timepoints $t,\tp$, with $\tp>t$, we compute the TDMI between RWs, c.o.m.\@, and RWs and c.o.m.\@
Referring to App.~\ref{app:N-RWcase_MI} for the mathematical details of the derivation, the functional expressions read
\begin{align}
\allowdisplaybreaks
    \label{eq:N_MI_xx}
    I(\Delta_i(t);\Delta_i(t^{\prime})) &= \frac{1}{2}\ln{\left(\frac{\tau}{\tau-1}\right)}-\gamma t +\mathcal{O}(\gamma^2)\,, \\
    \label{eq:N_MI_xixj} 
    I(\Delta_i(t);\Delta_{j}(t^{\prime})) &= \mathcal{O}(\gamma^2) \,,\quad i\neq j \\
    \label{eq:N_MI_vv}
    I(V(t);V(t^{\prime})) \,\,\, &= \frac{1}{2}\ln{\left(\frac{\tau}{\tau-1}\right)}+\mathcal{O}(\gamma^2) \, , \\ %
    \label{eq:N_MI_xv}
    I(\Delta_i(t);V(t^{\prime})) &= \frac{1}{2}\ln{\left(\frac{N\tau}{N\tau-1}\right)}+\gamma \frac{t-1}{N\tau-1}+\mathcal{O}(\gamma^2) \,, \\
    \label{eq:N_MI_vx}
    I(V(t);\Delta_i(t^{\prime})) &= \frac{1}{2}\ln{\left(\frac{N\tau}{N\tau-1}\right)}+\gamma \frac{t^{\prime}-1}{N\tau-1}+\mathcal{O}(\gamma^2) \,,
\end{align}
where $\tau\equiv t^{\prime}/t$.
Eqs.~\eqref{eq:N_MI_xx}-\eqref{eq:N_MI_vx} showcase the dependency of TDMI on the interaction strength of the system: for larger $\gamma$, ${I(\Delta_i(t),\Delta_i(t^{\prime}))}$ decreases, as the single RW predicts less of itself as its motion is increasingly affected by the other walkers, whereas $I(V(t),V(t^{\prime}))$ is independent of $\gamma$, as expected since $V$ behaves as an autonomous RW (c.f.\ Eq.~\eqref{eq:eom_V}). 
Both $I(\Delta_i(t),V(t^{\prime}))$ and $I(V(t),\Delta_i(t^{\prime}))$ increase with $\gamma$, as stronger interactions make the different parts more interdependent and therefore more explainable if one of them is known. 
We note that the exact calculations for the $N=2$ case lead to the same conclusions (App.~\ref{app:2-RWcase_MI}). 
Finally, the TDMI between different walkers vanishes. This is a spurious consequence of the perturbative approach at first order in $\gamma$. In fact, in the $N=2$ RW case $I(\Delta_i(t),\Delta_{j}(t^{\prime}))>0$ for $\gamma>0$, increasing with the interaction strength as one would expect (see Eq.~\eqref{eq:2_rw_mi_xixj}).

\subsection{Transfer Entropy}
Transfer entropy (TE) is an information-theoretic metric used to quantify how the past of one variable affects the future of another variable \cite{schreiber2000measuring}. 
It is often seen as a measure of the information transfer between two stochastic processes, capturing the directed flow of information over time from one variable to another. 

More precisely, given two stochastic processes $X$ and $Y$, TE quantifies the additional information that the past of $X$ contributes to predicting the future of $Y$, beyond what is already explained by the past of $Y$ alone.
For a system with Markovian dynamics -- such as the one discussed here -- TE can be expressed in terms of Shannon entropies as 
\begin{align}
    \mathcal{T}(X(t);Y(t+1))=& H(Y_{t+1}|Y_{t})-H(Y_{t+1}|Y_{t},X_{t}) \,. 
        \label{TE_def}
\end{align}

Referring to Apps.~\ref{app:N-RWcase_TE}-\ref{app:2-RWcase_TE} for the calculations, and denoting the future timestep as $\tp=t+1$, 
we obtain that the transfer entropy of the RWs and the c.o.m.\ vanish at first order in $\gamma$:
\begin{equation} \label{eq:N_case_TEs}
\begin{aligned}
    \mathcal{T}(\Delta_i(t); \Delta_j(t^{\prime})) & =
    \mathcal{T}(\Delta_j(t); V(t^{\prime})) \\
    & = \mathcal{T}(V(t); \Delta_j(t^{\prime})) = \mathcal{O}(\gamma^2) \,.
\end{aligned}
\end{equation}
This is related to the behaviour of Eq.~\eqref{eq:N_MI_xixj},
and it is a consequence of TE being a second-order quantity in $\gamma$ (as was the case for ${I(\Delta_i(t);\Delta_{j}(t^{\prime})), \,\, i\ne j}$). 
From a practical point of view, however, this suggests that TE may struggle to detect very weak interactions -- although this is an expected result of the geometric nature of TE~
(Sec.~\ref{sec:considerations}).

\begin{figure}[ht]
\centering
    \includegraphics[width=1\columnwidth]{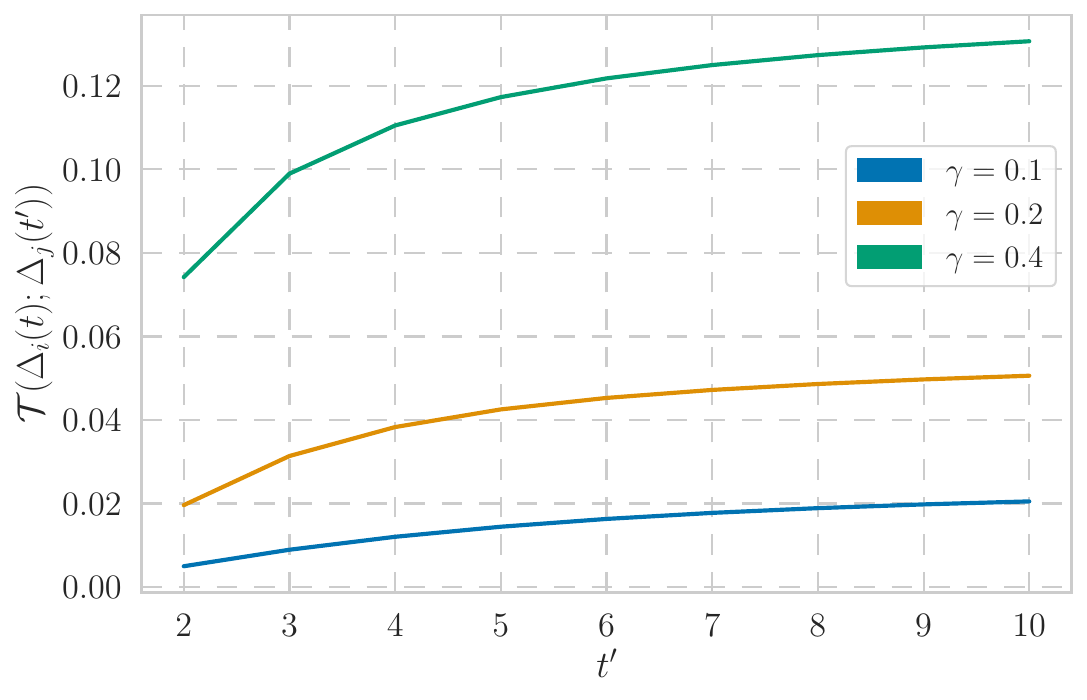}
    \caption{\textbf{Transfer entropy between random walkers increases with $\gamma$ as $\mathcal{O}(\gamma^2)$ for $N=2$ RWs.}
    For the system of $N=2$ RWs, the transfer entropy between the two random walkers $\mathcal{T}(\Delta_i(t); \Delta_j(t^{\prime}))$ increases with the system's interaction strength (App.~\ref{app:2-RWcase_TE}). This also corresponds to the TE between c.o.m.\ and a random walker $\mathcal{T}(V(t); \Delta_j(t^{\prime}))$. Colours represent different values of $\gamma$, and $\tp\equiv t+1$. 
    }
\label{fig:TE_img}
\end{figure}

As expected, if we consider the exact results for the system of $N=2$ random walkers, in which higher-order contributions in $\gamma$ are included, we obtain that the transfer entropies $\mathcal{T}(\Delta_i(t); \Delta_j(t^{\prime}))$ become non-zero (see Eqs.~\eqref{eq:2RW_te_xx}-\eqref{eq:2RW_te_vx}).
Moreover, we note that the information flow between the random walkers increases for larger coupling parameter $\gamma$ and for any timepoint (Fig.~\ref{fig:TE_img}). 

Additionally, this is also equal to the TE between the c.o.m.\ and the single walker, since the knowledge of the position of the two walkers completely determines the position of the c.o.m.\@, i.e.\ ${H(\Delta_j(t^{\prime})|\Delta_j(t), \Delta_i(t))=H(\Delta_j(t^{\prime})|\Delta_j(t), V(t))}$. This behaviour is expected, as it resembles the increase in the interconnectedness of the RWs due to stronger interactions. 

Finally, the TE from a RW to $V$ is always zero, as measuring $\Delta_j(t)$ does not provide any more information about $V(t^{\prime})$ than what is given by $V(t)$. In fact, we have that for all values of $\gamma$
\begin{equation}
    H(V(t')|V(t),\Delta(t))=H(V(t')|V(t)) \, , 
\end{equation}
and therefore 
\begin{equation}\label{eq:TE(D,V)}
    \mathcal{T}(\Delta_j(t); V(t^{\prime})) = 0 \quad \forall\gamma \,.
\end{equation}
This result emphasises the statistically autonomous dynamics of the c.o.m.\ as a free random walker. This property is also known as \textit{information closure} \cite{bertschinger2006information}, and is the foundation of some formulations of emergence \cite{barnett2023dynamical}. 

\subsection{Emergence estimators} \label{sec:results_emergence}

One promising aspect of information theory is that it allows us to quantify how \textit{emergent} a system is, i.e.\ how collective phenomena arise from the interaction of a group of elementary constituents. 
One novel approach to estimating causal emergence has been suggested by Rosas, Mediano \etal\ \cite{rosas_reconciling_2020,mediano2022greater}, who provide a mathematical framework and practical criteria to evaluate the presence and type of emergence. Here we briefly revisit some of the key concepts of the theory, slightly changing the original notation to match our scenario.

Given a macroscopic feature of the system $V$ and $N$ microscopic components $\{\Delta_j\}_{j=1}^N$, three measures 
are introduced to investigate whether $V$ is an emergent property of the system
(c.f.\ Eqs.~10a-10c in \cite{rosas_reconciling_2020}):
\begin{align} 
    \Psi_{t,t'}^{(1)} (V) &= I(V(t);V(t^{\prime})) - \sum_{j=1}^N I(\Delta_j(t); V(t')) \label{eq:psi_emergence} \\
    \Delta^{(1)}_{t,t'} (V) &= \begin{aligned}[t] 
    \max_i \Bigl( I(V(t);&\Delta_i(t')) + \\
    &- \sum_{j=1}^N I(\Delta_j(t);\Delta_i(t')) \Bigr) 
    \end{aligned} \\
    \Gamma^{(1)}_{t,t'} (V) &= \max_i \Bigl(I(V(t); \Delta_i(t'))\Bigr) \,.
\end{align}
For ease of notation, we refer to these quantities as $\Psi_{t\tp}$, $\Delta_{t\tp}$, $\Gamma_{t\tp}$, respectively.
Based on these estimators, the theory~\cite{rosas_reconciling_2020} provides three sufficient conditions to determine the emergent nature of $V$: 
\begin{enumerate}
    \item[(1)] $\Psi_{t\tp} > 0 $ is sufficient for $V$ to be causally emergent.
    \item[(2)] $\Delta_{t\tp} > 0 $ is sufficient for $V$ to have downward causation (i.e. influence on the microscopic components $\Delta_j$).
    \item[(3)] $\Psi_{t\tp} > 0 $ and $\Gamma_{t\tp} = 0 $ are sufficient for $V$ to be causally decoupled (i.e. $V$ has information about itself that is not present in the individual $\Delta_j$).
\end{enumerate}
Given the mutual information above (Eqs.~\eqref{eq:N_MI_xx}-\eqref{eq:N_MI_vx}), we can easily calculate $\Psi_{t \tp}, \Delta_{t \tp}, \Gamma_{t \tp}$ using the c.o.m.\ as $V$. 
Referring for the full calculations to App.~\ref{app:N-RWcase_emergence}, here we report the results for the infinite size limit $N\rightarrow\infty$: 
\begin{align}
\allowdisplaybreaks
    \Psi_{t\tp} & \underset{N\rightarrow\infty}\longrightarrow \underbrace{\frac12\left[\ln{\left(\frac\tau{\tau-1}\right)}-\frac1\tau\right]}_{\Psi^0_{t\tp}} - \dfrac{\gamma(t-1)}{\tau} \,, 
    \\ %
    \Delta_{t\tp} & \underset{N\rightarrow\infty}\longrightarrow \underbrace{-\frac12\ln{\left(\frac\tau{\tau-1}\right)}}_{\Delta^0_{t\tp}} + \gamma t \,,
    \\
    \Gamma_{t\tp} & \underset{N\rightarrow\infty}\longrightarrow \underbrace{0}_{\Gamma_{t\tp}^0} \,,
\end{align}
where we defined $\Psi^0_{t\tp}, \Delta^0_{t\tp}, \Gamma^0_{t\tp}$ as the three estimators in the non-interacting case ($\gamma=0$).
Interestingly, we notice that for large systems $\Psi^0_{t\tp}>0$ and $\Gamma^0_{t\tp}=0$, meaning that, according to the sufficient condition (3), $V$ is emergent and a causally decoupled feature of the system. 
This may seem like a surprising result, as the centre of mass of non-interacting constituents is not emergent in any sense that involves higher-order interactions, as there are none. Rather, this result seems to be reflecting the autonomous nature of $V(t)$, since it follows a dynamical evolution of a free RW (Eq.~\ref{eq:eom_V}). From this point of view, we can make sense of this result as $V(t)$ containing more information about $V(\tp)$ than any of the $\Delta_j(t)$ taken independently -- making it nearly causally decoupled for $N<\infty$ (c.f.\ Eq.~\eqref{eq:app_NRW_gamma}) and exactly decoupled for $N\rightarrow\infty$.
On the other hand, ${\Delta^0_{t\tp}<0} \,\,\text{for} \,\,{N>1}$ (Eq.~\eqref{eq:app_NRW_delta}) suggests no downward causation. 

Now turning on the interactions ($\gamma>0$), we observe that $\Psi_{t\tp}$ decreases while $\Delta_{t \tp}$ and $\Gamma_{t \tp}$ increase, but none changes sign, hence not affecting the conclusions above. However, if we interpret $\Psi_{t\tp}$ as the level of causal emergence of the feature $V$, the decrease of $\Psi_{t\tp}$ w.r.t.\ $\gamma$
implies that $V$ is in some way less emergent when the random walkers interact.
Formally, this is due to $I(V(t); V(t^{\prime}))$ not being dependent on $\gamma$, while $I(\Delta_j(t); V(t^{\prime}))$ increases with it. Following the original theory~\cite{rosas_reconciling_2020}, this is where one can see the limitations of $\Psi_{t\tp}$ as a sufficient (but not necessary) criterion for emergence -- one can interpret its sign as the presence of emergent (or, in this case, autonomous) behaviour, but not its magnitude as an overall level of emergence.

Overall, this is a clear example in which higher-order behaviours do not correspond to higher-order mechanisms: $\Psi_{t\tp}$ is capturing behaviour that is dynamically autonomous, but not generated by higher-order interactions.
The relationship between high-order interactions, emergence, and autonomy is a nuanced open problem, crucial for understanding results from real complex systems (Sec.~\ref{sec:autonomy}).

\subsection{Partial Information Decomposition} \label{sec:results_pid}
Partial Information Decomposition (PID) \cite{williams2010nonnegative} is a novel framework that provides a fine-grained characterisation of how information is distributed among multiple components of a system.
PID has been applied widely to the study of artificial neural networks \cite{beer2015information, wibral2017partial, ince2017measuring, tax2017partial, proca2024synergistic}, neural dynamics \cite{luppi2020synergisticCore}, cellular automata \cite{finn2018quantifying, rosas2018information}, and more. 

Formally, given three stochastic variables, of which two are denoted as sources ($X,Y$) and one as target ($Z$), the defining equations of PID can be written as
\begin{align}
    I(X;Z) & = \text{Un}(X;Z) + \text{Red}(X,Y;Z)\,,\label{eq:PID1} \\
    I(Y;Z) & = \text{Un}(Y;Z) + \text{Red}(X,Y;Z)\,, \label{eq:PID2} \\
    I(X,Y;Z) & \begin{aligned}[t] & 
        = \text{Un}(X;Z) + \text{Un}(Y;Z) + \\ 
        & + \text{Red}(X,Y;Z) + \text{Syn}(X,Y;Z) \,,\label{eq:PID3}
    \end{aligned}
\end{align}
where $\text{Un}(X;Z)$ and $\text{Un}(Y;Z)$ are the unique information about $Z$ only provided by $X$ and $Y$, respectively, $\text{Red}(X,Y;Z)$ is the redundant information provided by both, and $\text{Syn}(X,Y;Z)$ is the synergistic information only accessible when both $X$ and $Y$ are considered jointly. 
Since this is an undetermined system of 3 equations and 4 unknowns, an additional equation is needed. Hence, numerous definitions of synergistic, unique, or redundant information have been proposed in the literature~\cite{williams2010nonnegative, bertschinger2013shared, griffith2014quantifying, harder2013bivariate, barrett2015exploration, ince2017measuring, james2018unique, griffith2014intersection, griffith2015quantifying, quax2017quantifying, rosas2020operational}.

Due to its generality and simplicity, for the following analysis we employ the Minimal Mutual Information (MMI) formulation of PID \cite{barrett2015exploration}, in which the redundancy is defined as
\begin{equation} 
\begin{aligned}
    \text{Red}(X,Y;Z) & = \text{Red}_{\text{MMI}}(X,Y;Z) \\ 
    & \equiv \text{min}\Bigl(I(X;Z), I(Y;Z)\Bigr) . \label{eq:redMMI}
\end{aligned}
\end{equation}
Adding this to the system of equations in Eqs.~\eqref{eq:PID1}-\eqref{eq:PID3} leads to the following expression for synergy
\begin{equation}
\begin{aligned}
    \text{Syn}(&X,Y;Z) = \text{Syn}_{\text{MMI}}(X,Y;Z) \\ 
    & = I(X,Y;Z) - \text{max}\Bigl(I(X;Z), I(Y;Z)\Bigr) \,. \label{eq:synMMI}
\end{aligned}
\end{equation}
In practical data analysis settings, the MMI PID has been shown to agree with various other redundancy measures~\cite{mediano2021towards, rosas2020operational, liardi2024null}, and has been extensively employed to study real data in fields spanning from neuroscience \cite{luppi2020synergisticCore} to economics \cite{rajpal2023synergistic}. 

With this definition, we can calculate PID on the system of $N>2$ and $N=2$ RWs, examining in both cases how synergy and redundancy behave when the interaction strength varies. 
Leaving the computations for Apps.~\ref{app:N-RWcase_PID}-\ref{app:2-RWcase_PID}, here we present the results for 3 distinct PID scenarios, in which different sources and targets are considered. 

\begin{figure*}[ht]
\centering
\begin{subfigure}[]{.49\textwidth}
    \centering
    \figuretitle{RWs - c.o.m.}
\hspace*{-1.5mm}\includegraphics[width=1\textwidth]{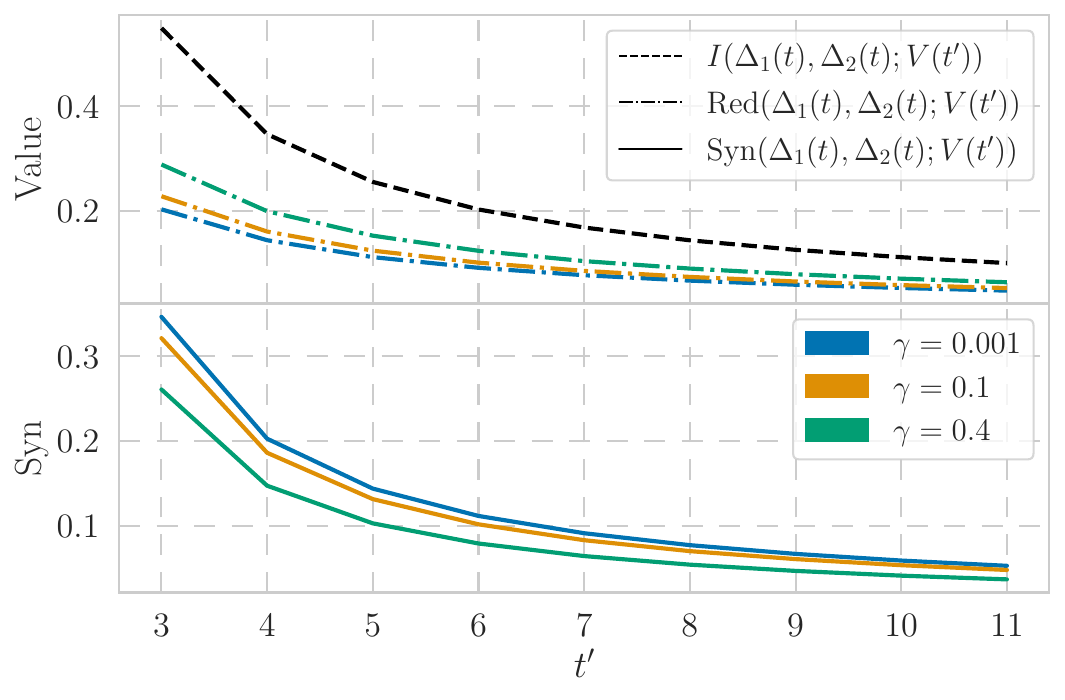}
    \caption{
    }
    \label{fig:PID1_img}
\end{subfigure}%
\hspace*{-1mm}
\begin{subfigure}[]{.49\textwidth} 
    \centering
    \figuretitle{RWs - RWs}
\hspace*{-1.5mm}\includegraphics[width=1\textwidth]{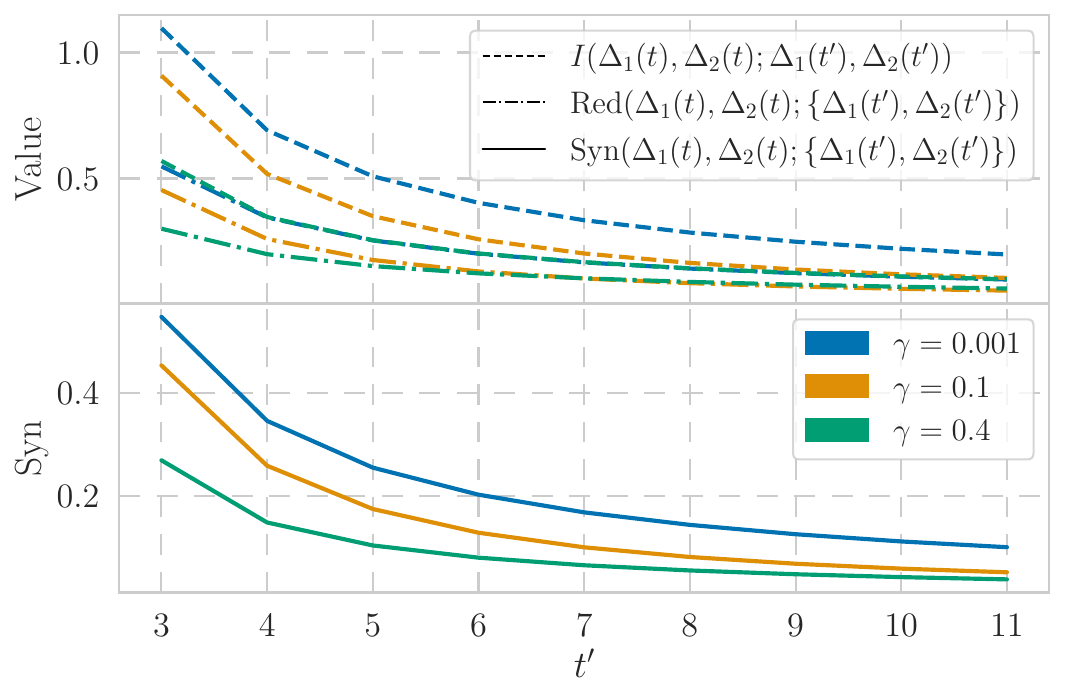}
    \caption{ 
    }
    \label{fig:PID2_img}
\end{subfigure}%
\caption{\textbf{Interaction strength increases correlations, thus reducing synergy in small systems.}
Partial Information Decomposition on the system of $N=2$ random walkers (App.~\ref{app:2-RWcase}). (a) PID with $X=\Delta_j(t)$, $Y=\Delta_i(t)$, $Z=V(t^{\prime})$, $i\ne j$ (case \textbf{(\underline{1})}): redundancy correlates with the interaction strength, while synergy anticorrelates. (b) PID with ${X=\Delta_j(t)}$, ${Y=V(t)}$, $Z=\{\Delta_j(t^{\prime}), V(t^{\prime})\}$, $i\ne j$ (case \textbf{(\underline{2})}), which also corresponds to the PID with ${X=\Delta_j(t)}$, ${Y=\Delta_i(t)}$, $Z=\{\Delta_j(t^{\prime}), \Delta_i(t^{\prime})\}$, $i\ne j$ (case \textbf{(\underline{3})}): TDMI, synergy, and redundancy anticorrelate with the interaction strength. 
The dashed lines correspond to the joint mutual information, the dash-dotted lines to redundancy, and the solid lines to synergy. Colours represent different values of $\gamma$, while the black line is independent of $\gamma$. Results are shown for $t=2$.}
\label{fig:PID_imgs}
\end{figure*}%

\textbf{(\underline{1})} We start by examining a PID with two random walkers as sources and the c.o.m.\ as target:
\begin{equation} \label{eq:pid_var_1}
    \begin{split}
        X&=\Delta_j(t), \quad Y=\Delta_i(t), \quad\text{and} \\
        Z&=V(t^{\prime}), \, i\ne j \,. 
    \end{split}
\end{equation}
Making use of Eq.~\eqref{eq:N_MI_xv}, for the $N>2$ random walkers we obtain
\begin{align}
    I(\Delta_j(t),\Delta_i(t);V(t^{\prime})) & 
    \begin{aligned}[t]
        = & \frac{1}{2}\ln{\left(\frac{N\tau}{N\tau-2}\right)}+\\
        & +\gamma \frac{t-1}{N\tau-2} 
    \end{aligned} \\
    \text{Red}(\Delta_j(t),\Delta_i(t);V(t^{\prime})) &
    \begin{aligned}[t]
        = & \frac{1}{2}\ln{\left(\frac{N\tau}{N\tau-1}\right)}+\\
        & + \gamma \frac{t-1}{N\tau-1}
    \end{aligned} \\
        \text{Syn}(\Delta_j(t),\Delta_i(t);V(t^{\prime})) & = 
        \begin{aligned}[t]
        & \frac{1}{2}\ln{\left(\frac{N\tau-1}{N\tau-2}\right)} +\\
        & +\gamma \left(\frac{t-1}{N\tau-2}- \frac{t-1}{N\tau-1}\right) \,. \label{eq:PID1_syn}
    \end{aligned}
\end{align}
We show the exact results for these quantities in the $N=2$-RW case in Fig.~\ref{fig:PID1_img}. 
Interestingly, we observe that the behaviour of synergy changes qualitatively in the two scenarios, increasing with $\gamma$ in the $N>2$ case (c.f.\ Eq.~\eqref{eq:PID1_syn}),
and decreasing for $N=2$.  
This suggests that considering a larger number of walkers in the system varies the relation between synergy and $\gamma$. 
On the other hand, redundancy grows with the coupling $\gamma$, both in the $N>2$ and $N=2$ cases.

These results slightly change if the PID atoms are normalised. 
In fact, even a simple normalisation procedure -- such as dividing each PID atom by the joint MI -- shows that normalised synergy decreases with the coupling strength in both $N>2$ and $N=2$ cases, while normalised redundancy still increases (App.~\ref{app:N-RWcase_PID}-\ref{app:2-RWcase_PID}).
The necessity of a normalisation is due to the joint mutual information increasing with $\gamma$, as the c.o.m.\ can be more easily predicted if the walkers are strongly coupled. Consequently, since a change in joint mutual information also affects the magnitude of the raw PID values, comparing PID quantities for systems with different $\gamma$ requires appropriate normalisation techniques~\cite{liardi2024null}.

\begin{figure*}[ht]
\centering
\begin{subfigure}[]{.49\textwidth}
    \centering
    \figuretitle{\hspace{6mm}Original system}
\hspace*{-1.5mm}\includegraphics[width=1\textwidth]{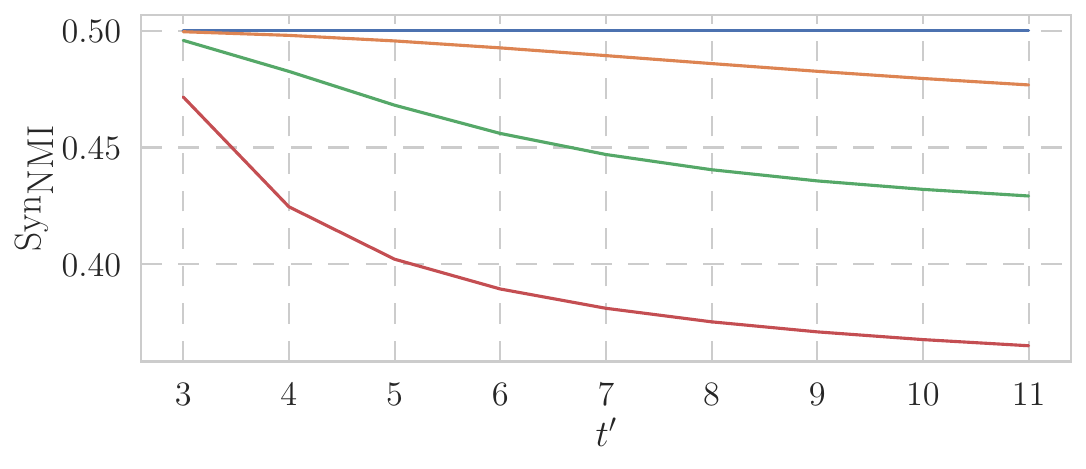}
    \caption{
    }
\end{subfigure}%
\hspace*{-1mm}
\begin{subfigure}[]{.49\textwidth} 
    \centering
    \figuretitle{\hspace{6mm}Perturbed system}
\hspace*{-1.5mm}\includegraphics[width=1\textwidth]{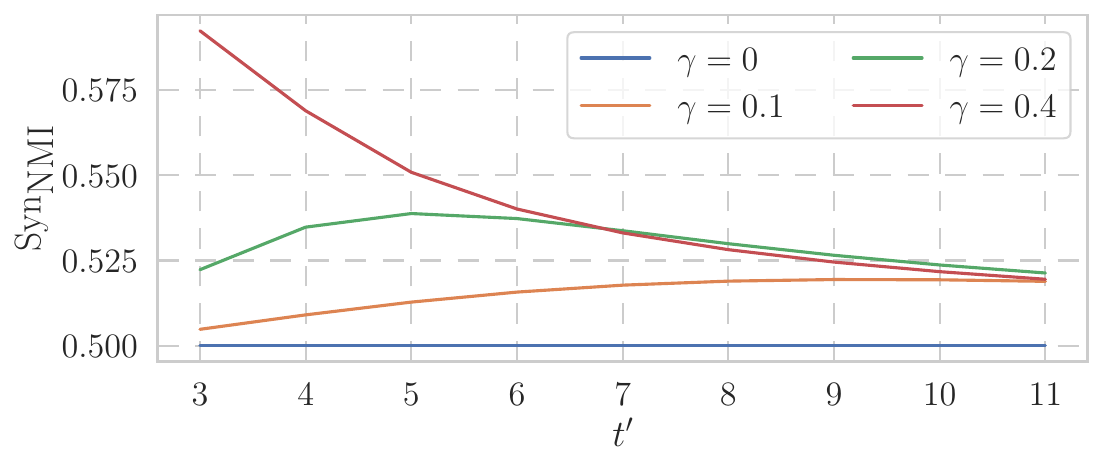}
    \caption{ 
    }
\end{subfigure}%
\caption{\textbf{Causal intervention can disentangle dynamical and steady-state informational contributions: normalised synergy increases with $\gamma$ in the perturbed system.}
Normalised synergy for the PID with ${X=\Delta_j(t)}$, ${Y=\Delta_i(t)}$, $Z=\{\Delta_j(t^{\prime}), \Delta_i(t^{\prime})\}$, $i\ne j$ on the system of $N=2$ random walkers (a) decreases in the original system (App.~\ref{app:2-RWcase}), and (b) increases in the causally perturbed system (App.~\ref{app:2-RWcase_intervention}). Results are shown for $t=2$.}
\label{fig:PID_imgs_NMI_intervention_synergy}
\end{figure*}%

\textbf{(\underline{2})} We continue our study considering a RW as one source, the c.o.m.\ as the other, and their joint future as the target. Mathematically, we have
\begin{equation} \label{eq:pid_var_3}
    \begin{split}
        X&=\Delta_j(t), \quad Y=V(t), \quad \text{and} \\
        Z&=\{\Delta_j(t^{\prime}), V(t^{\prime})\} \, .
    \end{split}
\end{equation}
This leads to
\begin{align}
\begin{split}
    I(\Delta_j(t), V(t);&\{\Delta_j(t^{\prime}),V(t^{\prime})\}) = \\
    = & \ln{\left(\frac{\tau}{\tau-1}\right)} - \gamma \frac{Nt}{N-1}
\end{split} \\
\begin{split}
    \text{Red}(\Delta_j(t), V(t);&\{\Delta_j(t^{\prime}),V(t^{\prime})\}) = \\
    = & \frac{1}{2}\ln{\left(\frac{\tau}{\tau-1}\right)}-\gamma t 
\end{split} \\
\begin{split}
    \text{Syn}(\Delta_j(t), V(t);&\{\Delta_j(t^{\prime}),V(t^{\prime})\}) = \\
    = & \frac{1}{2}\ln{\left(\frac{\tau}{\tau-1}\right)} -\gamma \frac{Nt}{N-1} \,. 
\end{split}
\end{align}
For both $N=2$ and $N>2$ RWs, synergy and redundancy decrease for larger $\gamma$, as does the joint mutual information (Fig.~\ref{fig:PID2_img}). 
If we examine the normalised PID atoms, we notice that normalised redundancy and synergy agree with the non-normalised quantities for the $N>2$ case, while normalised redundancy for $N=2$ becomes bigger with larger $\gamma$ (App.~\ref{app:N-RWcase_PID}-\ref{app:2-RWcase_PID}). As above, this result suggests that the size of the system also plays a role in the reorganisation of its information dynamics. 

\textbf{(\underline{3})} Finally, we turn to the more interesting case in which we have the random walkers as sources and their joint future as target variable:
\begin{equation} \label{eq:pid_var_2}
\begin{split}
    X&=\Delta_j(t), \quad {Y=\Delta_i(t)}, \quad\text{and}\\ Z&=\{\Delta_j(t^{\prime}), \Delta_i(t^{\prime})\}, \,i\ne j \,.
\end{split}
\end{equation} 
Since in the $N>2$ RW case there is no mutual information between different walkers at first order in $\gamma$ (Eq.~\eqref{eq:N_MI_xixj}), then the joint mutual information, redundancy, and synergy simply become
\begin{align}
    & I(\Delta_i(t),\Delta_j(t);\Delta_i(t^{\prime}),\Delta_j(t^{\prime})) = \ln{\left(\frac{\tau}{\tau-1}\right)}-2\gamma t \\
    \begin{split} \label{eq:NRW_red_syn_PID2}
    & \text{Red}(\Delta_i(t),\Delta_j(t);\{\Delta_i(t^{\prime}),\Delta_j(t^{\prime})\}) = \\
    & \quad\quad\quad\quad = \text{Syn}(\Delta_i(t),\Delta_j(t);\{\Delta_i(t^{\prime}),\Delta_j(t^{\prime})\}) \\
    & \quad\quad\quad\quad = \frac{1}{2}\ln{\left(\frac{\tau}{\tau-1}\right)}-\gamma t
    \end{split}
\end{align}
For the $N=2$ RWs, this case is analogous to \textbf{(\underline{2})}, as for $N=2$ the knowledge of $\{\Delta_j(t^{\prime}), V(t^{\prime})\}$ determines $\{\Delta_1(t^{\prime}), \Delta_2(t^{\prime})\}$, and viceversa. Thus, we refer again to Fig.~\ref{fig:PID2_img}. 
Note that the apparent symmetry between synergy and redundancy of Eq.~\eqref{eq:NRW_red_syn_PID2} is only true for small $\gamma$: when the interaction increases, then redundancy dominates. This behaviour can indeed be observed in the case of $N=2$ RWs (Fig.~\ref{fig:PID2_img}). 
In this and the $N>2$ scenario, we observe again that joint mutual information, synergy, and redundancy all decrease for larger coupling strength $\gamma$ and bigger timesteps. 

However, the normalised atoms reveal an important fundamental behaviour: normalised redundancy increases with the coupling strength, while normalised synergy decreases with it\footnote{The comment above refers exclusively to the $N=2$ case. For $N>2$, the first order approximation in $\gamma$ ensures symmetry between redundancy and synergy, and hence their normalised value is just $1/2$ (App.~\ref{app:N-RWcase_PID}).} (App.~\ref{app:2-RWcase_PID}). 
This can be well understood by looking at the two extreme cases: if two RWs are independent, then ${I(\Delta_1(t),\Delta_2(t);\Delta_1(t^{\prime}),\Delta_2(t^{\prime}))=2I(\Delta_j(t);\Delta_j(t^{\prime}))}$, thus the total mutual information is equally split between synergy and redundancy, both amounting to ${I(\Delta_j(t);\Delta_j(t^{\prime}))}$. 
On the other hand, if the walkers are rigidly connected, then the knowledge of one completely determines the other, hence ${I(\Delta_1(t),\Delta_2(t);\Delta_1(t^{\prime}),\Delta_2(t^{\prime}))=I(\Delta_j(t);\Delta_j(t^{\prime}))}$, meaning that all the information is redundant, and synergy vanishes. 
In other words, normalised synergy decreases with the coupling strength $\gamma$, while normalised redundancy increases. 
Therefore, within the MMI PID, synergy stands more as a measure of \textit{complementarity} \cite{bertschinger2014quantifying} than being related to interacting mechanisms, as recently also noted by Varley \cite{varley2024considering}.

It is important here to remark that, in dynamical systems, redundancy and synergy can arise as a consequence of both the system's dynamical evolution and the structure of the statistical distributions of the components (see discussion in Sec.~\ref{sec:warnings}). 
This scenario is different from the commonly studied examples using logic gates~\cite{james2018unique,ince2017partial,griffith2014quantifying,rosas2020operational}, in which there is no dynamical evolution and the only contributions to synergy and redundancy arise from the channel itself. 
Therefore, in the system of RWs under examination, stronger interactions have two parallel effects: making the walkers' dynamics more intertwined and increasing the correlation between them. However, disentangling these effects is in general non-trivial.

\begin{figure*}[ht]
\centering
\begin{subfigure}[]{.45\textwidth}
    \centering
    \figuretitle{\hspace*{5mm}Integration between RWs}
\hspace*{-1.5mm}\includegraphics[width=1\textwidth]{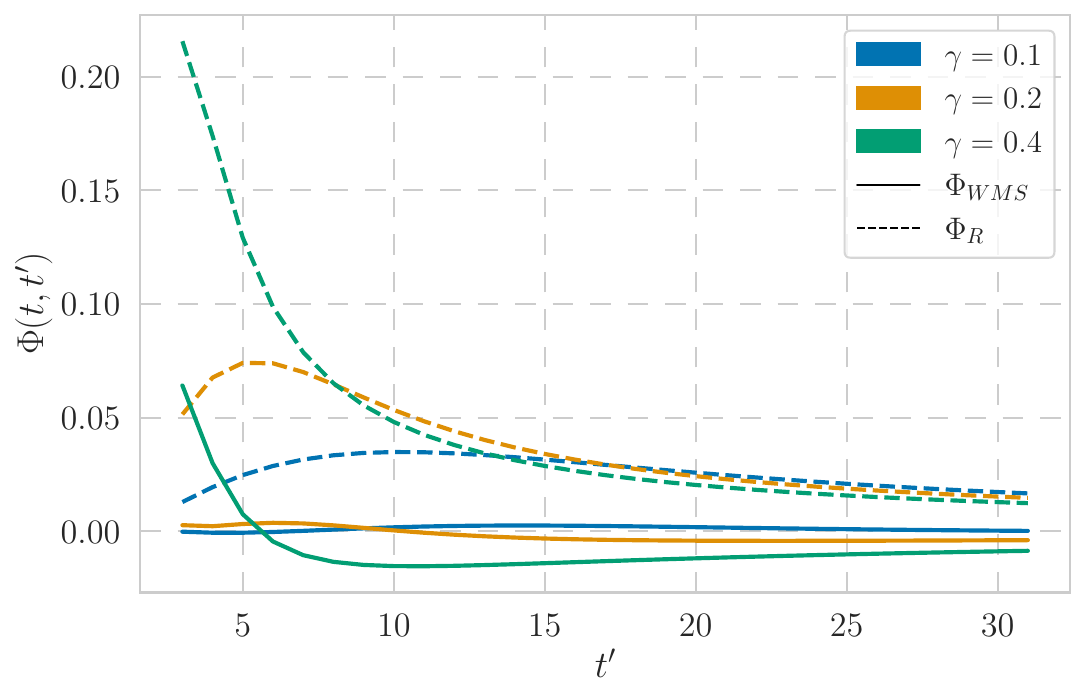}
    \caption{}
    \label{fig:Phi_2rws}
\end{subfigure}%
\hspace*{-1mm}
\begin{subfigure}[]{.45\textwidth} 
    \centering
    \figuretitle{\hspace*{10mm}Integration between RWs and c.o.m.\@}
    \hspace*{-1.5mm}\includegraphics[width=1\textwidth]{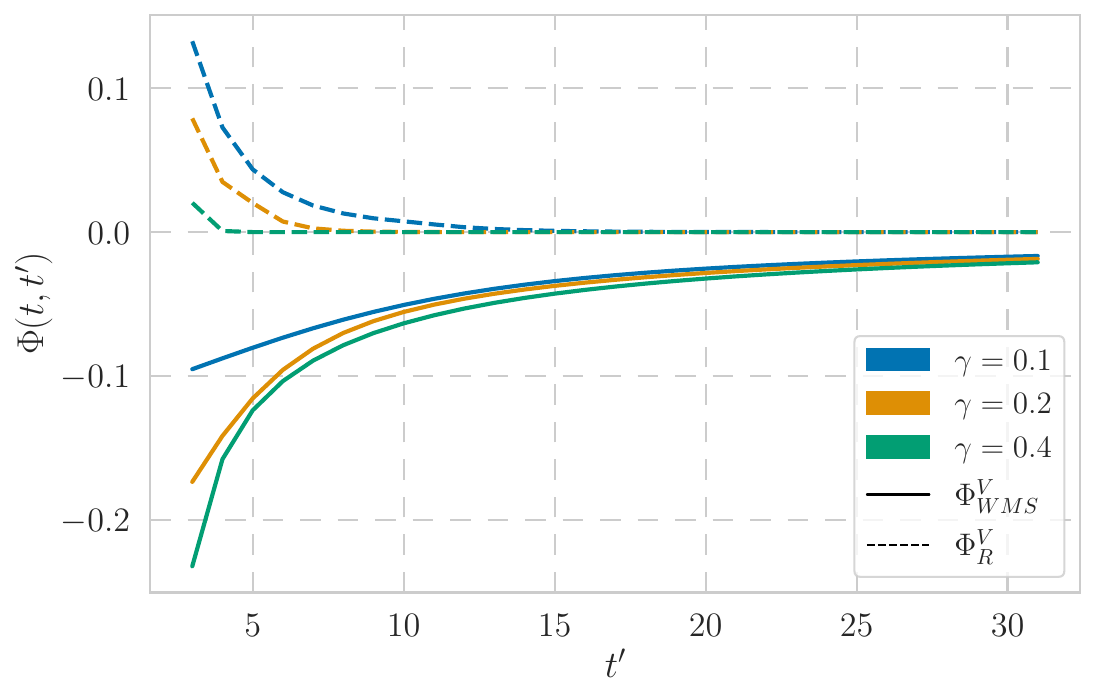}
    \caption{}
    \label{fig:Phiv_2rws}
\end{subfigure}%
\caption{\textbf{Integrated information between random walkers correlates with interaction strength at short timescales .} 
(a) Both whole-minus-sum (\wmsphi) and revised (\phiR) integrated information between random walkers correlate with interaction strength. (b) However, when the centre of mass is added to the calculation ($\wmsphi^V,\phiR^V$), this relationship is reversed. Results are shown for $t=2$.}
\label{fig:Phis_imgs}
\end{figure*}%

A valuable approach to unravelling mechanistic relations within a complex system is through causal intervention. Within this framework, the causal architecture of the system under study is determined through an analysis of the system's response to an external perturbation. 
In statistics, a possible formalisation of this approach is via the \textit{do-calculus} \cite{pearl1995causal, pearl2009causal, pearl2012calculus}. 
A natural way to implement this method is by examining the dynamical evolution of a system which unfolds from a maximum entropy state \cite{pearl2009causality, balduzzi2008integrated, rosas_information-theoretic_2018, ay2008information,orio2023dynamical}. 
By imposing a maximum entropy distribution onto the initial system, one can observe how the possible states of the system are constrained by its temporal evolution -- isolating its effect from that of the system's distribution at a given time.

Inspired by Ref.~\cite{barrett2011practical} (and leaving the details in App.~\ref{app:2-RWcase_intervention}), we apply this intervention paradigm to the system of $N=2$ RWs to isolate the informational contributions of the system's dynamics. In particular, we repeat the PID study above to compare how synergy and redundancy behave in the perturbed system. 
Interestingly, we notice that although the qualitative behaviour of the atoms does not significantly differ (App.~\ref{app:2-RWcase_intervention}), the normalised quantities %
do, as now synergy becomes higher for larger $\gamma$ (Fig.~\ref{fig:PID_imgs_NMI_intervention_synergy}), at least for short timescales ($\tp\lesssim7$). 
This example showcases the interplay between coupling and instantaneous distributions: although synergy induced by the dynamical evolution of the walkers increases with higher $\gamma$, in the unperturbed system such an effect is concealed by an increase in redundancy due to the RW positions becoming more correlated.

\subsection{Integrated information}

Finally, we conclude our analysis by looking at measures of integrated information.
The whole-minus-sum integrated information $\Phi_{\mathrm{WMS}}$ was introduced in \cite{balduzzi2008integrated, barrett2011practical} to quantify the information contained in the system as a whole, beyond what is present in the sum of its individual components.
Formally, given a multivariate process ${\bm{X}=(X^1,...,X^M)}$ of dimension $M$ and two timepoints $t,\tp,\,\tp>t$, $\Phi_{\mathrm{WMS}}$ is defined\footnote{In the original paper, $\Phi_{\mathrm{WMS}}$ is calculated with respect to the so-called minimum information partition~\cite{balduzzi2008integrated}. For simplicity, here we use the atomic partition -- see Ref.~\cite{mediano2018measuring} for details.} as
\begin{equation}
    \Phi_{\mathrm{WMS}}(\bm{X};t,\tp) = I(\bm{X}_t;\bm{X}_{t^{\prime}}) - \sum_{i=1}^M I(X^i_t;X^i_{t^{\prime}}) \,.
    \label{eq:phi_wms}
\end{equation}
Through an extension of PID, Mediano, Rosas \etal~\cite{mediano2021towards} showed that $\Phi_{\mathrm{WMS}}$ can be seen as a balance between synergy and redundancy in the system, such that $\Phi_{\mathrm{WMS}}$ is negative if the system is redundancy-dominated, and positive otherwise. Mediano, Rosas \etal\ later introduced $\Phi_{\mathrm{R}}$ \cite{mediano2021towards} as a refined version of $\Phi_{\mathrm{WMS}}$, in which the double-counted redundancy contributions are added back in the definition. In the case of 2 variables and using MMI, $\Phi_R$ reads  
\begin{align} \label{eq:phi_r_2}
    \Phi_{\mathrm{R}}(\bm{X};t,\tp) & = \Phi_{\mathrm{WMS}}(\bm{X};t,\tp) + \min_{i,j\in(1,2)} I(X^i_t,X^j_{t^{\prime}}) \,.
\end{align}
In our case study, we explore how $\Phi_{\mathrm{WMS}}$ and $\Phi_{\mathrm{R}}$ behave for $N=2$ random walkers, avoiding the $N>2$ case as results would be trivial due to the MI between walkers being zero at first order in $\gamma$ (Eq.~\eqref{eq:N_MI_xixj}).
Detailed computations can be found in App.~\ref{app:2-RWcase_Phi}. 

Considering only the deviations $\Delta_j$ for the computations of both $\Phi$, i.e.\ ${\Phi_{\mathrm{WMS}}\equiv\Phi_{\mathrm{WMS}}(\Delta_1,\Delta_2;t,\tp)}$ and ${\Phi_{\mathrm{R}}\equiv\Phi_{\mathrm{R}}(\Delta_1,\Delta_2;t,\tp)}$, 
we observe that at short timescales (${\tp\lesssim7}$) both quantities are larger for stronger interactions, as expected (Fig.~\ref{fig:Phi_2rws}). However, for bigger $\tp$ and in the limit ${\tp\to\infty}$ it can be seen that the ordering of these measures completely reverses, with higher integrated information for less interactive systems.

Since in real systems one may not know the exact relationship between microscopic and macroscopic quantities -- due to no a priori knowledge of the system's architecture or noisy measurements -- we now investigate how these measures vary if we consider the c.o.m.\ $V$ in place of one RW $\Delta_j$ in the $\Phi$ equations. Hence, we calculate ${\Phi_{\mathrm{WMS}}^V \equiv \Phi_{\mathrm{WMS}}(\Delta_j, V;t,\tp)}$ and ${\Phi_{\mathrm{R}}^V \equiv \Phi_{\mathrm{R}}(\Delta_j, V;t,\tp)}$ (Fig.~\ref{fig:Phiv_2rws}).

In this scenario, the introduction of $V$ increases the level of redundancy in the system, lowering the magnitude of both $\Phi^V$ measures. 
More importantly, we observe that both $\Phi_{\mathrm{WMS}}^V$ and $\Phi_{\mathrm{R}}^V$ display a fundamentally different behaviour than $\Phi_{\mathrm{WMS}}$ and $\Phi_{\mathrm{R}}$, now showing larger integrated information for weaker interactions, both on short and long timescales. 
This last finding is difficult to understand intuitively. A possible interpretation could rely on framing integrated information as a measure of complementarity between different parts, similar to the case of MMI PID synergy seen above. This could potentially explain why a more independent system shows more information integration.

Taken together, these results suggest that $\Phi$ measures correlate with the strength of interactions between RWs when two conditions hold: when they are computed in timescales comparable to those of the real dynamics of the system, and when only the microscopic constituents of the system (here, the $\Delta_j$) are included in the calculation. Furthermore, when these conditions are met, $\Phi_{\mathrm{R}}$ shows greater differentiation between interaction strengths than $\Phi_{\mathrm{WMS}}$.

\section{Discussion} \label{sec:discussion}

\subsection{Mechanistic interpretations of information-theoretic measures} \label{sec:mechanistic_interpr}

In this study, we analysed how various information metrics depend on the interactions present in a simple model of Gaussian random walkers (RWs), focusing on how these measures are related to the underlying interaction strength $\gamma$. 
When measured over a few timesteps, we found that both transfer entropy and integrated information which only include RWs positively correlated with the coupling $\gamma$. Somewhat counter-intuitively, we also found that synergy (measured with Minimum Mutual Information (MMI)~\cite{barrett2015exploration}) negatively correlated with interaction strength -- suggesting that, in some conditions, MMI synergy may be thought of as quantifying complementarity~\cite{milinkovic2024capturing} more than high-order interactions. Interestingly, however, synergy became positively correlated with interaction strength under a causal intervention, showing that the original negative correlation was due to the instantaneous interdependencies between walkers.

Taken together, our results suggest that, in this simple physical scenario, information measures correlate with the mechanistic properties of the system under specific conditions. More specifically, this seems to occur when: 
\begin{enumerate}  
    \item the measures are computed at timescales close to the intrinsic timescale of the system;  
    \item the informational contributions of steady-state and dynamics are disentangled, for example using causal intervention; and
    \item the measures are calculated solely on the RWs, ensuring that microscopic and macroscopic quantities are not mixed.  
\end{enumerate}

Condition 1 is motivated by the results in Figs.~\ref{fig:PID_imgs_NMI_intervention_synergy}-~\ref{fig:Phi_2rws} showing clearer correspondence with $\gamma$ at short timescales (as well as by a broader line of work exploring the behaviour of information measures across timescales~\cite{barnett2017detectability}). 
Condition 2 is supported by the results in Fig.~\ref{fig:PID_imgs_NMI_intervention_synergy}, showing that under a causal intervention the system is more synergistic for stronger interactions -- an effect which is usually concealed by the redundant contribution of the steady-state distribution of the RWs. 
Finally, condition 3 is motivated by the results in Figs.~\ref{fig:PID_imgs}-~\ref{fig:Phiv_2rws}, where introducing the centre of mass $V$ in the calculations inverted the relationship between information measures and coupling strength.

Although informed by our synthetic study, we do not know to what extent these conditions generalise to other systems. Verifying these properties in broader settings is a promising avenue for future work.

\subsection{Coarse-grainings and implications for empirical applications} \label{sec:coarse_graining}

In our analysis, we showed that macroscopic variables can have different information dynamics than their microscopic constituents. Our ability to interpret our findings was enabled by the fact that we are operating within a fully accounted for mathematical model that dictates the system's mechanics, and therefore the micro-macro relationship. However, in empirical scenarios, one is likely not afforded this luxury.

Specifically, we saw that the TDMI, PID, and $\Phi$ measures behave differently with respect to the interaction strength $\gamma$ depending on whether the walkers or the c.o.m.\ are considered. 
This phenomenon has direct implications for the application of such estimators to empirical data. For example, in electroencephalography (EEG) data, it can be the case due to volume conduction or electrolyte bridging that the signal read by a particular electrode is an average or mixed input from surrounding electrodes, rather than being the desired localised signal \cite{van1998volume}. Consequently, when analysing the information dynamics between two arbitrary electrodes, it is hard to tell whether these should be considered microscopic or macroscopic variables. Thus, such effects in empirical recordings can produce information-theoretic results leading to precisely opposing interpretations regarding the underlying mechanisms of the system (in this case, the activity of individual neurons).

More broadly, however, while remaining vigilant about the incidental mixing of scales may be considered a relatively easy task, relating results from reliably known scales is far more difficult \cite{barack2021two}. In fact, computing information-theoretic quantities on macroscopic structures whose generative processes are unknown might conceal unexpected statistical interdependencies, which will eventually be reflected in the measured information dynamics.
For example, it remains an open problem whether the BOLD signal obtained by functional magnetic resonance imaging (fMRI) is a faithful reflection of neural activity happening in a particular voxel \cite{logothetis2003underpinnings, ma2022gaining}. Accordingly, work linking BOLD signal fluctuations with features like ongoing calcium dynamics remains highly non-trivial, with evidence for both convergent and divergent functional organisations \cite{vafaii2024multimodal}. 

However, even if one can assume understanding of the data generative process, for instance that a local field potential (LFP) is the net voltage of the superposition of all surrounding synchronous membrane conductances \cite{reimann2013biophysically}, our results suggest that one should explicitly expect different information dynamics between these scales. For instance, the TDMI of the synaptic potentials may decrease with increased coupling, and the TDMI of the LFP may be independent of it (c.f.\ Eqs.~\eqref{eq:N_MI_xx}-\eqref{eq:N_MI_vv}).

Therefore, our findings highlight the difficulty of bridging empirical results from disciplines operating on complex systems at different levels of organisation.

\subsection{Causal emergence and autonomy}
\label{sec:autonomy}

In Sec.~\ref{sec:results} we showed that, according to Ref.~\cite{rosas_reconciling_2020}, the centre of mass (c.o.m.\@) of the random walkers satisfies the criteria for causal decoupling, and thus could be considered an emergent property of the system. This was shown by a positive value of the emergence criterion, $\Psi_{t \tp}>0$.
In other words, this indicates that the c.o.m.\ strongly determines its own dynamics, with a vanishingly small contribution of each RW taken individually.

This behaviour naturally matches the statistically autonomous nature of the c.o.m.\@, whose dynamics depend only on its previous states (Eq.~\eqref{eq:eom_V}), but clashes with the intuition of $\Psi_{t \tp}$ as quantifying higher-order mechanisms (since there are none, especially in the non-interacting system). This suggests that causal emergence -- defined by $\Psi_{t \tp}>0$ -- does not always depict higher-order mechanisms, as it can also be seen as a measure of autonomy~\cite{rosas2024software}. 
This result establishes an interesting connection between approaches to emergence based on synergy~\cite{rosas_reconciling_2020,mediano2022greater} and others based on dynamical independence~\cite{barnett2023dynamical, milinkovic2024capturing}, opening a promising avenue for future research. 

Moreover, we found that $\Psi_{t \tp}$ is larger when the walkers do not interact.\footnote{Formally, this is due to the c.o.m.\ $V$ acting as a free random walker, with $I(V(t);V(t^{\prime}))$ being independent of $\gamma$, whereas the term $I(\Delta_j(t);V(t^{\prime}))$ grows, thus decreasing the value of $\Psi_{t\tp}$ (Eq.~\eqref{eq:psi_emergence}).} 
Although this result may seem counter-intuitive, it highlights an important aspect of the original theory~\cite{rosas_reconciling_2020}: although $\Psi_{t \tp}>0$ is a sufficient condition for causal emergence, the magnitude of $\Psi_{t \tp}$ is not by itself indicative of the ``amount'' of emergence (or autonomy) in the system.

Hence, the notion of emergence provided by $\Psi_{t \tp}$ should not be interpreted in a mechanistic sense. 
In this case, thanks to the analytical tractability of the system, we could explain the emergent nature of the c.o.m.\ through its autonomy. Nonetheless, developing a measure that is sensitive only to emergence due to higher-order mechanisms (and not to autonomous dynamics) offers an interesting direction for future work.

\subsection{Technical considerations on stationarity and geometry}
\label{sec:considerations}

Although the model treated in the main text is non-stationary, the results presented above are qualitatively the same for the stationary Ornstein-Uhlenbeck (OU) process studied in Apps.~\ref{app:N-RWcase}-\ref{app:2-RWcase}. 
Interestingly, it is possible to show that the steady-state limit of a stationary OU process corresponds to a Vector Autoregressive (VAR) model, which has been studied extensively in the information theory literature~\cite{barnett2011behaviour,mediano2018measuring,chicharro2011spectral}. Solving the VAR model at first order in $\gamma$ yields qualitatively similar results to the ones presented above, and shows behaviours consistent with prior literature, both on transfer entropy~\cite{barnett2011behaviour} and integrated information~\cite{mediano2018measuring}.

Finally, it is worth discussing the implications of the fact that mutual information between RWs and transfer entropy are second-order quantities in the interaction parameter $\gamma$.
This result has a clear geometric interpretation: for any differentiable information measure with a minimum at $\gamma=0$, 
elementary calculus theorems guarantee that the first non-zero term in its expansion is of order $\mathcal{O}(\gamma^2)$. 
Thus, this also implies that any measure that can be written in terms of the Kullback-Leibler divergence (such as mutual information and transfer entropy) is a second-order measure in $\gamma$.
As a consequence, TE and MI might struggle to detect weak interactions, as their second-order dependence on the interaction strength increases the statistical power needed to distinguish weak interactions from noise. 

Following this reasoning, we hypothesise that information measures based on other probability distances (e.g.\ total variation distance~\cite{gibbs2002choosing}) may scale better with low values of $\gamma$, potentially resulting in estimators that are more reliable for large, very weakly interacting systems, which are common across the complex systems literature~\cite{garas2008structural, csermely2004strong, berlow1999strong, karlin1972polymorphisms, pichler2020machine}.

\section{Conclusion} \label{sec:conclusion} 
In this work, we provided an in-depth study of the relationship between information-theoretic measures and the mechanistic properties of a system of random walkers, investigating how the latter might be related to the former and developing some intuitions based on the model's physical parameters. 
To do this, we analysed some of the currently most established as well as novel frameworks in information theory, including transfer entropy~\cite{schreiber2000measuring}, PID~\cite{williams2010nonnegative} and emergence~\cite{rosas_reconciling_2020}.
The simplicity of the 1D Gaussian random walkers enabled us to derive analytical approximations of all information measures based on the model's dynamics, highlighting their functional dependence on the system's physical parameters, such as the interaction strength. 

Our results indicate that even in a simple scenario such as the one investigated here, establishing a direct link between information-theoretic quantities and the causal make-up of a complex system is challenging. 
However, we noticed that these statistical measures reflect more faithfully the system's mechanistic features when only microscopic quantities are used in the computations, when short timescales are considered, and when the system is subject to a maximum-entropy causal perturbation. In particular, we believe that future work on causal perturbations could help establish a deeper connection between behaviours and mechanisms~\cite{rosas2022disentangling}.

This study suffers from two important limitations: in the case of $N>2$ random walkers, we only consider interactions to the first order, effectively approximating the information measures to a regime of weak coupling, whereas with $N=2$ random walkers we obtain exact results, but lack a meaningful number of system components. Although we mitigated these shortcomings by investigating both scenarios together, future work should continue developing numerical and analytical treatments of information quantities.

We believe this study is a small step in a larger effort to link physical systems to statistical models, in the hope that physics can provide meaningful intuitions to guide the use and development of information-theoretic measures. 
We hope this analysis helps inform the interplay between statistical interdependence and causal mechanisms, and motivates future work pushing the boundaries of the promising interaction between physics and information theory.

\begin{acknowledgments}
We thank Hardik Rajpal, Fernando Rosas, and Borjan Milinkovic for useful discussions. AL and HJJ gratefully acknowledge support from the EPSRC through grant EP/W007142/1.
\end{acknowledgments}

\newpage

\bibliography{refs.bib}

\onecolumngrid
\clearpage

\appendix 
\counterwithout{equation}{section}

\section{Information theory background}    \label{app:MI}
In this section, we review some of the fundamental quantities of information theory used in this work.

\subsection{Mutual information}
Mutual information (MI) is a crucial measure in information theory \cite{cover1999elements}, being widely used by itself or as a building block for more sophisticated metrics \cite{beer2015information, mediano2019beyond}. 
MI measures the pairwise interdependence between two stochastic variables $X,Y$: assuming $X,Y$ are distributed according to the probabilities densities $p_X(x)$, $p_Y(y)$, and $p_{X,Y}(x;y)$, the mutual information between $X$ and $Y$ is defined as
\begin{equation}
    I(X;Y)=\sum_x\sum_yp_{X,Y}(x;y)\ln \left( \frac{p_{X,Y}(x;y)}{p_X(x)p_Y(y)}\right) \,
    \label{Mutual_Inf}
\end{equation}
where the two sums run over all possible outcomes $x,y$ of the random variables $X$ and $Y$, respectively.
Given the definition of the Shannon entropy of a random variable $X$
\begin{equation} \label{eq:shannon_entropy}
    H(X) = \sum_x p_X(x)\ln\frac{1}{p_X(x)} \,,
\end{equation}
it is illuminating to express MI in terms of entropies
\begin{equation}
\begin{aligned} \label{eq:MI_H4}
    I(X;Y) &= H(X)-H(X|Y)\\
           &=H(X)+H(Y)-H(X,Y) 
\end{aligned}
\end{equation}
where $H(X|Y)$ is the conditional entropy of $X$ given $Y$, and $H(X,Y)$ is the joint entropy of the two. 
In particular, if $X,Y$ are jointly Gaussian distributed with mean $\mu=0$ and covariance matrix $\Sigma$, then the entropy in Eq.~\eqref{eq:shannon_entropy} can be expressed as
\begin{align} \label{eq:entropy_gauss}
    H(X) &= \frac12 \ln \left(|2\pi e \Sigma_X| \right) \\
    H(X,Y) & = \frac12 \ln \left(|2\pi e \Sigma| \right) \label{eq:entropy_gauss_joint} \,,
\end{align}
where $\Sigma_X$ is the covariance matrix of $X$. An analogous expression holds for $H(Y)$. 
Thus, calculating the mutual information between Gaussian processes only involves estimating the covariance matrices of the distributions, and then using Eqs.~\eqref{eq:MI_H4}-\eqref{eq:entropy_gauss_joint} to obtain
\begin{equation} \label{eq:gen_MI_gaussian}
    I(X;Y) =  \logdet{|\Sigma_X|}{|\Sigma_Y|}{|\Sigma|} \,.
\end{equation}

\subsection{Transfer entropy}

Transfer entropy was introduced by Schreiber \cite{schreiber2000measuring} as a non-parametric statistical measure based on the concept of conditional mutual information. 
Within the study of complex systems, TE aims to uncover relationships between the different components, potentially probing causal and dynamical interactions \cite{lizier2010differentiating, hahs2011distinguishing}, and is equivalent to Granger Causality for Gaussian variables \cite{barnett2009granger, granger1963economic, geweke1982measurement}. 
Its applications span various fields, including neuroscience \cite{wibral2014transfer, vakorin2010exploring}, finance \cite{dimpfl2013using, korbel2019transfer}, climate science \cite{stramaglia2024disentangling, runge2012escaping}, and engineering \cite{bauer2006finding, overbey2009dynamic}, in which TE has profoundly contributed to the understanding of how information propagates through networks and affects the behaviour of interconnected elements.

Unlike simpler correlations, transfer entropy captures non-linear dependencies and is particularly effective in identifying causal relationships in complex and coupled systems. 
Specifically, it assesses how much the future state of one system (the target) can be predicted by knowing the past state of another (the source), beyond what the target’s past alone provides. 
Similarly to the expression of MI in Eq.~\eqref{eq:MI_H4}, it is convenient to look at how TE can be expressed in terms of Shannon entropies:
\begin{equation}
    \mathcal{T}(X(t-L);Y(t))= H(Y_t|Y_{t-1:t-L})-H(Y_t|Y_{t-1:t-L},X_{t-1:t-L}) \, ,
        \label{eq:TE_def2}
\end{equation}
where ${X_{t-1:t-L}\equiv(X_{t-1},X_{t-2},...,X_{t-L})}$. This expression highlights that TE is a measure of how much one time series $Y$ depends on the past history of another time series $X$: if $Y$ is independent of $X$, the two conditioned entropies on the right-hand side will be equal, and TE vanishes. Instead, if the past of $X$ fully determines $Y_t$, then $H(Y_t|Y_{t-1:t-L},X_{t-1:t-L})=0$ and the transfer entropy is maximum.

\section{Swapping Signal}   \label{app:swap_sig}
We briefly present the details of the simple swapping signal used in the Sec. \ref{sec:warnings} to illustrate the behaviour of the mutual information in a time series as the stochastic content is changed. 

We consider a function $f(t)$ on the positive real axis which can assume two values $\{-A,A\}$. At time $t=0$ we set
\begin{equation}
    f(0) =\left \{ \begin{array}{ll}
    A &\mbox{with probability $\alpha$}\\
    -A & \mbox{with probability $1-\alpha$.}
    \end{array}
    \right. 
\end{equation}
Additionally, we assume the probability that the function swaps from $-A$ to $A$ or from $A$ to $-A$ during an infinitesimal time interval $dt$ is given by $\lambda dt$. I.e., we have a Poisson process and the probability that $f(t)$ swaps $n$ times during the interval $[0,t]$ is 
\begin{equation}
    p\{n \;{\rm  swaps}\}=\frac{(\lambda t)^n}{n!}e^{-\lambda t}
\end{equation}
Since we are interested in the expression for the mutual information between $f(0)$ and $f(t)$, we need to compute the probabilities $p(f(t))$ and $p(f(0);f(t))$ for all possible combinations of events $\{A,-A\}$.
We can calculate $p(f(0);f(t))$ by keeping track of how many times the signal has swapped between $t=0$ and the time $t>0$. We have for example 
\begin{equation}
    \begin{aligned}
     p(f(0)=A; f(t)=A) &= \alpha\left( \,p\{ 0 \; {\rm swaps}\} + p\{ 2 \; {\rm swaps}\} + p\{ 4 \; {\rm swaps}\} + \ldots \right) \\
     & = \alpha\left(e^{-\lambda t} + \frac{(\lambda t)^2}{2!}e^{-\lambda t} + \frac{(\lambda t)^4}{4!}e^{-\lambda t} + \cdots \right) \\
     & = \alpha e^{-\lambda t} \cosh(\lambda t)
    \end{aligned}
\end{equation}
The other joint probabilities are computed the same way, reading
\begin{align}
        p(f(0)=-A;f(t)=A) &= (1-\alpha) e^{-\lambda t} \sinh(\lambda t) \\
        p(f(0)=A;f(t)=-A) &= \alpha e^{-\lambda t} \sinh(\lambda t) \\
        p(f(0)=-A;f(t)=-A) &= (1-\alpha) e^{-\lambda t} \cosh(\lambda t) \,.
\end{align}
Finally, $p(f(t))=\pm A$ can be computed by summing over the marginal probabilities:
\begin{align}
    p(f(t)=A) & \begin{aligned}[t]
    & = p(f(0)=A; f(t)=A) + p(f(0)=-A; f(t)=A) \\ 
    & = \alpha e^{-\lambda t} \cosh(\lambda t) + (1-\alpha) e^{-\lambda t} \sinh(\lambda t) \\
    & = e^{-\lambda t} \sinh(\lambda t) + \alpha e^{-2\lambda t} 
    \end{aligned} \\
    p(f(t)=-A) & \begin{aligned}[t]
    & = p(f(0)=A; f(t)=-A) + p(f(0)=-A; f(t)=-A) \\
    & = \alpha e^{-\lambda t} \sinh(\lambda t) + (1-\alpha) e^{-\lambda t} \cosh(\lambda t) \\
    & = e^{-\lambda t} \cosh(\lambda t) - \alpha e^{-2\lambda t}
    \end{aligned}
\end{align}
Now using Eq.~\eqref{Mutual_Inf} and doing some algebra, we obtain the mutual information between $f(0)$ and $f(t)$
\begin{gather}
    I(f(0); f(t)) = \frac{1}{2}\left[ a \log a + b \log b - (a - c) \log(a - c) - (b + c) \log(b + c) \right] \label{MI_Swap} \,,
\end{gather}
where we introduced the shorthand notation 
\begin{equation*}
    \begin{aligned}
        a &= 1 + \exp(-2\lambda t) \\
        b &= 1 - \exp(-2\lambda t) \\
        c &= 2\alpha \exp(-2\lambda t) \,.
    \end{aligned}
\end{equation*}
We use this expression to illustrate the behaviour of the mutual information with $\alpha$ fixed and varying the stochastic element, i.e. the rate of fluctuations $\gamma$, and for fixed $\gamma$ and changing $\alpha$ (Fig.~\ref{fig:Swap_Signal}).

\section{$N>2$ random walkers} \label{app:N-RWcase}

In this section, we present the mathematical calculations that support the findings in the main text. 
We perform the following derivation in the case of an interacting Ornstein-Uhlenbeck (OU) process \cite{uhlenbeck1930theory}, which generalises the results shown in Sec.~\ref{sec:results} to the case of a widely studied stationary dynamics. 

\subsection{Ornstein-Uhlenbeck processes}
Ornstein-Uhlenbeck processes are a type of stochastic process used to model systems that tend toward a stable equilibrium over time. 
Unlike a simple Brownian motion, which drifts indefinitely, the OU process introduces a restoring force that pulls values back toward a long-term mean, making it especially useful for modelling mean-reverting behaviour in time series data, such as interest rates in finance \cite{vasicek1977equilibrium, bjork2009arbitrage, barndorff2001non}, and evolution dynamics in biological systems \cite{hunt2007relative, martins1994estimating}. 

Specifically, given a stochastic process $x_t$, a general OU process in discrete time can be written as 
\begin{equation}
    x_t = (1-\theta)\,x_{t-1} + \sigma\, \eta_t \,,
\end{equation}
where $\theta\in(0,1)$ is the restoring force, $\eta_t$ a white noise term, and $\sigma>0$ a parameter that controls the amount of noise in the process. For convenience, we set the time interval $\Delta t$ to 1. 

We employ OU processes since their mathematical tractability allows us to study in detail how information-theoretic measures behave on a simple coupled system with stationary dynamics. 

\subsection{Model}
For our study, we consider the model given the transition matrix $\boldsymbol{M}$ (Eq.~\eqref{eq:M_OU_NRW}) and the equations of motion (Eqs.~\eqref{eq:delta_ev}-\eqref{eq:chi_ev}), which we repeat for convenience 
\begin{gather}    \label{eq:M_OU_NRW_app}
    \bv{M} = 
    \begin{pmatrix}
    1 - \theta - \gamma & \gamma & 0 & \cdots & 0 \\
    \gamma & 1 - \theta - 2\gamma & 0 & \cdots & 0 \\
    \vdots & \vdots & \ddots & \ddots & \vdots \\
    0 & \cdots & 0 & 1 - \theta - 2\gamma & \gamma \\
    0 & 0 & \cdots & \gamma & 1 - \theta - \gamma
\end{pmatrix}  \\   \label{eq:delta_ev_app}
    \bv{\Delta}(t^{\prime}) = \boldsymbol{M}^{t^{\prime}-t}\boldsymbol{\Delta}(t) + \boldsymbol{\rchi}_{tt^{\prime}} 
    \\ \label{eq:chi_ev_app} 
    \boldsymbol{\rchi}_{tt^{\prime}} =
    \boldsymbol{M}^{t^{\prime}-t-1}\boldsymbol{\eta}(t) + \boldsymbol{M}^{t^{\prime}-t-2}\boldsymbol{\eta}(t+1) + \cdots + \boldsymbol{M}\boldsymbol{\eta}(t^{\prime}-2) + \boldsymbol{\eta}(t^{\prime}-1) \,. 
\end{gather}
This model consists of a OU Gaussian process with restoring force $\theta$ in which the different components are coupled via a spring of constant $\gamma$. 

The aim of this section is to compute variances and covariances of random walkers and centre of mass, which we then use for the calculations of entropies and all information-theoretic measures analysed.
We do this by expanding Eqs.~\eqref{eq:chi_ev_app}-\eqref{eq:delta_ev_app} and writing explicitly their dependence on the interaction strength to first order in $\gamma$, and for any $\theta\in[0,1)$.
Moreover, we work in the limit $N\rightarrow\infty$, so that we only consider the central walkers of the chain and ignore finite-size boundary effects. In this regime we can approximate the matrix $\boldsymbol{M}$ as
\begin{equation}
    \boldsymbol{\tilde{M}} = 
    \begin{pmatrix}
    1 - \theta - 2\gamma & \gamma & 0 & \cdots & 0 \\
    \gamma & 1 - \theta - 2\gamma & 0 & \cdots & 0 \\
    \vdots & \vdots & \ddots & \ddots & \vdots \\
    0 & \cdots & 0 & 1 - \theta - 2\gamma & \gamma \\
    0 & 0 & \cdots & \gamma & 1 - \theta - 2\gamma
\end{pmatrix} \, .
\end{equation}
For ease of notation, we will hereafter we indicate $\boldsymbol{\tilde{M}}$ with just $\boldsymbol{M}$.

We start by computing the powers of the matrix $\boldsymbol{M}$ by noticing that
\begin{equation} \label{eq:NRW_Mpowers}
    \boldsymbol{M}^n = 
    \begin{pmatrix}
    (1-\theta)^{n-1}(1 - \theta - 2n\gamma) & n\gamma(1-\theta)^{n-1} & 0 & \cdots & 0 \\
    n\gamma(1-\theta)^{n-1} & (1-\theta)^{n-1}(1 - \theta - 2n\gamma) & 0 & \cdots & 0 \\
    \vdots & \vdots & \ddots & \ddots & \vdots \\
    0 & \cdots & 0 & (1-\theta)^{n-1}(1 - \theta - 2n\gamma) & n\gamma(1-\theta)^{n-1} \\
    0 & 0 & \cdots & n\gamma(1-\theta)^{n-1} & (1-\theta)^{n-1}(1 - \theta - 2n\gamma)
\end{pmatrix} \, .
\end{equation}
By substituting this into Eq.~\eqref{eq:chi_ev_app} we obtain:
\begin{equation} \label{eq:NRW_chi_ev_expl}
\begin{split}
\rchi_{tt^{\prime}_j} &= 
    \eta_j(t^{\prime}-1) +M_{jk}\eta_k(t^{\prime}-2) + \ldots +M_{jk}^{t^{\prime}-t-2}\eta_k(t+1) +M_{jk}^{t^{\prime}-t-1}\eta_k(t) = \\
    & = \eta_j(t^{\prime}-1)+ \\
    & +\gamma\Bigl(\eta_{j+1}(t^{\prime}-2)+\eta_{j-1}(t^{\prime}-2)\Bigr)
    +\Bigl(1-\theta-2\gamma\Bigr)\eta_j(t^{\prime}-2) + \\
    & + 2\gamma\Bigl(1-\theta\Bigr)\Bigl(\eta_{j+1}(t^{\prime}-3)
    +\eta_{j-1}(t^{\prime}-3)\Bigr)+\Bigl(1-\theta\Bigr)\Bigl(1-\theta-4\gamma\Bigr)\eta_j(t^{\prime}-3) +\\
    & + \ldots + \\
    & +\Bigl(t^{\prime}-t-2\Bigr)\gamma\Bigl(1-\theta\Bigr)^{t^{\prime}-t-3}\Bigl(\eta_{j+1}(t+1)+\eta_{j-1}(t+1)\Bigr)
    +\Bigl(1-\theta\Bigr)^{t^{\prime}-t-3}\Bigl(1-\theta-2(t^{\prime}-t-2)\gamma\Bigr)\eta_j(t+1) + \\
    & +\Bigl(t^{\prime}-t-1\Bigr)\gamma\Bigl(1-\theta\Bigr)^{t^{\prime}-t-2}\Bigl(\eta_{j+1}(t)+\eta_{j-1}(t)\Bigr)
    +\Bigl(1-\theta\Bigr)^{t^{\prime}-t-2}\Bigl(1-\theta-2(t^{\prime}-t-1)\gamma\Bigr)\eta_j(t) \,,
    \end{split}
\end{equation}
where we are using Einstein's notation to indicate summation over repeated indices (in this case, $k$).

Luckily, for the calculation of information-theoretic quantities, we are only interested in the variance of this term, which greatly simplifies the expression. 
Since all the stochastic terms $\eta_j(t)$ are Gaussian-distributed with variance $\sigma^2 \,\forall j$ and $\forall t>0$, we can write:
\begin{align}
\begin{split} \label{eq:var_chi_pt1_NRW}
\mathrm{Var} \left( \rchi_{tt^{\prime}_j} \right) &= 
    \sigma^2 + \\
    & +\gamma^2 2\sigma^2
    +(1-\theta-2\gamma)^2\sigma^2 + \\
    & + (2\gamma)^2(1-\theta)^2 2\sigma^2
    +(1-\theta)^2(1-\theta-4\gamma)^2\sigma^2 +\\
    & + \ldots + \\
    & +(t^{\prime}-t-2)^2\gamma^2(1-\theta)^{2(t^{\prime}-t-3)}2\sigma^2
    +(1-\theta)^{2(t^{\prime}-t-3)}\Bigl(1-\theta-2(t^{\prime}-t-2)\gamma\Bigr)^2\sigma^2 + \\
    & +(t^{\prime}-t-1)^2\gamma^2(1-\theta)^{2(t^{\prime}-t-2)}2\sigma^2
    +(1-\theta)^{2(t^{\prime}-t-2)}\Bigl(1-\theta-2(t^{\prime}-t-1)\gamma\Bigr)^2\sigma^2 \\
    &= \sigma^2 \left( 1 + \sum_{\tilde{t}=1}^{\tp-t-1} (1-\theta)^{2(\tilde{t}-1)}(1-\theta-2\tilde{t}\gamma)^2 + 2 \sum_{\tilde{t}=0}^{\tp-t-1} (1-\theta)^{\tp-t-1} \gamma^2 \right) \\
    &= \sigma^2 \Bigl( 1 + \sum_{\tilde{t}=1}^{\tp-t-1} (1-\theta)^{2(\tilde{t}-1)} \left( (1-\theta)^2 - 4(1-\theta)\gamma \Bigr) + o(\gamma) \right) \\
    &= \sigma^2 \left( 1 + \sum_{\tilde{t}=1}^{\tp-t-1} (1-\theta)^{2\tilde{t}} - \sum_{\tilde{t}=1}^{\tp-t-1} (1-\theta)^{2\tilde{t}-1} 4\tilde{t}\gamma + o(\gamma) \right) \\
    &= \sigma^2 \left( \sum_{\tilde{t}=0}^{\tp-t-1} (1-\theta)^{2\tilde{t}} - \sum_{\tilde{t}=1}^{\tp-t-1} (1-\theta)^{2(\tilde{t}-1)} 4\tilde{t}\gamma \right) \\
    & \alphaeq \sigma^2 \left( \frac{1-(1-\theta)^{2(t^{\prime}-t)}}{1-(1-\theta)} -\frac{4\gamma(1-\gamma)}{(1-(1-\theta)^2)^2}\Bigl(1-(\tp-t)(1-\theta)^{2(\tp-t-1)}+(\tp-t-1)(1-\theta)^{2(\tp-t)} \Bigr)\right) \\
    & \betaeq \sigma^2 \left( \frac{1-\alpha^{2(t^{\prime}-t)}}{1-\alpha} -\frac{4\gamma(1-\gamma)}{(1-\alpha^2)^2}\Bigl(1-(\tp-t)\alpha^{2(\tp-t-1)}+(\tp-t-1)\alpha^{2(\tp-t)} \Bigr)\right) \,.
\end{split}
\end{align}
In step (b) we simply introduced the shorthand $\alpha \equiv (1-\theta)$ for convenience, which we will employ from now on. In (a) we noticed that the first term is a geometric series, while the second is an arithmetico-geometric series. Making use of their closed-form expressions for finite sums, we have
\begin{gather} \label{eq:geom_series}
    \sum_{\tilde{t}=0}^{\tp-t-1} (1-\theta)^{2\tilde{t}} =
    \frac{1-(1-\theta)^{2(t^{\prime}-t)}}{1-(1-\theta)}, \quad  \\
    \begin{split}
    - \frac{4\gamma}{1-\theta}\sum_{\tilde{t}=1}^{\tp-t-1} (1-\theta)^{2\tilde{t}}\tilde{t} &= 
    - \frac{4\gamma}{1-\theta} \cdot \frac{(1-\theta)^2}{(1-(1-\theta)^2)^2}\Bigl(1-(\tp-t)(1-\theta)^{2(\tp-t-1)}+(\tp-t-1)(1-\theta)^{2(\tp-t)} \Bigr) \\ \label{eq:arit_geom_series}
    & = -\frac{4\gamma(1-\theta)}{(1-(1-\theta)^2)^2}\Bigl(1-(\tp-t)(1-\theta)^{2(\tp-t-1)}+(\tp-t-1)(1-\theta)^{2(\tp-t)} \Bigr) \,.
    \end{split}
\end{gather}
Notice that by taking the limit $\theta\rightarrow0$ in Eq.~\eqref{eq:geom_series} we can obtain the case of the non-OU random walkers presented in the main part of the work. 

The expression of $\rchi_{t\tp_j}$ obtained in Eq.~\eqref{eq:var_chi_pt1_NRW} is one of the key ingredients needed for the calculation of information measures, as it corresponds to the variance of the $j-\text{th}$ walker that evolves from time $t$ to time $\tp$. 
It is of particular interest for the computation of the evolution of the walker from its starting position, in which case we have:
\begin{equation} \label{eq:var_delta_j_NRW}
    \begin{split}
    \var{\Delta_j(t)} & = \var{\rchi_{0t_j}} \\ 
    & = \sigma^2 \left( \frac{1-\alpha^{2t}}{1-\alpha^2} -\frac{4\gamma\alpha}{(1-\alpha^2)^2}\Bigl(1-t\alpha^{2(t-1)}+(t-1)\alpha^{2t} \Bigr)\right) \\
    & = \sigma^2 \frac{1-\alpha^{2t}}{1-\alpha^2} \left( 1 -\frac{4\gamma\alpha}{(1-\alpha^2)(1-\alpha^{2t})}\Bigl(1-t\alpha^{2(t-1)}+(t-1)\alpha^{2t} \Bigr)\right)
    \end{split}
\end{equation}

We proceed to compute the covariance between the positions of a walker at different timesteps. We have
\begin{equation} \label{eq:NRW_cov_xjxj}
\begin{split}
    \cov{\Delta_j(t);\Delta_j(\tp)} & = \cov{\Delta_j(t); M_{jk}^{\tp-t} \Delta_k(t) + \rchi_{t\tp_j}} \\
    &\alphaeq \cov{\Delta_j(t); (1-\theta - 2(\tp-t)\gamma) \Delta_j(t)(1-\theta)^{\tp-t-1}} +\\
&\quad + \gamma (\tp-t) (1-\theta)^{\tp-t-1}\,\cov{\Delta_j(t); \Delta_{j-1}(t) + \Delta_{j+1}(t)} \\
&\betaeq \Bigl(1-\theta - 2(\tp-t)\gamma\Bigr) (1-\theta)^{\tp-t-1} \,\var{\Delta_j(t)} + o(\gamma)\\
&\gammaeq \Bigl(\alpha - 2(\tp-t)\gamma\Bigr) \alpha^{\tp-t-1} \sigma^2 \left( \frac{1-\alpha^{2t}}{1-\alpha^2} - \frac{4\gamma \alpha}{(1-\alpha^2)^2} \Bigl( 1 - t\alpha^{2(t-1)} + (t-1)\alpha^{2t} \Bigr) \right) \\
& = \sigma^2 \alpha^{\tp-t-1} \left[ \frac{1-\alpha^{2t}}{1-\alpha^2}\alpha 
-2\gamma\left((\tp-t)\frac{1-\alpha^{2t}}{1-\alpha^2} + \frac{2\alpha^2}{(1-\alpha^2)^2} \Bigl( 1 - t\alpha^{2(t-1)} + (t-1)\alpha^{2t}\Bigr) \right) \right] \\
& = \sigma^2 \frac{1-\alpha^{2t}}{1-\alpha^2} \alpha^{\tp-t} \left[ 1
-2\gamma\left(\frac{(\tp-t)}{\alpha}+  \frac{2\alpha}{(1-\alpha^2)(1-\alpha^{2t})} \Bigl( 1 - t\alpha^{2(t-1)} + (t-1)\alpha^{2t}\Bigr) \right) \right]
\end{split}
\end{equation}
Step (a) follows from Eq.~\eqref{eq:M_OU_NRW_app}, in (b) we neglected the second order term in $\gamma$, in (c) we made use of Eq.~\eqref{eq:var_delta_j_NRW}, and in the last steps we rearranged terms so as to isolate the non-interacting and interacting parts. 
Note that this result is coherent with Eq.~\eqref{eq:var_delta_j_NRW} in the case $t=\tp$.

Focusing now on neighbouring random walkers, we can calculate their covariance at the same time as
\begin{equation} \label{eq:NRW_cov_xjxi}
    \begin{split}
        \cov{\Delta_j(t);\Delta_{j+1}(t)} & = \cov{\rchi_{0t_j}; \rchi_{0t_{j\pm1}}} \\
        & = \begin{aligned}[t]
        \mathrm{Cov}\Bigl( & \eta_{j}(t-1) + \Bigl(1-\theta-2\gamma\Bigr) \eta_{j}(t-2) + \gamma \eta_{j+1}(t-2)+ \ldots + \\
        & + \alpha^{t-2} \Bigl(1-\theta-2(t-1)\gamma\Bigr) \eta_{j}(0) + \alpha^{t-2}(t-1) \gamma \eta_{j+1}(0) + \zeta_{j-1}; \\
        & \eta_{j+1}(t-1) + \Bigl(1-\theta-2\gamma\Bigr) \eta_{j+1}(t-2) + \gamma \eta_{j}(t-2)+ \ldots + \\
        & + \alpha^{t-2} \Bigl(1-\theta-2(t-1)\gamma\Bigr) \eta_{j+1}(0) + \alpha^{t-2}(t-1) \gamma \eta_{j}(0) + \zeta_{j+2}; \end{aligned} \\
        &= \begin{aligned}[t]
        \mathrm{Cov}\Biggl(& \sum_{\tilde{t}=0}^{t-1}  \left( \alpha^{\tilde{t}-1} \Bigl(1-\theta-2\tilde{t}\gamma\Bigr) \eta_{j}(t-\tilde{t}-1)\right) +  \sum_{\tilde{t}=0}^{t-1} \left(\tilde{t}\gamma \alpha^{\tilde{t}-1} \eta_{j+1}(t-\tilde{t}-1)\right); \\
        & \sum_{\tilde{t}=0}^{t-1}  \left( \alpha^{\tilde{t}-1} \Bigl(1-\theta-2\tilde{t}\gamma\Bigr) \eta_{j+1}(t-\tilde{t}-1)\right) + \sum_{\tilde{t}=0}^{t-1} \left(\tilde{t}\gamma \alpha^{\tilde{t}-1} \eta_{j}(t-\tilde{t}-1)\right)\Biggr) \end{aligned} \\
        &= 2\, \mathrm{Cov}\Biggl(\sum_{\tilde{t}=0}^{t-1}  \left( \alpha^{\tilde{t}-1} \Bigl(1-\theta-2\tilde{t}\gamma\Bigr) \eta_{j+1}(t-\tilde{t}-1)\right) ; \sum_{\tilde{t}=0}^{t-1} \left(\tilde{t}\gamma \alpha^{\tilde{t}-1} \eta_{j}(t-\tilde{t}-1)\right)\Biggr) \\
        &= 2 \sum_{\tilde{t}=0}^{t-1} \alpha^{2(\tilde{t}-1)} \Bigl(1-\theta-2\tilde{t}\gamma\Bigr)\tilde{t} \gamma \sigma^2 \\
        & = \sigma^2 \frac{2\gamma}{\alpha} \sum_{\tilde{t}=0}^{t-1} \alpha^{2(\tilde{t}-1)}\tilde{t} \\
        &= \sigma^2 \frac{2\gamma\alpha}{(1-\alpha^2)^2} \left( 1 - t \alpha^{2(t-1)} + (t-1)\alpha^{2t} \right) \,.
    \end{split}
\end{equation}
where at the beginning we made ude of Eq.~\eqref{eq:NRW_chi_ev_expl}.

In general, without diving into detailed calculations, we notice that for different times the covariances between different walkers are of first order in $\gamma$ only for neighbouring walkers, and of higher order for the remaining ones: 
\begin{equation} \label{eq:cov_ij_gamma}
    \cov{\Delta_j(t); \Delta_i(\tp)} = 
    \begin{cases*}
      \mathcal{O}(\gamma) & if $i=j\pm1$ \\
      o(\gamma)       & if $i\ne j\pm1$
    \end{cases*}
    , i\ne j
\end{equation}
Thus, this means that since we limit our calculations to first order effects, only the interactions between neighbouring sites are included.

We now want to calculate the covariances involving the centre of mass of the system, defined as in Eq.~\eqref{eq:com_def_NRW}. 
We first notice that we can express the position of c.o.m.\ at time $\tp$ in terms of that at time $t$ as
\begin{equation} \label{eq:NRW_vv'}
    \begin{split}
        V(\tp) &= \frac{1}{N}\sum_{j=1}^N\Delta_j(\tp) \\
        &\alphaeq \frac{1}{N}\sum_{j=1}^N \alpha \Delta_j(\tp-1) + \rchi_{ \tp-1 \tp_V}\\
        & = \ldots \\
        &=  \frac{1}{N}\sum_{j=1}^N\alpha^{\tp-t}\Delta_j(t) + \rchi_{t\tp_V} \\
        &= \alpha^{\tp-t}V(t) + \rchi_{t\tp_V} \,,
    \end{split}
\end{equation}
where in step (a) we took advantage of the symmetry of the system to cancel all interaction terms. %
$\rchi_{t\tp_V}$ is the evolution term of the centre of mass due to the stochastic component, which can be obtained by using Eqs.~\eqref{eq:delta_ev_app}-\eqref{eq:chi_ev_app} and reads
\begin{equation}
    \rchi_{t\tp_V} = \sum_{j=1}^N\sum_{\tilde{t}=t}^{{\tp}-1} \dfrac{\alpha^{\tp-\tilde{t}-1}}{N} \eta_j(\tilde{t}) \,.
\end{equation}
Therefore, we can proceed to calculate:
\begin{equation} \label{eq:NRW_cov_vv}
    \begin{split}
        \cov{V(t); V(\tp)} &\alphaeq \cov{V(t); \alpha^{\tp-t}V(t) + \rchi_{t\tp_V}} \\
        &= \frac{\alpha^{\tp-t}}{N^2}\covs{\sum_{j=1}^N\Delta_j(t);\sum_{j=1}^N\Delta_j(t)} \\
        &\betaeq \frac{\alpha^{\tp-t}}{N^2}\left( N\cov{\Delta_i(t); \Delta_i(t)} + 2(N-1)\cov{\Delta_i(t); \Delta_{i\pm1}(t)} \right) \\
        & \gammaeq \begin{aligned}[t] \sigma^2 \frac{\alpha^{\tp-t}}{N}\Biggl[\Biggl( 
        \frac{1-\alpha^{2t}}{1-\alpha^2}  & 
        - \frac{4\gamma \alpha}{(1-\alpha^2)^2}\left( 1-t\alpha^{2(t-1)}+(t-1)\alpha^{2t}\right) + \\ & +
        2\cdot\frac{2\gamma \alpha}{(1-\alpha^2)^2}\left( 1-t\alpha^{2(t-1)}+(t-1)\alpha^{2t}\right)
        \Biggr)\Biggr] \end{aligned} \\
        & = \sigma^2 \frac{\alpha^{\tp-t}}{N} \cdot \frac{1-\alpha^{2t}}{1-\alpha^2} 
    \end{split} \,,
\end{equation}
where in (a) we used Eq.~\eqref{eq:NRW_vv'}, in (b) we neglected the term -1 term in (N-1) as we work in the limit of infinite size, and in (c) we employed Eqs.~\eqref{eq:NRW_cov_xjxj}-\eqref{eq:NRW_cov_xjxi}.

The covariance between a random walker and the c.o.m.\ reads
\begin{equation}
    \begin{split}
        \cov{\Delta_j(t);V(\tp)} & = \cov{\Delta_j(t); \alpha^{\tp-t}V(t)+\rchi_{t\tp_V}} \\
        & = \alpha^{\tp-t} \,\cov{\Delta_j(t);\frac{\Delta_{j-1}(t)+\Delta_j(t)+\Delta_{j+1}(t)}{N} } \\
        & = \frac{\alpha^{\tp-t}}{N} \left(\var{\Delta_j(t)} + 2\,\cov{\Delta_{j}(t); \Delta_{j\pm1}(t)}\right) \\
        & = \begin{aligned}[t] \sigma^2 \frac{\alpha^{\tp-t}}{N}\Biggl[\Biggl( 
        \frac{1-\alpha^{2t}}{1-\alpha^2}  & 
        - \frac{4\gamma \alpha}{(1-\alpha^2)^2}\left( 1-t\alpha^{2(t-1)}+(t-1)\alpha^{2t}\right) + \\ & +
        2\cdot\frac{2\gamma \alpha}{(1-\alpha^2)^2}\left( 1-t\alpha^{2(t-1)}+(t-1)\alpha^{2t}\right)
        \Biggr)\Biggr] \end{aligned} \\
        & = \sigma^2 \frac{\alpha^{\tp-t}}{N} \cdot \frac{1-\alpha^{2t}}{1-\alpha^2} \,,
    \end{split}
\end{equation}
where we substituted Eqs.~\eqref{eq:var_chi_pt1_NRW}-\eqref{eq:NRW_cov_xjxi} in the expression.
Similarly, for the covariance between the c.o.m.\ and a RW we obtain
\begin{equation}
\begin{split}
\cov{V(t); \Delta_j(\tp))} &\alphaeq \cov{V(t); M^{\tp-t}_{jk}\Delta_k(t)) + \rchi_{t\tp_j}} \\
&\betaeq \begin{aligned}[t] \cov{ & \frac{1}{N}\Bigl(\Delta_{j-2}(t) + \Delta_{j-1}(t) + \Delta_{j}(t) + \Delta_{j+1}(t) + \Delta_{j+2}(t)\Bigr); \\
& \alpha^{\tp-t-1}\Bigl(1 - \theta - 2(\tp - t) \gamma\Bigr)\Delta_j(t)+\alpha^{\tp-t-1}\gamma(\tp-t)\Bigl(\Delta_{j+1}(t)+\Delta_{j-1}(t) \Bigr) } \end{aligned} \\
&= \begin{aligned}[t] \frac{\alpha^{\tp-t-1}}{N} \Biggl[ & \begin{aligned}[t] \gamma(\tp - t)\Bigl[ 
& \cov{\Delta_{j - 2}(t); \Delta_{j - 1}(t)} + \cov{\Delta_{j + 2}(t), \Delta_{j + 1}(t)}+ \\
& + \var{\Delta_j(t)} + \var{\Delta_{j+1}(t)} + \\
& + \cov{\Delta_{j}(t); \Delta_{j - 1}(t)} + \cov{\Delta_{j}(t), \Delta_{j + 1}(t)}\Bigr]+ \end{aligned} \\
& + \Bigl(1 - \theta - 2(\tp - t-1)\gamma\Bigr)\Bigl[ \var{\Delta_j(t)} + \\
& \quad \quad \quad \quad + \cov{\Delta_{j-1}(t); \Delta_j(t))} + \cov{\Delta_{j+ 1}(t), \Delta_j(t)}\Bigr] \Biggr] \end{aligned} \\
& \gammaeq \frac{\alpha^{\tp-t-1}}{N} \Bigl[\gamma(\tp-t)\left[2\mu+2\nu+2\mu \right] +  \Bigl(1 - \theta - 2(\tp - t-1)\gamma\Bigr) \left[\nu+2\mu\right]\Bigr] \\
& = \frac{\alpha^{\tp-t-1}}{N} \Bigl[(1-\theta)\nu + (1-\theta)2\mu\Bigr] \\
& \deltaeq  \begin{aligned}[t] \sigma^2 \frac{\alpha^{\tp-t}}{N}\Biggl[\Biggl( 
\frac{1-\alpha^{2t}}{1-\alpha^2}  & 
- \frac{4\gamma \alpha}{(1-\alpha^2)^2}\left( 1-t\alpha^{2(t-1)}+(t-1)\alpha^{2t}\right) + \\ & +
2\cdot\frac{2\gamma \alpha}{(1-\alpha^2)^2}\left( 1-t\alpha^{2(t-1)}+(t-1)\alpha^{2t}\right)
\Biggr)\Biggr] \end{aligned} \\
& = \sigma^2 \frac{\alpha^{\tp-t}}{N} \cdot \frac{1-\alpha^{2t}}{1-\alpha^2} \,.
\end{split}
\end{equation}
In step (a) and (b) we employed Eqs.~\eqref{eq:delta_ev_app}-\eqref{eq:NRW_Mpowers}, in (c) we used the shorthand notations ${\mu=\cov{\Delta_j(t);\Delta_{j\pm1}(t)}}$ and ${\nu=\var{\Djt}}$, and finally in (d) we substituted Eqs.~\eqref{eq:NRW_cov_xjxi}-\eqref{eq:var_chi_pt1_NRW}.

\subsection{Mutual information} \label{app:N-RWcase_MI}
We are now finally ready to calculate the mutual information between the different terms. The following manipulations consist of substituting the variances and covariances found above in Eq.~\eqref{eq:gen_MI_gaussian}, and then rearranging terms to have an expression to first order in $\gamma$. In doing so, we make use of two useful expressions Eqs.~\eqref{eq:NRW_trick1}-\eqref{eq:NRW_trick2}.

We begin with the MI of a walker between its positions at time $t$ and $\tp$, for which we can write:
\begin{equation} \label{eq:NRW_Ixx}
    \begin{split}
        & I(\Delta_j(t);\Delta_j(\tp)) = \dfrac12 \ln{\dfrac{ \var{\Djt} \var{\Djtp} }{|\Sigma_{\Delta_j\Delta_j^{\prime}}|}} \\
        & = \dfrac12 \ln{\left( \dfrac{\sigma^2 \dfrac{1-\alpha^{2t}}{1-\alpha^2} \left( 1 -\dfrac{4\gamma\alpha\left(1-t\alpha^{2(t-1)}+(t-1)\alpha^{2t} \right)}{(1-\alpha^2)(1-\alpha^{2t})}\right) \cdot 
        \sigma^2 \dfrac{1-\alpha^{2\tp}}{1-\alpha^2} \left( 1 -\dfrac{4\gamma\alpha\left(1-\tp\alpha^{2(\tp-1)}+(\tp-1)\alpha^{2\tp} \right)}{(1-\alpha^2)(1-\alpha^{2\tp})}\right)}
        {\splitfrac{\sigma^2 \dfrac{1-\alpha^{2t}}{1-\alpha^2} \left( 1 -\dfrac{4\gamma\alpha\left(1-t\alpha^{2(t-1)}+(t-1)\alpha^{2t} \right)}{(1-\alpha^2)(1-\alpha^{2t})}\right) \cdot 
        \sigma^2 \dfrac{1-\alpha^{2\tp}}{1-\alpha^2} \left( 1 -\dfrac{4\gamma\alpha\left(1-\tp\alpha^{2(\tp-1)}+(\tp-1)\alpha^{2\tp} \right)}{(1-\alpha^2)(1-\alpha^{2\tp})}\right)+}
        {-\Biggl( \sigma^2 \dfrac{1-\alpha^{2t}}{1-\alpha^2} \alpha^{\tp-t} \left[ 1-2\gamma\left(\dfrac{(\tp-t)}{\alpha}+  \dfrac{2\alpha\left( 1 - t\alpha^{2(t-1)} + (t-1)\alpha^{2t}\right)}{(1-\alpha^2)(1-\alpha^{2t})}  \right) \right]\Biggr)^2 }}\right)} \\
        & = \dfrac12 \ln{\left( \dfrac{ \dfrac{(1-\alpha^{2t})(1-\alpha^{2\tp})}{(1-\alpha^2)^2}\left(1- 
        \dfrac{4\gamma\alpha}{1-\alpha^2} \left(\dfrac{\left(1-t\alpha^{2(t-1)}+(t-1)\alpha^{2t} \right)}{1-\alpha^{2t}}+\dfrac{\left(1-\tp\alpha^{2(\tp-1)}+(\tp-1)\alpha^{2\tp} \right)}{1-\alpha^{2\tp}} \right)\right)}
        {\splitfrac{\dfrac{(1-\alpha^{2t})(1-\alpha^{2\tp})}{(1-\alpha^2)^2}\left(1- 
        \dfrac{4\gamma\alpha}{1-\alpha^2} \left(\dfrac{\left(1-t\alpha^{2(t-1)}+(t-1)\alpha^{2t} \right)}{1-\alpha^{2t}}+\dfrac{\left(1-\tp\alpha^{2(\tp-1)}+(\tp-1)\alpha^{2\tp} \right)}{1-\alpha^{2\tp}} \right)\right)+}
        {-\Biggl( \left(\dfrac{1-\alpha^{2t}}{1-\alpha^2}\right)^2 \alpha^{2(\tp-t)} \left[ 1-4\gamma\left(\dfrac{(\tp-t)}{\alpha}+  \dfrac{2\alpha\left( 1 - t\alpha^{2(t-1)} + (t-1)\alpha^{2t}\right)}{(1-\alpha^2)(1-\alpha^{2t})}  \right) \right]\Biggr) }}\right)} \\
        & = \frac12 \ln{\left(\dfrac{\dfrac{(1-\alpha^{2t})(1-\alpha^{2\tp})}{(1-\alpha^{2})^2}} {\dfrac{(1-\alpha^{2t})(1-\alpha^{2\tp})}{(1-\alpha^{2})^2}-\left(\dfrac{1-\alpha^{2t}}{1-\alpha^{2}}\right)^2 \alpha^{2(\tp-t)}} \right)} + \\
        & \quad\quad\quad\quad + \frac12 \ln{\left(1+ \dfrac{\splitfrac{\left(\dfrac{1-\alpha^{2t}}{1-\alpha^{2}}\right)^2 \alpha^{2(\tp-t)}\Biggl[
        -4\gamma\left(\dfrac{(\tp-t)}{\alpha}+  \dfrac{2\alpha\left( 1 - t\alpha^{2(t-1)} + (t-1)\alpha^{2t}\right)}{(1-\alpha^2)(1-\alpha^{2t})}  \right)+}{ - 
        \left(\dfrac{4\gamma\alpha}{1-\alpha^2} \left(\dfrac{\left(1-t\alpha^{2(t-1)}+(t-1)\alpha^{2t} \right)}{1-\alpha^{2t}}+\dfrac{\left(1-\tp\alpha^{2(\tp-1)}+(\tp-1)\alpha^{2\tp} \right)}{1-\alpha^{2\tp}} \right) \right)
        \Biggr]}}
        {\dfrac{(1-\alpha^{2t})(1-\alpha^{2\tp})}{(1-\alpha^{2})^2}-\left(\dfrac{1-\alpha^{2t}}{1-\alpha^{2}}\right)^2 \alpha^{2(\tp-t)}}\right)} \\
        & = \frac12 \ln{\left( \dfrac{1-\alpha^{2\tp}}{1-\alpha^{2(\tp-t)}}\right)} + \\
        & \quad\quad\quad\quad + \frac12 \ln{\left(1- \dfrac{4\gamma(1-\alpha^{2t})\alpha^{2(\tp-t)}}{1-\alpha^{2(\tp-t)}}\left(\dfrac{\tp-t}{\alpha}
        + \dfrac{\alpha\left(1-t\alpha^{2(t-1)}+(t-1)\alpha^{2t} \right)}{(1-\alpha^{2t})(1- \alpha^2)}
        -\dfrac{\alpha\left(1-\tp\alpha^{2(\tp-1)}+(\tp-1)\alpha^{2\tp} \right)}{(1-\alpha^{2\tp})(1- \alpha^2)} \right) \right) } \\
        & = \frac12 \ln{\left( \dfrac{1-\alpha^{2\tp}}{1-\alpha^{2(\tp-t)}}\right)}  -\dfrac{2\gamma(1-\alpha^{2t})\alpha^{2(\tp-t)}}{1-\alpha^{2(\tp-t)}}\left(\dfrac{\tp-t}{\alpha}
        + \dfrac{\alpha\left(1-t\alpha^{2(t-1)}+(t-1)\alpha^{2t} \right)}{(1-\alpha^{2t})(1- \alpha^2)}
        -\dfrac{\alpha\left(1-\tp\alpha^{2(\tp-1)}+(\tp-1)\alpha^{2\tp} \right)}{(1-\alpha^{2\tp})(1- \alpha^2)} \right) \,.
    \end{split}
\end{equation}

As for $I(\Djt, \Delta_{j\pm1}(\tp)$, since the off-diagonal terms in the covariance matrix $\Sigma_{\Delta_j\Delta_i^{\prime}}$ are ${\mathcal{O}(\gamma)}$ (Eq.~\eqref{eq:cov_ij_gamma}, in the covariance determinant their product will be $o(\gamma^2)$, and therefore a negligible term. Hence, the joint entropy acts as the sum of the marginal entropies, leading to $I(\Djt, \Delta_{j\pm1}(\tp)=o(\gamma)$.

We continue with the mutual information of the centre of mass at different times:
\begin{equation}\label{eq:NRW_Ivv}
    \begin{split}
        I(V(t);V(\tp)) & = \dfrac12 \ln{\left(\dfrac{\var{V(t)}\var{V(t)}}{\left|\Sigma_{VV^{\prime}}\right|} \right)} \\
        & = \dfrac12 \ln{\left( \dfrac{\dfrac{\sigma^2}{N} \dfrac{1-\alpha^{2t}}{1-\alpha^2} \cdot \dfrac{\sigma^2}{N} \dfrac{1-\alpha^{2\tp}}{1-\alpha^2}}
        {\left(\dfrac{\sigma^2}{N}\right)^2\left(\dfrac{1-\alpha^{2t}}{1-\alpha^2} \cdot \dfrac{1-\alpha^{2\tp}}{1-\alpha^2} - \alpha^{2(\tp-t)}\left(\dfrac{1-\alpha^{2t}}{1-\alpha^2} \right)^2 \right)}\right)} \\
        & = \dfrac12 \ln{\left(\dfrac{1-\alpha^{2\tp}}{1-\alpha^{2(\tp-t)}}\right)} \,.
    \end{split}
\end{equation}
Note that, as expected, the centre of mass behaves as a free random walker.

The MI between a random walker and the c.o.m.\ reads
\begin{equation}\label{eq:NRW_Ixv}
    \begin{split}
        I(\Djt; V(\tp)) & = \frac12 \ln{\left( \frac{\var{\Djt}\var{V(\tp)}}{|\Sigma_{\Delta V^{\prime}}|}\right)} \\
        & = \frac12 \ln{\left( \dfrac{
        \sigma^2 \dfrac{1-\alpha^{2t}}{1-\alpha^2} \left( 1 -\dfrac{4\gamma\alpha\left(1-t\alpha^{2(t-1)}+(t-1)\alpha^{2t} \right)}{(1-\alpha^2)(1-\alpha^{2t})}\right)
        \cdot
        \dfrac{\sigma^2}{N} \dfrac{1-\alpha^{2\tp}}{1-\alpha^2}}
        {\sigma^2 \dfrac{1-\alpha^{2t}}{1-\alpha^2} \left( 1 -\dfrac{4\gamma\alpha\left(1-t\alpha^{2(t-1)}+(t-1)\alpha^{2t} \right)}{(1-\alpha^2)(1-\alpha^{2t})}\right)
        \cdot
        \dfrac{\sigma^2}{N} \dfrac{1-\alpha^{2\tp}}{1-\alpha^2}
        -\left(\dfrac{\sigma^2}{N}\right)^2 \left(\dfrac{1-\alpha^{2t}}{1-\alpha^2}\right)^2 \alpha^{2(\tp-t)} }
        \right)} \\
        & = \frac12 \ln{\left( 
        \dfrac{N(1-\alpha^{2\tp})}{N(1-\alpha^{2\tp})-\alpha^{2(\tp-t)}(1-\alpha^{2t})} \cdot \left(
        1+\dfrac{4\gamma\alpha\dfrac{\alpha^{2(\tp-t)}}{1-\alpha^2}\left(1-t\alpha^{2(t-1)}+(t-1)\alpha^{2t} \right)}{N(1-\alpha^{2\tp})-\alpha^{2(\tp-t)}(1-\alpha^{2t})}
        \right) \right)} \\
        & = \frac12 \ln{\left( 
        \dfrac{N(1-\alpha^{2\tp})}{N(1-\alpha^{2\tp})-\alpha^{2(\tp-t)}(1-\alpha^{2t})}\right)} + \frac12 \ln{\left(
        1+\dfrac{4\gamma\alpha\dfrac{\alpha^{2(\tp-t)}}{1-\alpha^2}\left(1-t\alpha^{2(t-1)}+(t-1)\alpha^{2t} \right)}{N(1-\alpha^{2\tp})-\alpha^{2(\tp-t)}(1-\alpha^{2t})}
        \right)} \\
        & = \frac12 \ln{\left( 
        \dfrac{N(1-\alpha^{2\tp})}{N(1-\alpha^{2\tp})-\alpha^{2(\tp-t)}(1-\alpha^{2t})}\right)} +
        \dfrac{2\gamma\alpha\dfrac{\alpha^{2(\tp-t)}}{1-\alpha^2}\left(1-t\alpha^{2(t-1)}+(t-1)\alpha^{2t} \right)}{N(1-\alpha^{2\tp})-\alpha^{2(\tp-t)}(1-\alpha^{2t})} \,.
    \end{split}
\end{equation}

\begin{figure}[ht]
    \centering
    \includegraphics[width=1\linewidth]{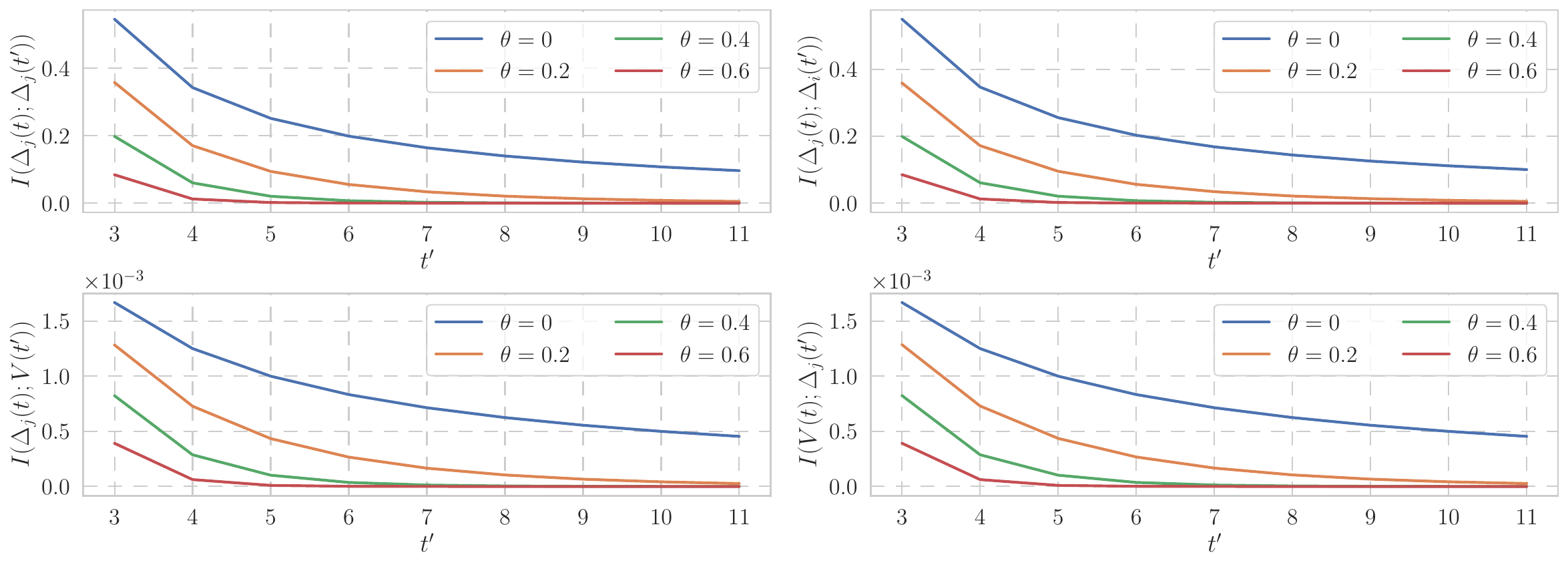}
    \caption{\textbf{Mutual information behaves qualitatively the same as in the non-stationary case.}
    Mutual information of the $N>2$ random walkers at first order in $\gamma$ (Eqs.~\eqref{eq:NRW_Ixx}-\eqref{eq:NRW_Ivx}), for various $\theta$ and $\gamma=10^{-3}$. Results are shown for $t=2$.}
    \label{fig:NRWS_MI_theta}
\end{figure}

Finally, the mutual information between the centre of mass at time $t$ and the position of a RW at time $\tp$ is
\begin{equation} \label{eq:NRW_Ivx}
    \begin{split}
        I(V(t);\Djtp) & = \frac12 \ln{\left( \frac{\var{V(t)}\var{\Djtp}}{|\Sigma_{V \Delta^{\prime}}|}\right)} \\
        & = \frac12 \ln{\left( \dfrac{
        \dfrac{\sigma^2}{N} \dfrac{1-\alpha^{2t}}{1-\alpha^2}
        \cdot
        \sigma^2 \dfrac{1-\alpha^{2\tp}}{1-\alpha^2} \left( 1 -\dfrac{4\gamma\alpha\left(1-\tp\alpha^{2(\tp-1)}+(\tp-1)\alpha^{2\tp} \right)}{(1-\alpha^2)(1-\alpha^{2\tp})}\right) }
        {\dfrac{\sigma^2}{N} \dfrac{1-\alpha^{2t}}{1-\alpha^2}
        \cdot
        \sigma^2 \dfrac{1-\alpha^{2\tp}}{1-\alpha^2} \left( 1 -\dfrac{4\gamma\alpha\left(1-\tp\alpha^{2(\tp-1)}+(\tp-1)\alpha^{2\tp} \right)}{(1-\alpha^2)(1-\alpha^{2\tp})}\right) 
        -\left(\dfrac{\sigma^2}{N}\right)^2 \left(\dfrac{1-\alpha^{2t}}{1-\alpha^2}\right)^2 \alpha^{2(\tp-t)} }
        \right)} \\
        & = \frac12 \ln{\left( 
        \dfrac{N(1-\alpha^{2\tp})}{N(1-\alpha^{2\tp})-\alpha^{2(\tp-t)}(1-\alpha^{2t})} \cdot \left(
        1+\dfrac{4\gamma\alpha\dfrac{\alpha^{2(\tp-t)}}{1-\alpha^2}\dfrac{1-\alpha^{2t}}{1-\alpha^{2\tp}}\left(1-\tp\alpha^{2(\tp-1)}+(\tp-1)\alpha^{2\tp} \right)}{N(1-\alpha^{2\tp})-\alpha^{2(\tp-t)}(1-\alpha^{2t})}
        \right) \right)} \\
        & = \frac12 \ln{\left( 
        \dfrac{N(1-\alpha^{2\tp})}{N(1-\alpha^{2\tp})-\alpha^{2(\tp-t)}(1-\alpha^{2t})}\right)} + \frac12 \ln{\left(
        1+\dfrac{4\gamma\alpha\dfrac{\alpha^{2(\tp-t)}}{1-\alpha^2}\dfrac{1-\alpha^{2t}}{1-\alpha^{2\tp}}\left(1-\tp\alpha^{2(\tp-1)}+(\tp-1)\alpha^{2\tp} \right)}{N(1-\alpha^{2\tp})-\alpha^{2(\tp-t)}(1-\alpha^{2t})}
        \right)} \\
        & = \frac12 \ln{\left( 
        \dfrac{N(1-\alpha^{2\tp})}{N(1-\alpha^{2\tp})-\alpha^{2(\tp-t)}(1-\alpha^{2t})}\right)} 
        + \dfrac{2\gamma\alpha\dfrac{\alpha^{2(\tp-t)}}{1-\alpha^2}\dfrac{1-\alpha^{2t}}{1-\alpha^{2\tp}}\left(1-\tp\alpha^{2(\tp-1)}+(\tp-1)\alpha^{2\tp} \right)}{N(1-\alpha^{2\tp})-\alpha^{2(\tp-t)}(1-\alpha^{2t})}
    \end{split} \,.
\end{equation}
Although these expressions could seem quite intricate, the key takeaway is simple: stronger restoring forces, i.e.\ smaller values of $\alpha\in(0,1]$, suppress the magnitude of the mutual information by erasing the memory of the system, and thus decrease the information transfer. 
More quantitatively, remembering $\alpha=1-\theta$, we show the exact behaviour of the MI for various $\theta$ and $\gamma=10^{-3}$ in Fig.~\ref{fig:NRWS_MI_theta}.

As expected, higher $\theta$ reduce the mutual information, while different $\gamma$ values have the same effect as in the non-stationary case studied in the main article. %

\subsubsection{$\theta\rightarrow0$ limit}
We can recover the expressions outlined in the main part of the work by taking the limit $\theta\rightarrow0$ (or equivalently $\alpha\rightarrow1$) in the equations above. 
This can be achieved by simply noting that:
\begin{gather} \label{eq:1-a_limit}
    1-\alpha^{2x} = 1-(1-\theta)^{2x} \underrel{\theta\to0}{=} 2x\theta + o(\theta) \\
    1-x\alpha^{2(x-1)}+(x-1)\alpha^{2x} = 1-x(1-\theta)^{2(x-1)}+(x-1)(1-\theta)^{2x} \underrel{\theta\to0}{=}  2x(x-1)\theta^2 + o(\theta^2)
\end{gather}
Thus, Eqs.~\eqref{eq:NRW_Ixx}-\eqref{eq:NRW_Ivx} become
\begin{align}
    \begin{split}
         I(\Delta_j(t);\Delta_j(\tp)) & \underrel{\alpha\to1}{=} \frac12 \ln{\left( \dfrac{2\tp}{2(\tp-t)}\right)} 
         -\dfrac{2\gamma \,2t}{2(\tp-t)}\left(\tp-t
         + \dfrac{2t(t-1)}{4t}-\dfrac{2\tp(\tp-1)}{4\tp} \right) \\
         & \,\,\, = \frac12 \ln{\left( \dfrac{\tp}{\tp-t}\right)} 
         -\dfrac{2\gamma t}{\tp-t}\left(\dfrac{\tp-t}{2}\right) \\
        & \,\,\, = \frac12\ln{\left(\frac{\tau}{\tau-1}\right)}-\gamma t \,,
    \end{split} \\
    \begin{split}
        I(V(t);V(\tp)) & \underrel{\alpha\to1}{=} \dfrac12 \ln{\left(\dfrac{2\tp}{2(\tp-t)}\right)} \\
        & \,\,\, = \frac12\ln{\left(\frac{\tau}{\tau-1}\right)}\,,
    \end{split} \\
    \begin{split}
        I(\Djt; V(\tp)) & \underrel{\alpha\to1}{=} \frac12 \ln{\left( 
        \dfrac{N\,2\tp}{N\,2\tp-2t}\right)} +
        \dfrac{2\gamma\dfrac{2t(t-1)}{2}}{N\,2\tp-2t} \\
        & \,\,\, \frac12\ln{\left(\frac{N\tau}{N\tau-1}\right)}+\gamma \frac{t-1}{N\tau-1} \,,
    \end{split} \\
    \begin{split}
        I(V(t);\Djtp) & \underrel{\alpha\to1}{=} \frac12 \ln{\left( 
        \dfrac{N\,2\tp}{N\,2\tp-2t}\right)} +
        \dfrac{2\gamma\dfrac{2t}{2\tp}\dfrac{2\tp(\tp-1)}{2}}{N\,2\tp-2t} \\
        & \,\,\, = \frac{1}{2}\ln{\left(\frac{N\tau}{N\tau-1}\right)}+\gamma \frac{\tp-1}{N\tau-1} \,,
    \end{split} 
\end{align}
where $\tau\equiv t^{\prime}/t$.

\subsection{Transfer entropy}  \label{app:N-RWcase_TE}
Here we report the calculations for TE between walkers and between one walker and the c.o.m.\, in both directions. 
Since the calculations for TE in the general OU case would be somewhat cumbersome, we focus on the simpler case in which $\theta\to0$.  
Making use of Eq.\eqref{eq:TE_def2}, the TE from one walker to another reads
\begin{equation}
\begin{split}
    \mathcal{T}(\Dit;\Djtp) & = H(\Djtp|\Djt) - H(\Djtp|\Djt,\Dit) \\
    & = \dfrac12 \lnp{2\pi e\, \sigma^2 t(\tau-1)(1-2\gamma(\tp-t-1))} + \\
    & \quad - \dfrac12 \lnp{2\pi e\, \sigma^2 t(\tau-1)(1-2\gamma(\tp-t-1))} \\
    & = 0 \,,
\end{split}
\end{equation}
where we employed
\begin{align}
    \begin{split}
        H(\Djtp|\Djt) & = H(\Djtp) - I(\Djt;\Djtp) \\
        & = \dfrac12 \lnp{2\pi e\, \sigma^2 \tp(1-2\gamma(\tp-1))} - \frac12\ln{\left(\frac{\tau}{\tau-1}(1-2\gamma t)\right)} \\
        & = \dfrac12 \lnp{2\pi e\, \sigma^2 t(\tau-1)(1-2\gamma(\tp-t-1))}
    \end{split} \\
    \begin{split}
        H(\Djtp|\Djt,\Dit) & =  H(\Djtp,\Djt,\Dit) -  H(\Djt,\Dit) \\
        & = \dfrac12 \lnp{\lrpars{2\pi e\,\sigma^2}^3 t^3(\tau-1) (1-2\gamma (\tp + t-3))} - \dfrac12 \lnp{\lrpars{2\pi e\,\sigma^2}^2 t^2 (1-2\gamma (2t-2))} \\
        & = \dfrac12 \lnp{2\pi e\, \sigma^2 t(\tau-1)(1-2\gamma(\tp-t-1))} \,.
    \end{split}
\end{align}
The transfer entropy from the c.o.m. to one walker is to first order in $\gamma$
\begin{equation}
\begin{split}
    \mathcal{T}(V(t);\Djtp) & = H(\Djtp|\Djt) - H(\Djtp|\Djt,V(t)) \\
    & = \dfrac12 \lnp{2\pi e\, \sigma^2 t(\tau-1)(1-2\gamma (\tp-t-1))} + \\
    & \quad - \dfrac12 \lnp{2\pi e\, \sigma^2 t(\tau-1) \lrpars{1-2\gamma(\tp-t-1)}} \\
    & = 0 \,,
\end{split}
\end{equation}
where we used
\begin{align}
    \begin{split}
        H(\Djtp|\Djt,V(t)) & = H(\Djtp,\Djt,V(t)) - H(\Djt,V(t)) \\
        & =  \dfrac12 \lnp{\left(2\pi e\, \sigma^2\right)^3 \lrpars{\dfrac{t}{N}}^3 N(\tau-1)\Bigl(N(1-2\gamma(\tp-2))-1+2\gamma(\tp-t-1)\Bigr)} + \\
        & \quad - \dfrac12 \lnp{\left(2\pi e\, \sigma^2\right)^2 \lrpars{\dfrac{t}{N}}^2 (N(1-2\gamma(t-1))-1)}  \\
        & = \dfrac12 \lnp{2\pi e\, \sigma^2 t(\tau-1) \dfrac{N(1-2\gamma(\tp-2))-1+2\gamma(\tp-t-1)}{N(1-2\gamma(t-1))-1}} \\
        & = \dfrac12 \lnp{2\pi e\, \sigma^2 t(\tau-1) \dfrac{(N-1)\lrpars{1-2\gamma\dfrac{N}{N-1}(\tp-2)+2\gamma\dfrac{1}{N-1}(\tp-t-1)}\lrpars{1+2\gamma \dfrac{N}{N-1}(t-1)}}{N-1}} \\
        & = \dfrac12 \lnp{2\pi e\, \sigma^2 t(\tau-1) \lrpars{1-2\gamma(\tp-t-1)}} \,.
    \end{split}
\end{align}
Finally, the TE from one random walker to the centre of mass reads 
\begin{flushleft}
\begin{equation}
\begin{split}
    \mathcal{T}(\Djt;V(\tp)) & = H(V(\tp)|V(t)) - H(V(\tp)|V(t),\Djt) \\
    & = \dfrac12 \ln\left(2\pi e\, \sigma^2 t\dfrac{\tau-1}{N} \right) - \dfrac12 \ln\left(2\pi e\, \sigma^2 t\dfrac{\tau-1}{N} \right) \\
    & = 0 \,,
\end{split}
\end{equation}
\end{flushleft}
where we used
\begin{align}
    \begin{split}
    H(V(\tp)|V(t)) &= H(V(\tp)) - I(V(t);V(\tp)) \\
    & = \dfrac12 \lnp{2\pi e\, \frac{\sigma^2}{N} \tp} - \dfrac12 \lnp{2\pi e\, \frac{\sigma^2}{N} \frac{\tau}{\tau-1}} \\
    & = \dfrac12 \ln\left(2\pi e\, \sigma^2 t\dfrac{\tau-1}{N} \right)
    \end{split} \\
    \begin{split}
    H(V(\tp)|V(t),\Djt) & = H(V(\tp),V(t),\Djt) - H(V(t),\Djt) \\
    & = \dfrac12 \lnp{\left(2\pi e\, \sigma^2\right)^3 \lrpars{\dfrac{t}{N}}^3 (\tau-1)(N(1-2\gamma(t-1))-1)} + \\
    & \quad - \dfrac12 \lnp{\left(2\pi e\, \sigma^2\right)^2 \lrpars{\dfrac{t}{N}}^2 (N(1-2\gamma(t-1))-1)}\\
    & = \dfrac12 \ln\left(2\pi e\, \sigma^2 t\dfrac{\tau-1}{N} \right)
    \end{split} \,.
\end{align}
In all three cases above, we obtained that the transfer entropy vanishes. This result could have been expected as TE is a second-order quantity in $\gamma$.  %

\subsection{Causal emergence quantities}  \label{app:N-RWcase_emergence}
\begin{figure}[ht]
    \centering
    \includegraphics[width=1\linewidth]{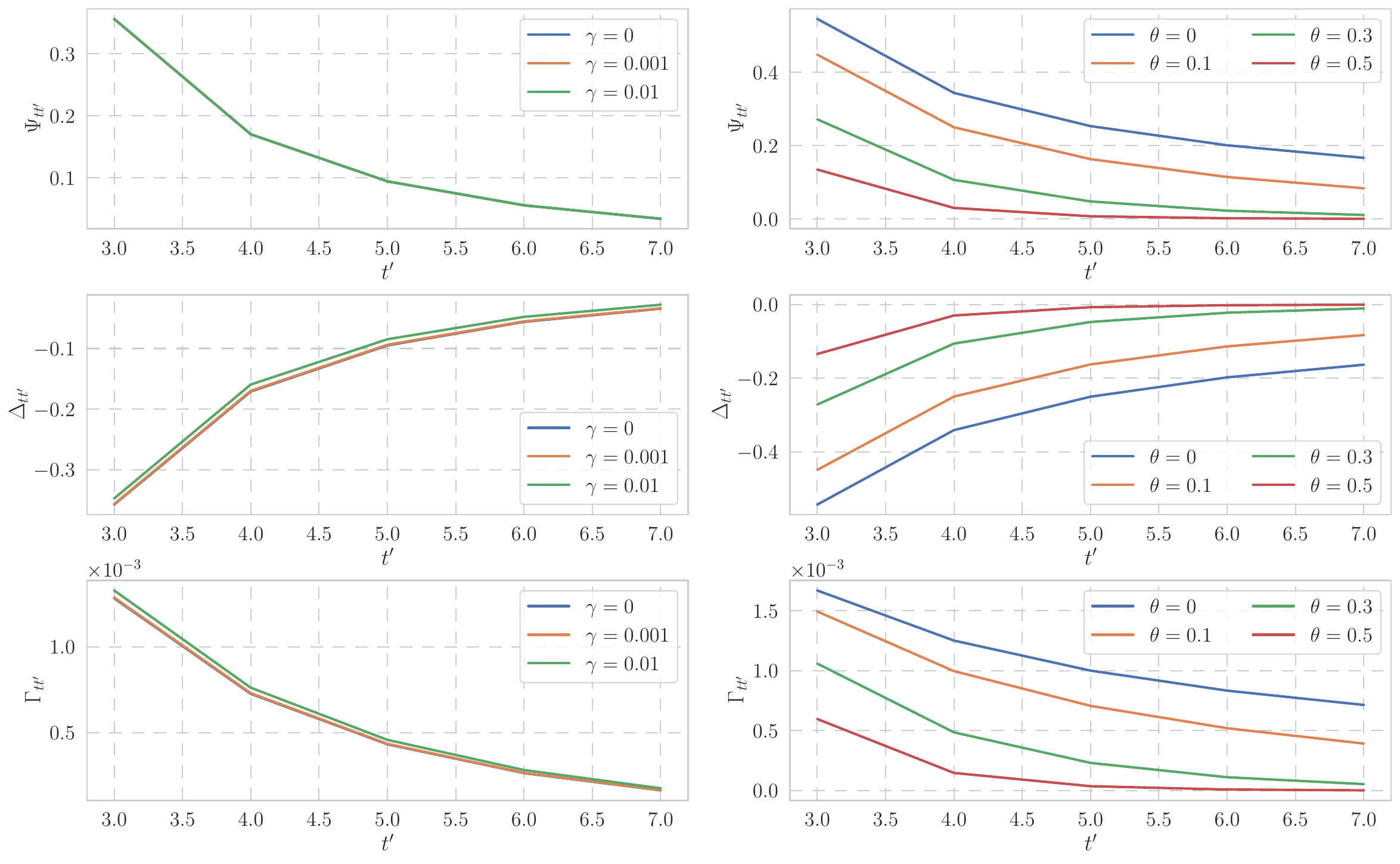}
    \caption{\textbf{Emergence measures behave qualitatively the same as in the non-stationary case.}
    $\Psi_{t\tp}, \Delta_{t\tp}, \Gamma_{t\tp}$ in the case of $N$ random walkers for various choices of $(\theta, \gamma)$. Left column: various $\gamma$ for fixed $\theta=0.2$. Right column: Various $\theta$ for fixed $\gamma=0.001$. Results are shown for $t=2$.} %
    \label{fig:NRWs_Psi_Delta_Gamma}
\end{figure}

In this section we provide the results for the $\Psi, \Gamma, \Delta$ causal emergent quantities in the more general case in which $\theta\ne0$. 
\begin{align}
\allowdisplaybreaks
    \begin{split}
        \Psi_{t\tp} & = I(V(t); V(t^{\prime})) - \sum_{j=1}^N I(\Delta_j(t); V(t')) \\
        & = \dfrac12 \ln{\left(\dfrac{1-\alpha^{2\tp}}{1-\alpha^{2(\tp-t)}}\right)} - \frac{N}{2} \ln{\left( 
        \dfrac{N(1-\alpha^{2\tp})}{N(1-\alpha^{2\tp})-\alpha^{2(\tp-t)}(1-\alpha^{2t})}\right)} -
        N\dfrac{2\gamma\alpha\dfrac{\alpha^{2(\tp-t)}}{1-\alpha^2}\left(1-t\alpha^{2(t-1)}+(t-1)\alpha^{2t} \right)}{N(1-\alpha^{2\tp})-\alpha^{2(\tp-t)}(1-\alpha^{2t})} \\
        & \underrel{N\to+\infty}{=} \dfrac12 \ln{\left(\dfrac{1-\alpha^{2\tp}}{1-\alpha^{2(\tp-t)}}\right)}-\dfrac{\alpha^{2(\tp-t)}(1-\alpha^{2t})}{(1-\alpha^{2\tp})} -
        \dfrac{2\gamma\alpha\,\alpha^{2(\tp-t)}\left(1-t\alpha^{2(t-1)}+(t-1)\alpha^{2t} \right)}{(1-\alpha^2)(1-\alpha^{2\tp})}
    \end{split} \\
    \begin{split} 
    \Delta_{t\tp} & = \max_i \Bigl( I(V(t);\Delta_i(t')) - \sum_{j=1}^N I(\Delta_j(t);\Delta_i(t')) \Bigr) \\
    & = \frac12 \ln{\left( 
    \dfrac{N(1-\alpha^{2\tp})}{N(1-\alpha^{2\tp})-\alpha^{2(\tp-t)}(1-\alpha^{2t})}\right)} 
    + \dfrac{2\gamma\alpha\dfrac{\alpha^{2(\tp-t)}}{1-\alpha^2}\dfrac{1-\alpha^{2t}}{1-\alpha^{2\tp}}\left(1-\tp\alpha^{2(\tp-1)}+(\tp-1)\alpha^{2\tp} \right)}{N(1-\alpha^{2\tp})-\alpha^{2(\tp-t)}(1-\alpha^{2t})} - \frac12 \ln{\left( \dfrac{1-\alpha^{2\tp}}{1-\alpha^{2(\tp-t)}}\right)} + \\
    & \quad \quad +\dfrac{2\gamma(1-\alpha^{2t})\alpha^{2(\tp-t)}}{1-\alpha^{2(\tp-t)}}\left(\dfrac{\tp-t}{\alpha}
    + \dfrac{\alpha\left(1-t\alpha^{2(t-1)}+(t-1)\alpha^{2t} \right)}{1-\alpha^{2t}(1- \alpha^2)}
    -\dfrac{\alpha\left(1-\tp\alpha^{2(\tp-1)}+(\tp-1)\alpha^{2\tp} \right)}{(1-\alpha^{2\tp})(1- \alpha^2)} \right) \\
    & \underrel{N\to+\infty}{=} - \frac12 \ln{\left( \dfrac{1-\alpha^{2\tp}}{1-\alpha^{2(\tp-t)}}\right)} + \\
    & \quad \quad +\dfrac{2\gamma(1-\alpha^{2t})\alpha^{2(\tp-t)}}{1-\alpha^{2(\tp-t)}}\left(\dfrac{\tp-t}{\alpha}
    + \dfrac{\alpha\left(1-t\alpha^{2(t-1)}+(t-1)\alpha^{2t} \right)}{1-\alpha^{2t}(1- \alpha^2)}
    -\dfrac{\alpha\left(1-\tp\alpha^{2(\tp-1)}+(\tp-1)\alpha^{2\tp} \right)}{(1-\alpha^{2\tp})(1- \alpha^2)} \right) 
    \end{split} \displaybreak \\
    \begin{split} 
        \Gamma_{t\tp} &= \max_i \Bigl(I(V(t); \Delta_i(t'))\Bigr) \\
        & = \frac12 \ln{\left( 
    \dfrac{N(1-\alpha^{2\tp})}{N(1-\alpha^{2\tp})-\alpha^{2(\tp-t)}(1-\alpha^{2t})}\right)} 
    + \dfrac{2\gamma\alpha\dfrac{\alpha^{2(\tp-t)}}{1-\alpha^2}\dfrac{1-\alpha^{2t}}{1-\alpha^{2\tp}}\left(1-\tp\alpha^{2(\tp-1)}+(\tp-1)\alpha^{2\tp} \right)}{N(1-\alpha^{2\tp})-\alpha^{2(\tp-t)}(1-\alpha^{2t})} \\
    & \underrel{N\to+\infty}{=} 0
    \end{split}
\end{align}
From analytical considerations, we already observe the same qualitative results as for the non-OU case: $\Psi_{t\tp}$ is positive and decreases for stronger interactions, $\Delta_{t\tp}$ is negative, and $\Gamma_{t\tp}$ is negative and vanishes in the limit $N\to+\infty$. 
Thus, we obtain that also in the stationary case $V$ is an emergent property of the system, and that causal decoupling emerges at $N\to+\infty$.
More quantitatively, Fig.~\ref{fig:NRWs_Psi_Delta_Gamma} shows how these quantities behave for various parameters $(\theta, \gamma)$, confirming the comments above.

Finally, by taking the limit $\theta\rightarrow0$, we obtain the equations shown in Sec.~\ref{sec:results_emergence}:
\begin{align}
\allowdisplaybreaks
    \begin{split}
    \Psi_{t\tp} & = \underbrace{\frac{1}{2}\ln{\left(\frac{\tau}{\tau-1}\right)}+\frac{1}{2}\ln{\left(1-\frac{1}{N\tau}\right)}^{N}}_{\Psi^0_{t\tp}} - \dfrac{\gamma(t-1)}{\tau}  \underset{N\rightarrow\infty}\longrightarrow \frac12\left[\ln{\left(\frac\tau{\tau-1}\right)}-\frac1\tau\right] - \dfrac{\gamma(t-1)}{\tau} \,, 
    \end{split} \\ \displaybreak %
    \begin{split} \label{eq:app_NRW_delta}
    \Delta_{t\tp} & = \underbrace{\frac{1}{2}\ln{\left(\frac{N(\tau-1)}{N\tau-1}\right)}}_{\Delta^0_{t\tp}} + \gamma \left( \frac{t^{\prime}-1}{N\tau-1}+t\right)  \underset{N\rightarrow\infty}\longrightarrow -\frac12\ln{\left(\frac\tau{\tau-1}\right)} + \gamma t \,,
    \end{split} \\
    \begin{split} \label{eq:app_NRW_gamma}
    \Gamma_{t\tp} & = \underbrace{\frac12\ln{\left(\frac{N\tau}{N\tau-1}\right)}}_{\Gamma^0_{t\tp}}  + \gamma \frac{t^{\prime}-1}{N\tau-1}  \underset{N\rightarrow\infty}\longrightarrow 0 \,.
    \end{split}
\end{align}

\subsection{Partial Information Decomposition}  \label{app:N-RWcase_PID}

We present here the computations for the three PID cases outlined in Sec.~\ref{sec:results_pid}.

\textbf{(\underline{1})}
Recalling 
\begin{equation}
    \begin{split}
        X&=\Delta_j(t), \quad Y=\Delta_i(t), \quad\text{and} \\
        Z&=V(t^{\prime}), \, i\ne j \,,
    \end{split}
\end{equation}
We compute the necessary mutual information:
\begin{equation} \label{eq:N_rw_Ixxv}
    \begin{split}
         I(\Delta_j(t),\Delta_i(t);V(t^{\prime})) & = \dfrac{1}{2} \lnp{\dfrac{|\Sigma_{\Delta_j \Delta_i}| \, \var{V(\tp)}}{|\Sigma_{\Delta_j \Delta_i V^{\prime}}|}} \\
         & = \dfrac{1}{2} \lnp{\dfrac{\left(\sigma^2 t\right)^2 \left(1 - 4\gamma(t-1)\right) \dfrac{\sigma^2 \tp}{N}}{\left(\dfrac{\sigma^2 t}{N}\right)^3 N \left(N\tau - 2 - 2\gamma(t-1)\left(2N\tau - 3\right)\right)}} \\
         & = \dfrac12 \lnp{\dfrac{N\tau(1 - 4\gamma(t-1))}{(N\tau-2)\left(1-2\gamma (t-1)\dfrac{2N\tau-3}{N\tau-2}\right)}} \\
         & = \dfrac12 \lnp{\dfrac{N\tau}{N\tau-2}} + \dfrac12 \lnp{(1 - 4\gamma(t-1))\left(1+2\gamma (t-1)\dfrac{2N\tau-3}{N\tau-2}\right)} \\
         & = \dfrac{1}{2} \lnp{\dfrac{N\tau}{N\tau - 2}} + \gamma \dfrac{t-1}{N\tau - 2} \,.
    \end{split}
\end{equation}
Using Eqs.~\eqref{eq:N_MI_xv}-\eqref{eq:N_rw_Ixxv} we then obtain
\begin{align}
    \begin{split}
        \text{Red}(\Delta_j(t),\Delta_i(t);V(t^{\prime})) & = I(\Djt; V(\tp)) \\
        & = \frac{1}{2}\ln{\left(\frac{N\tau}{N\tau-1}\right)}+\gamma \frac{t-1}{N\tau-1}
    \end{split} \\
    \begin{split}
        \text{Syn}(\Delta_j(t),\Delta_i(t);V(t^{\prime})) & = I(\Delta_j(t),\Delta_i(t);V(t^{\prime})) - I(\Djt; V(\tp)) \\
        & = \frac{1}{2}\ln{\left(\frac{N\tau-1}{N\tau-2}\right)}+\gamma \frac{t-1}{N\tau-2}- \gamma \frac{t-1}{N\tau-1} 
    \end{split}
\end{align}

To obtain the normalised quantities, we simply divide each atom by the joint mutual information. We indicate these normalised quantities as \emph{NMI} (Normalised by Mutual Information). 
\begin{align}
    \begin{split}
        \text{Red}_{\text{NMI}}(\Delta_j(t),\Delta_i(t);V(t^{\prime})) & \equiv \dfrac{\text{Red}(\Delta_j(t),\Delta_i(t);V(t^{\prime}))}{I(\Delta_j(t),\Delta_i(t);V(t^{\prime}))} \\
        & = \dfrac{\dfrac{1}{2}\ln{\left(\dfrac{N\tau}{N\tau-1}\right)}+\gamma \dfrac{t-1}{N\tau-1}}{\dfrac{1}{2} \lnp{\dfrac{N\tau}{N\tau - 2}} + \gamma \dfrac{t-1}{N\tau - 2}} \\
        & = 
        \dfrac{\dfrac{1}{2}\ln{\left(\dfrac{N\tau}{N\tau-1}\right)}+\gamma(t-1)\left(\dfrac{1}{N\tau-1}-\dfrac{1}{N\tau-2}\dfrac{\ln{\left(\dfrac{N\tau}{N\tau-1}\right)}}{\ln{\left(\dfrac{N\tau}{N\tau-2}\right)}}\right)}
        {\dfrac{1}{2}\ln{\left(\dfrac{N\tau}{N\tau-2}\right)}}
    \end{split} \\
    \begin{split}
        \text{Syn}_{\text{NMI}}(\Delta_j(t),\Delta_i(t);V(t^{\prime})) & \equiv \dfrac{\text{Syn}(\Delta_j(t),\Delta_i(t);V(t^{\prime}))}{I(\Delta_j(t),\Delta_i(t);V(t^{\prime}))} \\
        & = \dfrac{\dfrac{1}{2}\ln{\left(\dfrac{N\tau-1}{N\tau-2}\right)}+\gamma \dfrac{t-1}{N\tau-2}- \gamma \dfrac{t-1}{N\tau-1}}
        {\dfrac{1}{2} \lnp{\dfrac{N\tau}{N\tau - 2}} + \gamma \dfrac{t-1}{N\tau - 2}} \\
        & = \dfrac{\dfrac{1}{2}\ln{\left(\dfrac{N\tau-1}{N\tau-2}\right)}-\gamma (t-1)\left(\dfrac{1}{N\tau-1}+\dfrac{1}{N\tau - 2}\dfrac{\ln{\left(\dfrac{N\tau-1}{N\tau-2}\right)}}{\lnp{\dfrac{N\tau}{N\tau - 2}}}-\dfrac{1}{N\tau-2}\right)}
        {\dfrac{1}{2} \lnp{\dfrac{N\tau}{N\tau - 2}} } \,,
    \end{split}
\end{align}
where we factorised the terms in $\gamma$ so that the multiplying parentheses are positive. Thus $\text{Red}_{\text{NMI}}$ increases with the interaction strength, while $\text{Syn}_{\text{NMI}}$ decreases. 

\textbf{(\underline{2})}
In this case we look at

\begin{equation} 
    \begin{split}
        X&=\Delta_j(t), \quad Y=V(t), \quad \text{and} \\
        Z&=\{\Delta_j(t^{\prime}), V(t^{\prime})\} \, .
    \end{split}
\end{equation}
In this situation we have
\begin{align}
\begin{split}
\allowdisplaybreaks
    I(\Delta_j(t);\{\Delta_j(t^{\prime}),V(t^{\prime})\}) & = \logdet{\var{\Djt}}{|\Sigma_{\Delta^{\prime}V^{\prime}}|}{|\Sigma_{\Delta \Delta^{\prime}V^{\prime}}|} \\
    & = \dfrac12 \lnp{\dfrac{\lrpars{\dfrac{\sigma^2 t}{N}} t(1-2\gamma(t-1)) \,\lrpars{\dfrac{\sigma^2 t}{N}}^2 \tau^2 (N-1)\lrpars{1-2\gamma \dfrac{N(\tp-1)}{N-1}}}{\lrpars{\dfrac{\sigma^2 t}{N}}^3 \tau (\tau-1)N(N-1)\lrpars{1-2\gamma \dfrac{N(\tp-2)+1}{N-1}}}} \\
    & = \dfrac12 \lnp{\dfrac{\tau}{\tau-1}} + \lnp{\lrpars{1-2\gamma(t-1)}\lrpars{1-\dfrac{2\gamma N(\tp-1)}{N-1}}\lrpars{1+\dfrac{2\gamma(N(\tp-2)+1)}{N-1}}} \\
    & = \dfrac12 \lnp{\dfrac{\tau}{\tau-1}} + \lnp{1+\dfrac{-2\gamma(t-1)(N-1)-2\gamma N+2\gamma}{N-1}} \\
    & = \frac{1}{2}\ln{\left(\frac{\tau}{\tau-1}\right)}-\gamma t \\
    & = I(\Delta_j(t);\Delta_j(t^{\prime})) 
    \end{split}  \displaybreak \\
    \begin{split}
    I(V(t);\{\Delta_j(t^{\prime}),V(t^{\prime})\}) & = \logdet{\var{V(t)}}{|\Sigma_{\Delta^{\prime}V^{\prime}}|}{|\Sigma_{V \Delta^{\prime}V^{\prime}}|} \\
    & = \dfrac12 \lnp{\dfrac{\lrpars{\dfrac{\sigma^2 t}{N}} \,\lrpars{\dfrac{\sigma^2 t}{N}}^2 \tau^2 (N-1)\lrpars{1-2\gamma \dfrac{N(\tp-1)}{N-1}}}{\lrpars{\dfrac{\sigma^2 t}{N}}^3 \tau (\tau-1)(N-1)\lrpars{1-2\gamma \dfrac{N(\tp-1)}{N-1}}}} \\
    & = \frac{1}{2}\ln{\left(\frac{\tau}{\tau-1}\right)} \\
    & = I(V(t);V(t^{\prime})) 
    \end{split}  \\
    \begin{split}
    I(\Delta_j(t), V(t);\{\Delta_j(t^{\prime}),V(t^{\prime})\}) & = \logdet{|\Sigma_{\Delta V}|}{|\Sigma_{\Delta^{\prime}V^{\prime}}|}{|\Sigma_{\Delta V \Delta^{\prime}V^{\prime}}|} \\
    & = \dfrac12 \lnp{\dfrac{\lrpars{\dfrac{\sigma^2 t}{N}}^2 \lrpars{1-2\gamma \dfrac{N(t-1)}{N-1}} \,\lrpars{\dfrac{\sigma^2 t}{N}}^2 \tau^2 (N-1)\lrpars{1-2\gamma \dfrac{N(\tp-1)}{N-1}}}{\lrpars{\dfrac{\sigma^2 t}{N}}^4 (\tau-1)^2(N-1)^2\lrpars{1-2\gamma \dfrac{N(\tp-2)}{N-1}}}} \\
    & = \lnp{\dfrac{\tau}{\tau-1}} + \lnp{1-\dfrac{2\gamma N(t-1)+2\gamma N(\tp-1)-2\gamma N (\tp-2)}{N-1}} \\
    & = \ln{\left(\frac{\tau}{\tau-1}\right)} - \gamma \frac{Nt}{N-1}\,.
    \end{split}
\end{align}
Redundancy and synergy then read
\begin{align}
\begin{split}
    \text{Red}(\Delta_j(t), V(t);\{\Delta_j(t^{\prime}),V(t^{\prime})\}) = & I(\Delta_j(t);\{\Delta_j(t^{\prime}),V(t^{\prime})\}) \\
    = & \frac{1}{2}\ln{\left(\frac{\tau}{\tau-1}\right)}-\gamma t 
\end{split} \\
\begin{split}
    \text{Syn}(\Delta_j(t), V(t);\{\Delta_j(t^{\prime}),V(t^{\prime})\}) = & I(\Delta_j(t), V(t);\{\Delta_j(t^{\prime}),V(t^{\prime})\}) - I(V(t);\Delta_j(t^{\prime})) \\
    = & \frac{1}{2}\ln{\left(\frac{\tau}{\tau-1}\right)} -\gamma \frac{Nt}{N-1} \,. 
\end{split}
\end{align}

The normalised quantities can be computed as
\begin{align}
\allowdisplaybreaks
\begin{split}
        \text{Red}_{\text{NMI}}(\Delta_j(t), V(t);\{\Delta_j(t^{\prime}),V(t^{\prime})\}) & \equiv \dfrac{\text{Red}(\Delta_j(t), V(t);\{\Delta_j(t^{\prime}),V(t^{\prime})\})}{I(\Delta_j(t), V(t);\{\Delta_j(t^{\prime}),V(t^{\prime})\})} \\
        & = \dfrac{\dfrac{1}{2}\ln{\left(\dfrac{\tau}{\tau-1}\right)} -\gamma t}{\ln{\left(\dfrac{\tau}{\tau-1}\right)} - \gamma \dfrac{Nt}{N-1}} \\
        & = \dfrac{\dfrac{1}{2}\ln{\left(\dfrac{\tau}{\tau-1}\right)} -\gamma t\left(1-\dfrac12\dfrac{N}{N-1}\right)}{\ln{\left(\dfrac{\tau}{\tau-1}\right)}}
    \end{split} \displaybreak \\
    \begin{split}
        \text{Syn}_{\text{NMI}}(\Delta_j(t), V(t);\{\Delta_j(t^{\prime}),V(t^{\prime})\}) & \equiv \dfrac{\text{Syn}(\Delta_j(t), V(t);\{\Delta_j(t^{\prime}),V(t^{\prime})\})}{I(\Delta_j(t), V(t);\{\Delta_j(t^{\prime}),V(t^{\prime})\})} \\
        & = \dfrac{\dfrac{1}{2}\ln{\left(\dfrac{\tau}{\tau-1}\right)} -\gamma \dfrac{Nt}{N-1}}{\ln{\left(\dfrac{\tau}{\tau-1}\right)} - \gamma \dfrac{Nt}{N-1}} \\
        & = \dfrac{\dfrac{1}{2}\ln{\left(\dfrac{\tau}{\tau-1}\right)} -\gamma \dfrac12\dfrac{Nt}{N-1}}{\ln{\left(\dfrac{\tau}{\tau-1}\right)}} \,,
    \end{split}
\end{align}
showing that both normalised redundancy and synergy decrease with $\gamma$.

\textbf{(\underline{3})}
Finally, we focus on
\begin{equation} 
\begin{split}
    X&=\Delta_j(t), \quad {Y=\Delta_i(t)}, \quad\text{and}\\ Z&=\{\Delta_j(t^{\prime}), \Delta_i(t^{\prime})\}, \,i\ne j \,.
\end{split}
\end{equation}
The mutual information read
\begin{align}
    \begin{split}
        I(\Djt; \Djtp, \Ditp) & = \dfrac12 \lnp{\dfrac{\var{\Djt} |\Sigma_{\Delta_j^{\prime} \Delta_i^{\prime}}|}{|\Sigma_{\Delta_j \Delta_j^{\prime} \Delta_i^{\prime}}|}} \\
        & = \dfrac12 \lnp{\dfrac{\sigma^2 t(1-2\gamma(t-1)) \cdot (\sigma^2 \tp)^2 (1-4\gamma(\tp-1))}{(\sigma^2 t)^3 \tau (\tau-1) (1-2\gamma (2\tp-3))}} \\
        & = \dfrac12 \lnp{\dfrac{\tau}{\tau-1}} + \dfrac12 \ln \Bigl(1-2\gamma(t-1)-4\gamma(\tp-1)+2\gamma(2\tp-3)\Bigr) \\
        & = \dfrac12 \lnp{\dfrac{\tau}{\tau-1}} - \gamma t \\
        & = I(\Djt; \Djtp)
    \end{split} \\
    \begin{split}
        I(\Djt, \Dit; \Djtp, \Ditp) & = \dfrac12 \lnp{\dfrac{|\Sigma_{\Delta_j \Delta_i}| |\Sigma_{\Delta_j^{\prime} \Delta_i^{\prime}}|}{|\Sigma_{\Delta_j\Delta_i \Delta_j^{\prime} \Delta_i^{\prime}}|}} \\
        & = \dfrac12 \lnp{\dfrac{(\sigma^2 t)^2(1-4\gamma(t-1)) \cdot (\sigma^2 \tp)^2 (1-4\gamma(\tp-1))}{(\sigma^2 t)^4 (\tau-1)^2 (1-4\gamma (\tp-2))}} \\
        & = \lnp{\dfrac{\tau}{\tau-1}} + \dfrac12 \ln \Bigl(1-4\gamma(t-1)-4\gamma(\tp-1)+4\gamma(\tp-2)\Bigr) \\
        & = \lnp{\dfrac{\tau}{\tau-1}} - 2\gamma t \\
        & = 2I(\Djt; \Djtp) \,.
    \end{split}
\end{align}
Thus, redundancy and synergy are
\begin{align}
    \begin{split}
        \text{Red}(\Delta_i(t),\Delta_j(t);\{\Delta_i(t^{\prime}),\Delta_j(t^{\prime})\}) & = I(\Delta_j(t);\{\Delta_i(t^{\prime}),\Delta_j(t^{\prime})\}) \\
        & = I(\Delta_j(t);\Delta_j(t^{\prime})\}) \\
        & = \frac{1}{2}\ln{\left(\frac{\tau}{\tau-1}\right)}-\gamma t 
    \end{split} \\
    \begin{split}
        \text{Syn}(\Delta_i(t),\Delta_j(t);\{\Delta_i(t^{\prime}),\Delta_j(t^{\prime})\}) & = I(\Delta_i(t),\Delta_j(t);\Delta_i(t^{\prime}),\Delta_j(t^{\prime})) - I(\Delta_j(t);\Delta_j(t^{\prime})) \\
        & = 2I(\Delta_j(t);\Delta_j(t^{\prime}))-I(\Delta_j(t);\Delta_j(t^{\prime})) \\
        & = I(\Delta_j(t);\Delta_j(t^{\prime})) \\
        & = \frac{1}{2}\ln{\left(\frac{\tau}{\tau-1}\right)}-\gamma t \,.
    \end{split} 
\end{align}

Finally, we turn to the normalised atoms:
\begin{align}
    \begin{split}
        \text{Red}_{\text{NMI}}(\Delta_i(t), \Delta_j(t);\{\Delta_i(t^{\prime}),\Delta_j(t^{\prime})\}) & \equiv \dfrac{\text{Red}(\Delta_i(t), \Delta_j(t);\{\Delta_i(t^{\prime}),\Delta_j(t^{\prime})\})}{I(\Delta_i(t), \Delta_j(t);\{\Delta_i(t^{\prime}),\Delta_j(t^{\prime})\})} \\
        & =  \dfrac{\dfrac{1}{2}\ln{\left(\dfrac{\tau}{\tau-1}\right)}-\gamma t}{\lnp{\dfrac{\tau}{\tau-1}} - 2\gamma t} \\
        & = \dfrac12
    \end{split} \\
    \begin{split}
        \text{Syn}_{\text{NMI}}(\Delta_i(t), \Delta_j(t);\{\Delta_i(t^{\prime}),\Delta_j(t^{\prime})\}) & =  \text{Red}_{\text{NMI}}(\Delta_i(t), \Delta_j(t);\{\Delta_i(t^{\prime}),\Delta_j(t^{\prime})\}) \\
        & = \dfrac12 \,.
    \end{split} 
\end{align}
We obtained that both atoms are constant in $\gamma$, which is a consequence of the first order approximation in $\gamma$. Since for small coupling strength there is complete symmetry between synergy and redundancy, each contributes to half the information in the system, hence resulting in $\text{Red}_{\text{NMI}}=\text{Syn}_{\text{NMI}}=1/2$.  

The discussion of these equations and their broader implications are discussed in the main body of the work (Secs.~\ref{sec:results_pid}-\ref{sec:discussion}).

\subsection{Useful identities}
We report here two identities which are helpful when computing the mutual information at first order.
Considering $F_v, F_c$ as the terms coming from variance and covariance expressions which \textit{do not} contain $\gamma$, and $\epsilon_v, \epsilon_c$ the parts that \textit{do}, the following holds
\begin{equation} \label{eq:NRW_trick1}
    \begin{split}
        \dfrac{F_v(1+\epsilon_v)}{F_v(1+\epsilon_v)-F_c(1+\epsilon_c)} & = 
        \dfrac{F_v(1+\epsilon_v)}{(F_v-F_c)\left(1+\dfrac{F_v \epsilon_v-F_c \epsilon_c}{F_v-F_c}\right)} \\
        & = \dfrac{F_v(1+\epsilon_v)\left(1-\dfrac{F_v \epsilon_v-F_c \epsilon_c}{F_v-F_c}\right)}{F_v-F_c} + o(\epsilon) \\
        & = \dfrac{F_v}{F_v-F_c} \cdot \left(1+\epsilon_v-\dfrac{F_v \epsilon_v-F_c \epsilon_c}{F_v-F_c}\right) + o(\epsilon) \\
        & = \dfrac{F_v}{F_v-F_c} \cdot \left(1+ \dfrac{F_v\epsilon_v-F_c\epsilon_v-F_v\epsilon_v+F_c \epsilon_c}{F_v-F_c}\right) + o(\epsilon) \\
        & = \dfrac{F_v}{F_v-F_c} \cdot \left(1+\dfrac{F_c (\epsilon_c-\epsilon_v)}{F_v-F_c} \right) + o(\epsilon) \,.
    \end{split}
\end{equation}
where we expanded everything at first order in $\epsilon_v, \epsilon_c$ (i.e. $\gamma$).
By setting $\epsilon_c=0$, we obtain another useful expression
\begin{equation} \label{eq:NRW_trick2}
    \dfrac{F_v(1+\epsilon_v)}{F_v(1+\epsilon_v)-F_c} = \dfrac{F_v}{F_v-F_c} \cdot \left(1-\dfrac{F_c \epsilon_v}{F_v-F_c} \right) + o(\epsilon) \,.
\end{equation}

\section{$N=2$ random walkers} \label{app:2-RWcase}
In this section, we focus on a simpler system composed of only 2 random walkers. Although the dynamic is the same as the one studied in the $N$ case, the presence of only two particles allows us to perform exact calculations at all orders in $\theta, \gamma$.  
Clearly, no multi-component mechanical interactions can be present in such a system. However, it is still interesting to see how information theoretic quantities depend on multivariate terms, and if their behaviour changes when a larger number of components is then considered. 
Drawing frequent connections with the $N$ case studied above, we show this scenario of 2 RWs complements the previous analysis by tackling information measures from a slightly different angle.

\subsection{Model} \label{app:2-RWcase_model}
    \begin{numcases}{}
        x_1(t+1) = \bigl(1-\theta\bigr)x_1(t) + \gamma \bigl(x_2(t)-x_1(t)-a\bigr) + \eta_1(t) \\
        x_2(t+1) = \bigl(1-\theta\bigr)x_2(t) - \gamma \bigl(x_2(t)-x_1(t)-a\bigr) + \eta_2(t) \,,
    \end{numcases}
as before, we express the equations of motion in terms of the displacement from their resting position $\Delta_i(t) = x_i(t)-a$:
    \begin{numcases}{} \label{eq:2RWs_delta1_tt1}
        \Delta_1(t+1) = \bigl(1-\theta\bigr)\Delta_1(t) + \gamma \bigl(\Delta_2(t)-\Delta_1(t)\bigr) + \eta_1(t) \\
        \Delta_2(t+1) = \bigl(1-\theta\bigr)\Delta_2(t) - \gamma \bigl(\Delta_2(t)-\Delta_1(t)\bigr) + \eta_2(t) \,. \label{eq:2RWs_delta2_tt1}
    \end{numcases}
We can decouple the system by using the coordinates in the center-of-mass frame
\begin{gather}
    \nu(t) \equiv \dfrac{1}{\sqrt{2}} (\Delta_1(t)+\Delta_2(t)) = \dfrac{1}{\sqrt{2}} (x_1(t)+x_2(t)-a) \\
    \rho(t) \equiv \dfrac{1}{\sqrt{2}} (\Delta_2(t)-\Delta_1(t)) = \dfrac{1}{\sqrt{2}} (x_2(t)+x_1(t)-a) \,,
\end{gather}
and their inverse
\begin{gather}
    \Delta_1(t) = \dfrac{\nu(t)-\rho(t)}{\sqrt{2}} \\
    \Delta_2(t) = \dfrac{\nu(t)+\rho(t)}{\sqrt{2}} \,.
\end{gather}
Note that $\rho$ and $\nu$ are not exactly the relative positions and center of mass of the displacements $\Delta$, as they differ by a constant and additive factors: ${V(t) = \nu(t)/\sqrt{2}+a}$ and ${r(t) = \sqrt{2}\rho(t)+a}$. 

However, using the new coordinates, the equations of motion become
\begin{numcases}{} 
\displaystyle
    \begin{split}
        \nu(t+1) & = \dfrac{1}{\sqrt{2}} (\Delta_1(t+1)+\Delta_2(t+1)) \\
        & = \dfrac{1}{\sqrt{2}} \Bigl( \bigl(1-\theta\bigr)\Delta_1(t) + \gamma \bigl(\Delta_2(t)-\Delta_1(t)\bigr) + \eta_1(t) + \bigl(1-\theta\bigr)\Delta_2(t) - \gamma \bigl(\Delta_2(t)-\Delta_1(t)\bigr) + \eta_2(t) \Bigr) \\
        & =  \dfrac{1}{\sqrt{2}} \Bigl((1-\theta)\bigl(\Delta_1(t)+\Delta_2(t)\bigr) + \eta_1(t) + \eta_2(t)\Bigr) \\
        & = (1-\theta)\nu(t) + \eta_+(t) \,,
    \end{split} \label{eq:2RW_nu_1step} \\
    \displaystyle
    \begin{split}
        \rho(t+1) & = \dfrac{1}{\sqrt{2}} (\Delta_2(t+1)-\Delta_1(t+1)) \\
        & = \dfrac{1}{\sqrt{2}} \Bigl(\bigl(1-\theta\bigr)\Delta_2(t) - \gamma \bigl(\Delta_2(t)-\Delta_1(t)\bigr) + \eta_2(t) -  \bigl(1-\theta\bigr)\Delta_1(t) - \gamma \bigl(\Delta_2(t)-\Delta_1(t)\bigr) - \eta_1(t) \Bigr) \\
        & =  \dfrac{1}{\sqrt{2}}  \Bigl((1-\theta-2\gamma)(\Delta_2(t)-\Delta_1(t)) + \eta_2(t) - \eta_1(t)\Bigr) \\
        & = (1-\theta-2\gamma)\rho(t) + \eta_-(t) \,,
    \end{split} \label{eq:2RW_rho_1step}
\end{numcases}
where ${\eta_+(t) = (\eta_1(t)+\eta_2(t))/\sqrt{2}}$ and ${\eta_-(t) = (\eta_2(t)-\eta_1(t))/\sqrt{2}}$ are Gaussian variables with mean zero and variance $\sigma^2$. In other words, $\eta_\pm(t)$ behaves just like the usual terms ${\eta_i(t), i=1,2,...,N}$ employed in the N-case. 
Moreover, we have ${\cov{\eta_j(t_m); \eta_i(t_n)} = \sigma^2 \delta_{ij} \delta_{mn}}, \,\,\text{with } i,j = \pm, \, m,n>0$, where $\delta_{ij}$ is the Kronecker delta, confirming that $\nu$ and $\rho$ are indeed decoupled.    

Recursively applying Eqs.~\eqref{eq:2RW_nu_1step}-\eqref{eq:2RW_rho_1step} gives the system's evolution between two generic timepoints $t,\tp$ ($\tp>t$):
    \begin{numcases}{}
    \displaystyle
    \begin{split}
        \nu(\tp) & = (1-\theta)^{\tp-t}\nu(t) + \sum_{\tilde{t}=t}^{\tp-1}(1-\theta)^{\tp-\tilde{t}-1}\eta_+(\tilde{t}) \\
        & = \alpha^{\tp-t}\nu(t) + \sum_{\tilde{t}=t}^{\tp-1}\alpha^{\tp-\tilde{t}-1}\eta_+(\tilde{t}) 
    \end{split}  \label{eq:2RW_nu_eom} \\
    \displaystyle
    \begin{split}
        \rho(\tp) & = (1-\theta-2\gamma)^{\tp-t}\rho(t) + \sum_{\tilde{t}=t}^{\tp-1}(1-\theta-2\gamma)^{\tp-\tilde{t}-1}\eta_-(\tilde{t}) \\
        & = \beta^{\tp-t}\rho(t) + \sum_{\tilde{t}=t}^{\tp-1}\beta^{\tp-\tilde{t}-1}\eta_-(\tilde{t}) \,,
    \end{split} \label{eq:2RW_rho_eom}
    \end{numcases}
where we introduced the shorthands ${\alpha\equiv(1-\theta)}$ and ${\beta\equiv(1-\theta-2\gamma)}$. 
Choosing ${\Delta_{1,2}(0)=0}$ and therefore ${\nu(0)=\rho(0)=0}$ as starting conditions, from Eqs.~\eqref{eq:2RW_nu_eom}-\eqref{eq:2RW_rho_eom} we can obtain variances and time-lagged covariances of these terms:  
\begin{gather}
    \var{\nu(t)} = \vars{\sum_{\tilde{t}=0}^{t-1}\alpha^{t-\tilde{t}-1}\eta_+(\tilde{t})} = \sigma^2 \sum_{\tilde{t}=0}^{t-1}\alpha^{2(t-\tilde{t}-1)} = \sigma^2 \dfrac{1-\alpha^{2t}}{1-\alpha^2} \\
    \var{\rho(t)} = \vars{\sum_{\tilde{t}=0}^{t-1}\beta^{t-\tilde{t}-1}\eta_-(\tilde{t})} = \sigma^2 \sum_{\tilde{t}=0}^{t-1}\beta^{2(t-\tilde{t}-1)} = \sigma^2 \dfrac{1-\beta^{2t}}{1-\beta^2} \\
    \cov{\nu(t);\nu(\tp)} = \cov{\nu(t);\alpha^{\tp-t}\nu(t)+\sum_{\tilde{t}=t}^{\tp-1}\alpha^{\tp-\tilde{t}-1}\eta_+(\tilde{t})} = \alpha^{\tp-t}\,\var{\nu(t)} = \sigma^2 \dfrac{1-\alpha^{2t}}{1-\alpha^2} \alpha^{\tp-t} \\
    \cov{\rho(t);\rho(\tp)} = \cov{\rho(t);\beta^{\tp-t}\rho(t)+\sum_{\tilde{t}=t}^{\tp-1}\beta^{\tp-\tilde{t}-1}\eta_-(\tilde{t})} = \beta^{\tp-t}\,\var{\rho(t)} = \sigma^2 \dfrac{1-\beta^{2t}}{1-\beta^2} \beta^{\tp-t} \,.
\end{gather}

We are now finally ready to compute the variances and covariances of the positions of the walkers $\Delta_{1,2}$ and the center of mass $V$. 
\begin{align}
    \var{\Delta_j(t)} & = \vars{\dfrac{\nu(t)\pm\rho(t)}{\sqrt{2}}} = \dfrac12 \left(\var{\nu(t)}+\var{\rho(t)}\right) = \dfrac{\sigma^2}{2} \left(\geom{\alpha}{t}+\geom{\beta}{t} \right) \quad j=1,2  
    \label{eq:2RW_var_xjxj}\\
    \var{V(t)} & = \vars{\dfrac{1}{\sqrt{2}}\nu(t)} = \dfrac{\sigma^2}{2} \geom{\alpha}{t} \\
    \begin{split}
        \cov{\Delta_j(t); \Delta_j(\tp)} & = \covs{\dfrac{\nu(t)\pm\rho(t)}{\sqrt{2}};\dfrac{\nu(\tp)\pm\rho(\tp)}{\sqrt{2}}} = \dfrac12 \left(\cov{\nu(t);\nu(\tp)}+\cov{\rho(t);\rho(\tp)} \right) \\
        & = \dfrac{\sigma^2}{2} \left(\alpha^{\tp-t}\geom{\alpha}{t}+\beta^{\tp-t}\geom{\beta}{t} \right)
    \end{split} \\
    \begin{split}
        \cov{\Delta_j(t); \Delta_i(\tp)} & = \covs{\dfrac{\nu(t)\pm\rho(t)}{\sqrt{2}};\dfrac{\nu(\tp)\mp\rho(\tp)}{\sqrt{2}}} = \dfrac12 \left(\cov{\nu(t);\nu(\tp)}-\cov{\rho(t);\rho(\tp)} \right) \\
        & = \dfrac{\sigma^2}{2} \left(\alpha^{\tp-t}\geom{\alpha}{t}-\beta^{\tp-t}\geom{\beta}{t} \right) \quad i,j=1,2,\,\,i\ne j \label{eq:2RW_cov_xjxi}
    \end{split} \\
    \covs{V(t);V(\tp)} & = \covs{\dfrac{1}{\sqrt{2}}\nu(t);\dfrac{1}{\sqrt{2}}\nu(\tp)} = \dfrac12 \cov{\nu(t);\nu(\tp)} =  \dfrac{\sigma^2}{2} \geom{\alpha}{t} \alpha^{\tp-t} \\
    \covs{\Delta_j(t);V(\tp)} & = \covs{\dfrac{\nu(t)\pm\rho(t)}{\sqrt{2}};\dfrac{1}{\sqrt{2}}\nu(\tp)} = \dfrac12 \cov{\nu(t);\nu(\tp)} =  \dfrac{\sigma^2}{2} \geom{\alpha}{t} \alpha^{\tp-t} \\
    \covs{V(t);\Delta_j(\tp)} & = \covs{\dfrac{1}{\sqrt{2}}\nu(t);\dfrac{\nu(\tp)\pm\rho(\tp)}{\sqrt{2}}} = \dfrac12 \cov{\nu(t);\nu(\tp)} =  \dfrac{\sigma^2}{2} \geom{\alpha}{t} \alpha^{\tp-t} \,. \label{eq:2RW_cov_vx}
\end{align}

\subsection{Mutual information} \label{app:2-RWcase_MI}
We can now use Eqs.~\eqref{eq:2RW_var_xjxj}-\eqref{eq:2RW_cov_vx} to conveniently calculate the mutual information, analogously to Sec.~\ref{app:N-RWcase_MI}.
\begin{align}
 \allowdisplaybreaks
    \begin{split} \label{eq:2_rw_mi_xjxj}
        I(\Delta_j(t);\Delta_j(\tp)) & = \dfrac12 \ln{\left( \dfrac{\var{\Djt}\var{\Djtp}}{|\Sigma_{x_j x_j^{\prime}}|} \right)} \\
        & = \dfrac12 \lnp{ \dfrac{\left(\sumgeom{t}\right)\left(\sumgeom{\tp}\right)}
        {\left(\sumgeom{t}\right)\left(\sumgeom{\tp}\right)-\left(\alpha^{\tp-t}\geom{\alpha}{t}+\beta^{\tp-t}\geom{\beta}{t} \right)^2}}
    \end{split} \\[0.5em]
    \begin{split} \label{eq:2_rw_mi_xixj}
        I(\Delta_j(t);\Delta_i(\tp)) & = \dfrac12 \ln{\left( \dfrac{\var{\Djt}\var{\Delta_i(\tp)}}{|\Sigma_{x_j x_i^{\prime}}|} \right)} \\
        & = \dfrac12 \lnp{\dfrac{\left(\sumgeom{t}\right)\left(\sumgeom{\tp}\right)}
        {\left(\sumgeom{t}\right)\left(\sumgeom{\tp}\right)-\left(\alpha^{\tp-t}\geom{\alpha}{t}-\beta^{\tp-t}\geom{\beta}{t} \right)^2}} \,,\quad i\ne j
    \end{split} \\[0.5em]
    \begin{split}
        I(V(t);V(\tp)) & = \dfrac12 \ln{\left( \dfrac{\var{V(t)}\var{V(\tp)}}{|\Sigma_{vv^{\prime}}|} \right)} \\
        & = \dfrac12 \ln{\left( \dfrac{\geom{\alpha}{t} \cdot\geom{\alpha}{\tp}}{\geom{\alpha}{t} \cdot\geom{\alpha}{\tp}-\alpha^{2(\tp-t)}\left(\geom{\alpha}{t} \right)^2} \right)} \\ 
        & = \dfrac12 \ln{\left(\dfrac{1-\alpha^{2\tp}}{1-\alpha^{2\tp}-(\alpha^{2(\tp-t)}-\alpha^{2\tp})} \right)} \\
        & = \dfrac12 \ln{\left(\dfrac{1-\alpha^{2\tp}}{1-\alpha^{2(\tp-t)}} \right)}
    \end{split} \\[0.5em]
    \begin{split}
    I(\Delta_j(t);V(\tp)) & = \dfrac12 \ln{\left( \dfrac{\var{\Djt}\var{V(\tp)}}{|\Sigma_{xv^{\prime}}|} \right)} \\
    & =\dfrac12 \ln{\left(\dfrac{\lrpars{\sumgeom{t}}\geom{\alpha}{\tp}}
    {\lrpars{\sumgeom{t}}\geom{\alpha}{\tp}-\alpha^{2(\tp-t)}\lrpars{\geom{\alpha}{t}}^2} \right)}  
    \end{split} \displaybreak  \\[0.5em]
    \begin{split} \label{eq:2_rw_mi_vx}
        I(V(t);\Delta_j(\tp)) & = \dfrac12 \ln{\left( \dfrac{\var{V(t)}\var{\Djtp}}{|\Sigma_{vx^{\prime}}|} \right)} \\
        & =\dfrac12 \ln{\left(\dfrac{\geom{\alpha}{t}\lrpars{\sumgeom{\tp}}}
        {\geom{\alpha}{t}\lrpars{\sumgeom{\tp}}-\alpha^{2(\tp-t)}\lrpars{\geom{\alpha}{t}}^2} \right)} \\
        & = \dfrac12 \ln{\left(\dfrac{\sumgeom{\tp}}
        {\sumgeom{\tp}-\alpha^{2(\tp-t)}\lrpars{\geom{\alpha}{t}}} \right)} \\
        & = \dfrac12 \ln{\left(\dfrac{\sumgeom{\tp}}
        {\sumgeom{\tp}-\dfrac{\alpha^{2(\tp-t)}-\alpha^{2\tp}}{1-\alpha^2}} \right)} \\
        & = \dfrac12 \ln{\left(\dfrac{\sumgeom{\tp}}
        {\geom{\alpha}{\tp-t}+\geom{\beta}{\tp}} \right)}
    \end{split} 
\end{align}

\begin{figure}[ht]
    \centering
    \includegraphics[width=1\linewidth]{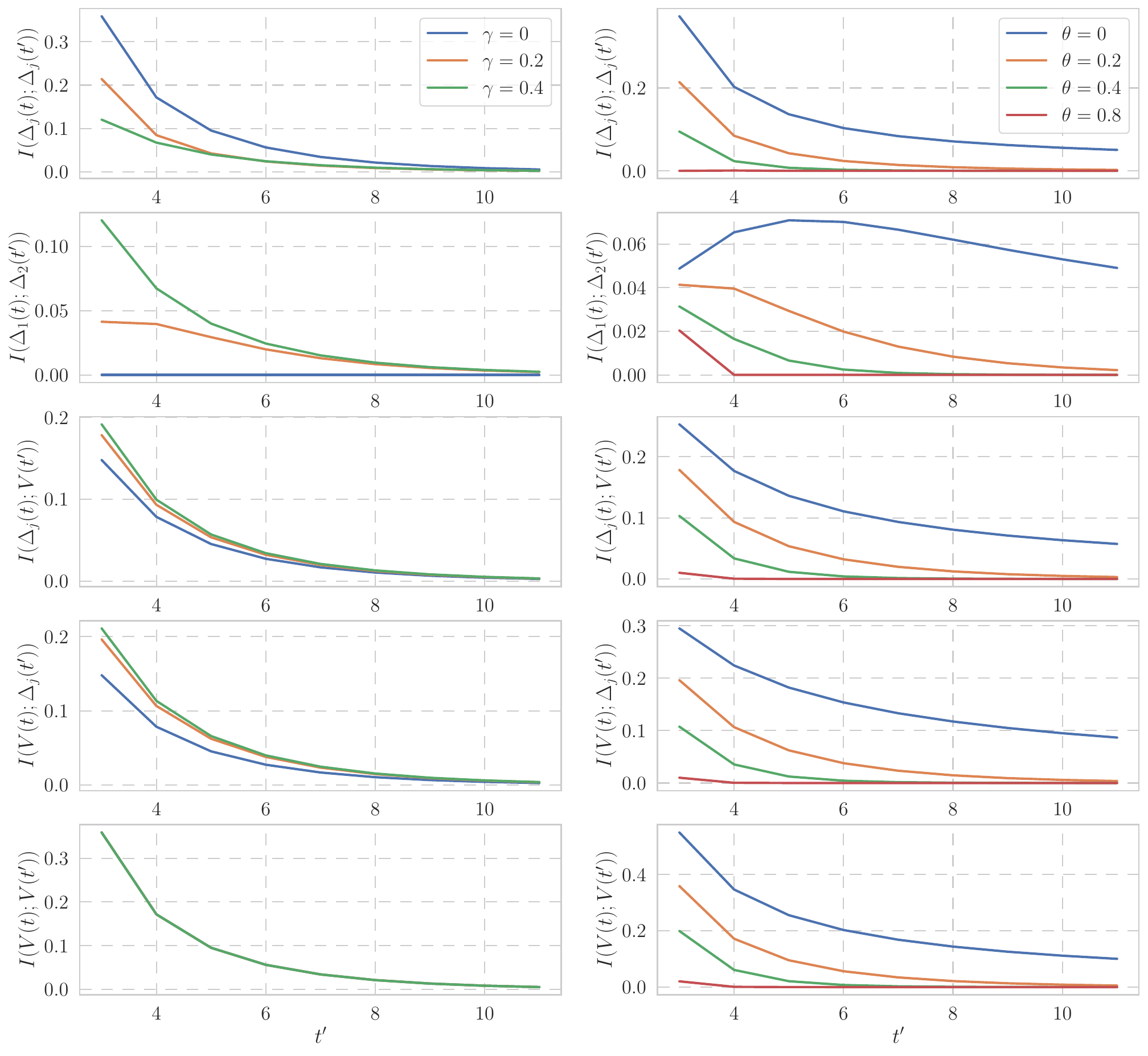}
    \caption{\textbf{Mutual information in higher orders of $\gamma$ behave as expected in the stationary case.}
    Mutual information of Eqs.~\eqref{eq:2_rw_mi_xjxj}-\eqref{eq:2_rw_mi_vx} for the case of $N=2$ random walkers with various choice of the parameters $(\theta, \gamma)$. First column is for fixed $\theta=0.2$ and varying $\gamma$, second column is for fixed $\gamma=0.2$ and varying $\theta$. Results are shown for $t=2$.}
    \label{fig:2RWS_MI}
\end{figure}

To help understand the meaning of these quantities, in Fig.~\ref{fig:2RWS_MI} we show the behaviours of these mutual information for different choices of the parameters $(\theta, \gamma)$. 

Fixing $\gamma=0.2$, we can note that for increasing $\theta$ all the mutual information decrease, as the random walkers are brought back more strongly by the $\theta$ term. On the other hand, keeping $\theta=0.2$ fixed and varying $\gamma$ we obtain the results already observed in the case of $N$ RWs, with the addition of the mutual information between walkers increasing with $\gamma$.

\subsubsection{$\theta\to0$ limit}
It is easy to obtain the non-OU mutual information by simply taking the limit ${\theta\to0}$, i.e. ${\alpha\to1}$ and ${\beta\to1-2\gamma}$, and using Eq.~\eqref{eq:1-a_limit}. However, since the geometrical sum in $\beta$ would remain, this would not provide a more intuitive understanding of the formula obtained above. For this reason, we will not provide the explicit equations for this case.

\subsubsection{$\gamma\to0$ limit}
It is interesting to see that the mutual information between walkers is indeed a second-order quantity in the interaction $\gamma$, as already noted in the $N>2$ case.
We first notice that
\begin{align}
    \geom{\beta}{x} & \underrel{\gamma\to0}{=} \geom{\alpha}{x} -\dfrac{4\alpha\gamma\left(1-x\alpha^{2(x-1)}+(x-1)\alpha^{2x}\right)}{(1-\alpha)^2} +o(\gamma) \\[1em]
    \beta^{x} & \underrel{\gamma\to0}{=} \alpha^x \,,
\end{align}
therefore, in the denominator of Eq.~\eqref{eq:2_rw_mi_xixj} the second part becomes
\begin{equation}
    \left(\alpha^{\tp-t}\geom{\alpha}{t}-\beta^{\tp-t}\geom{\beta}{t} \right)^2 \underrel{\gamma\to0}{=} \lrpars{-\alpha^{\tp-t}\cdot \gamma\dfrac{4\alpha\left(1-t\alpha^{2(t-1)}+(t-1)\alpha^{2t}\right)}{(1-\alpha)^2}}^2 \sim O(\gamma^2) \,.
\end{equation}
This second order quantity is needed for the expression to not be zero.
Performing a more detailed calculation, we obtain
\begin{equation}
\begin{split}
    I(\Delta_j(t);\Delta_i(\tp)) & \underrel{\gamma\to0}{=} \dfrac12 \lnp{ 1-\gamma^2\frac{\alpha^{2(\tp-t-1)} \left(\tp-t - \alpha^2 (\tp - t -2) + \alpha^{2(t+1)} (\tp+t-2) - \alpha^{2t} (\tp+t)\right)^2 }{(1 - \alpha^2)^2 (1 - \alpha^{2t}) (1 - \alpha^{2\tp})}} \\
    & \,\,\,= -\gamma^2\frac{\alpha^{2(\tp-t-1)} \left(\tp-t - \alpha^2 (\tp - t -2) + \alpha^{2(t+1)} (\tp+t-2) - \alpha^{2t} (\tp+t)\right)^2 }{2(1 - \alpha^2)^2 (1 - \alpha^{2t}) (1 - \alpha^{2\tp})} \,.
\end{split}
\end{equation}

\subsection{Transfer entropy}  \label{app:2-RWcase_TE}
 
In this section, we report the computations of transfer entropy across walkers and between walkers and centre of mass.
Analogously to the $N$-RW case, here we compute transfer entropy in the limit $\theta\to0$.
For the TE between walkers we have
\begin{equation} \label{eq:2RW_te_xx}
    \begin{split}
         \mathcal{T}(\Dit;\Djtp) & = H(\Djtp|\Djt) - H(\Djtp|\Djt,\Dit) \\
         & = \dfrac{1}{2} \ln \left( 2 \pi e \dfrac{\sigma^2}{2} \cdot \dfrac{ \left( \left( t + \geom{\beta}{t} \right)\left( \tp + \geom{\beta}{\tp} \right) - \left( t + \beta^{\tp-t} \geom{\beta}{t} \right)^2 \right)}{t + \geom{\beta}{t}} \right) + \\
         & \quad - \dfrac{1}{2} \ln \left( 2 \pi e \dfrac{\sigma^2}{2} \cdot \left( \tp - t + \geom{\beta}{\tp} - \beta^{2(\tp-t)} \geom{\beta}{t} \right) \right) \\
         & = \dfrac{1}{2} \ln \left( 2 \pi e \dfrac{\sigma^2}{2} \cdot \dfrac{ \left( \left( t + \geom{\beta}{t} \right)\left( \tp + \geom{\beta}{\tp} \right) - \left( t + \beta^{\tp-t} \geom{\beta}{t} \right)^2 \right)}{\lrpars{t + \geom{\beta}{t}}\left( \tp - t + \geom{\beta}{\tp} - \beta^{2(\tp-t)} \geom{\beta}{t} \right)} \right) \,,
    \end{split}
\end{equation}
where we used
\begin{align}
    \begin{split}
        H(\Djtp|\Djt) &= H(\Djtp) - I(\Djt; \Djtp)  \\
        & = \dfrac12 \lnp{2\pi e\, \dfrac{\sigma^2}{2} \lrpars{\tp+\bgeom{\tp}}} + \\
        & \quad - \dfrac12 \lnp{ \dfrac{\left(t+\bgeom{t}\right)\left(\tp+\bgeom{\tp}\right)}
        {\left(t+\bgeom{t}\right)\left(\tp+\bgeom{\tp}\right)-\left(t+\beta^{\tp-t}\geom{\beta}{t} \right)^2}} \\
        & = \dfrac{1}{2} \ln \left( 2 \pi e \dfrac{\sigma^2}{2} \cdot \dfrac{ \left( \left( t + \geom{\beta}{t} \right)\left( \tp + \geom{\beta}{\tp} \right) - \left( t + \beta^{\tp-t} \geom{\beta}{t} \right)^2 \right)}{t + \geom{\beta}{t}} \right)
    \end{split} \\
    \begin{split}
        H(\Djtp|\Djt,\Dit) & = H(\Djtp,\Djt,\Dit) - H(\Djt,\Dit) \\
         & = \dfrac12 \lnp{\lrpars{2\pi e\, \dfrac{\sigma^2}{2}}^3 4t\bgeom{t}\left( \tp - t + \geom{\beta}{\tp} - \beta^{2(\tp-t)} \geom{\beta}{t} \right)} + \\
         & \quad - \dfrac12 \lnp{\lrpars{2\pi e\, \dfrac{\sigma^2}{2}}^2 4t \lrpars{\bgeom{t}}} \\
         & = \dfrac{1}{2} \ln \left( 2 \pi e \dfrac{\sigma^2}{2} \cdot \left( \tp - t + \geom{\beta}{\tp} - \beta^{2(\tp-t)} \geom{\beta}{t} \right) \right) \,.
    \end{split}
\end{align}
We observe that, in this simple case of 2 particles, $H(\Djtp|\Djt,\Dit) = H(\Djtp|\Djt,V(t))$, as measuring the position of one walker and the c.o.m.\ or the positions of both walkers provides the same information. 
Thus the transfer entropy from the center of mass to one walker must be the same as Eq.~\eqref{eq:2RW_te_xx}:
\begin{equation} \label{eq:2RW_te_vx}
    \begin{split}
        \mathcal{T}(V(t);\Djtp) & = H(\Djtp|\Djt) - H(\Djtp|\Djt,V(t)) \\
        & = H(\Djtp|\Djt) - H(\Djtp|\Djt,\Dit) \\
        & = \mathcal{T}(\Dit;\Djtp) \,.
    \end{split}
\end{equation}
Finally, the transfer entropy from one walker to the c.o.m.\ reads 
\begin{equation} \label{eq:2RW_te_xv}
\begin{split}
    \mathcal{T}(\Djt;V(\tp)) & = H(V(\tp)|V(t)) - H(V(\tp)|V(t),\Djt) \\
    & = \dfrac12 \ln\left(2\pi e\, \sigma^2 t\dfrac{\tau-1}{2} \right) - \dfrac12 \ln{\left(2\pi e\, \sigma^2 t\dfrac{\tau-1}{2}\right)} \\
    & = 0 \,,
\end{split}
\end{equation}
where we used
\begin{align}
    \begin{split}
    H(V(\tp)|V(t)) &= H(V(\tp)) - I(V(t);V(\tp)) \\
    & = \dfrac12 \lnp{2\pi e\, \frac{\sigma^2}{2} \tp} - \dfrac12 \lnp{2\pi e\, \frac{\sigma^2}{2} \frac{\tau}{\tau-1}} \\
    & = \dfrac12 \ln\left(2\pi e\, \sigma^2 t\dfrac{\tau-1}{2} \right)
    \end{split} \\
    \begin{split}
    H(V(\tp)|V(t),\Djt) & = H(V(\tp),V(t),\Djt) - H(V(t),\Djt) \\
    & = \dfrac12 \ln\left(\lrpars{2\pi e\, \dfrac{\sigma^2}{2}}^3 (\tau-1)\,t^2\geom{\beta}{t} \right)-\dfrac12 \ln\left(\lrpars{2\pi e\, \dfrac{\sigma^2}{2}}^2 t\geom{\beta}{t} \right) \\
    & = \dfrac12 \ln{\left(2\pi e\, \dfrac{\sigma^2}{2} t(\tau-1)\right)} \,.
    \end{split}
\end{align}
The joint entropies above are calculated using Eq.~\eqref{eq:entropy_gauss_joint} and the results obtained in Sec.~\ref{app:2-RWcase_model}.

We see that TE between walkers increases with the interaction strength $\gamma$, and is zero if and only if $\gamma=0$, confirming that $\mathcal{T}(\Dit;\Djtp)=0$ obtained in the N case was a confound due to the calculations at first order in $\gamma$. 
This also accounts for the transfer entropy from the center of mass to one random walker.  
On the other hand, we obtained that $\mathcal{T}(\Djt;V(\tp))$ vanishes, generalising the result obtained in the $N$ case to higher orders of $\gamma$. This is a consequence of the c.o.m.\ of the system being an independent RW by itself (Eq.~\eqref{eq:eom_V}). 
This can also be seen as an extreme case of statistical causal decoupling, in which a macroscopic variable becomes statistically completely detached from the single microscopic components that generate it.

\subsection{Causal emergence quantities}  \label{app:2-RWcase_emergence}

\begin{figure}[ht]
    \centering
    \includegraphics[width=1\linewidth]{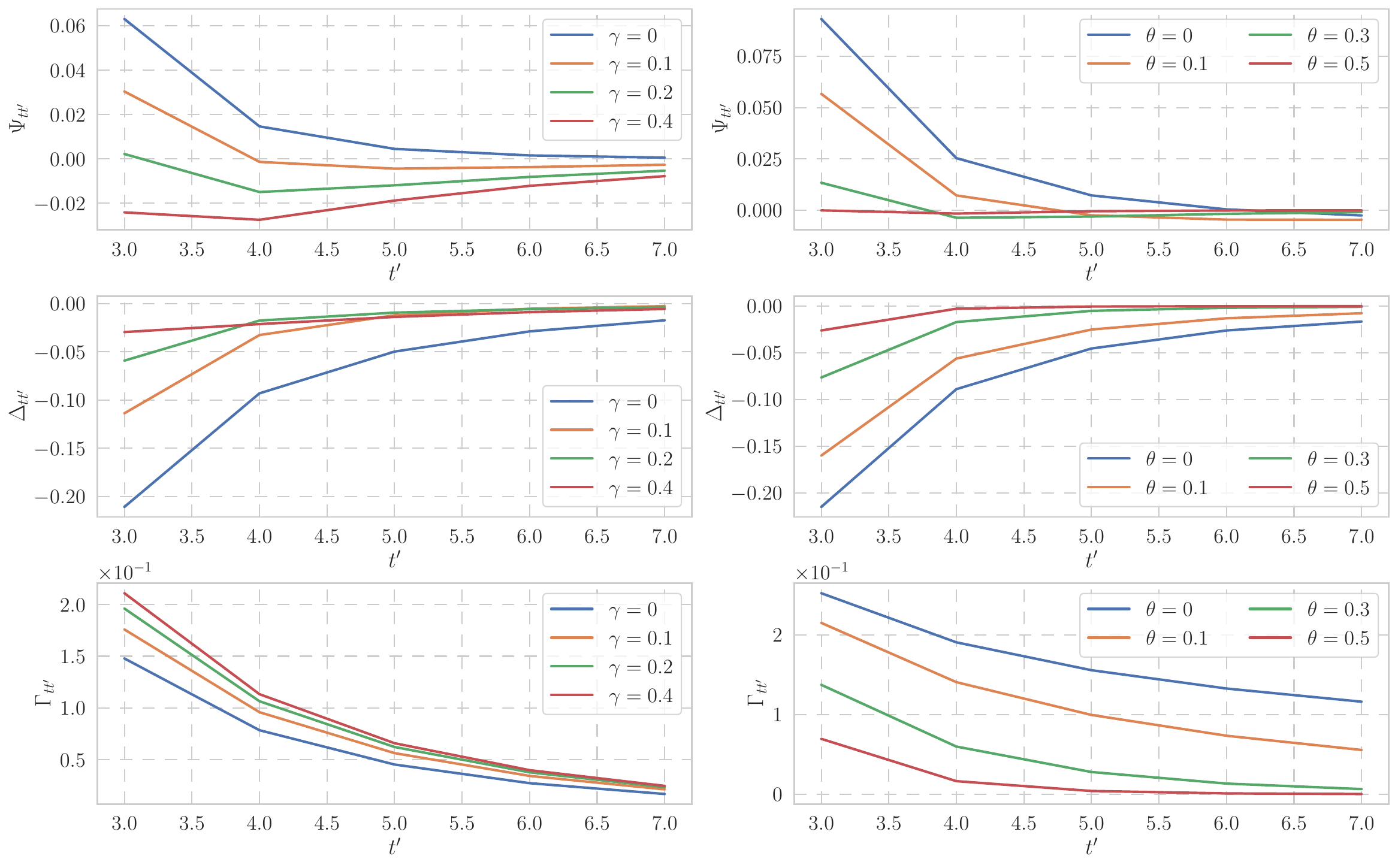}
    \caption{{Emergence measures for $N=2$ RWs follow the same patterns of the $N>2$ case.}
    $\Psi_{t\tp}, \Delta_{t\tp}, \Gamma_{t\tp}$ quantities in the case of 2 RWs for various parameters $(\theta, \gamma)$. Left column: $\gamma\in[0,0.1,0.2, 0.4]$ for $\theta=0.2$. Right column: $\theta\in[0,0.1,0.3, 0.5]$ for $\gamma=0.001$. Results are shown for $t=2$.}
    \label{fig:2RWs_Psi_Delta_Gamma}
\end{figure}

$\Psi, \Gamma, \Delta$ can be easily calculated using Eqs.~\eqref{eq:2_rw_mi_xjxj}-\eqref{eq:2_rw_mi_vx}:
\begin{align}
\begin{split}
    \Psi_{t\tp} & = I(V(t); V(t^{\prime})) - \sum_{j=1}^2 I(\Delta_j(t); V(t')) \\
    & = I(V(t); V(t^{\prime})) - 2I(\Delta_j(t); V(t'))
\end{split} \\
\begin{split}
    \Delta_{t\tp} & = \max_i \Bigl( I(V(t);\Delta_i(t')) - \sum_{j=1}^2 I(\Delta_j(t);\Delta_i(t')) \Bigr) \\
    & = I(V(t);\Delta_j(t')) - I(\Delta_j(t);\Delta_j(t')) - I(\Delta_j(t);\Delta_i(t')) 
\end{split} \\
\begin{split}
    \Gamma_{t\tp} &= \max_i \Bigl(I(V(t); \Delta_i(t'))\Bigr) = I(V(t); \Delta_j(t')) \,.
\end{split}
\end{align}
For various choices of the parameters $(\theta, \gamma)$, we plot the behaviour of these quantities in Fig.~\ref{fig:2RWs_Psi_Delta_Gamma}.

We can note that for all cases $\Delta$ and $\Gamma$ increase with the interaction strength, while $\Psi$ decreases, as already noted in the $N$-RWs case. However, importantly, we notice that for high enough values of $\gamma$ $\Psi$ becomes negative, i.e. the sufficient condition for causal emergence fails when the system is more cohesively interdependent.

\subsection{Partial Information Decomposition}  \label{app:2-RWcase_PID}
\begin{figure}[ht]
    \centering
    \begin{subfigure}[]{1\textwidth} 
        \centering
        \includegraphics[width=1\linewidth]{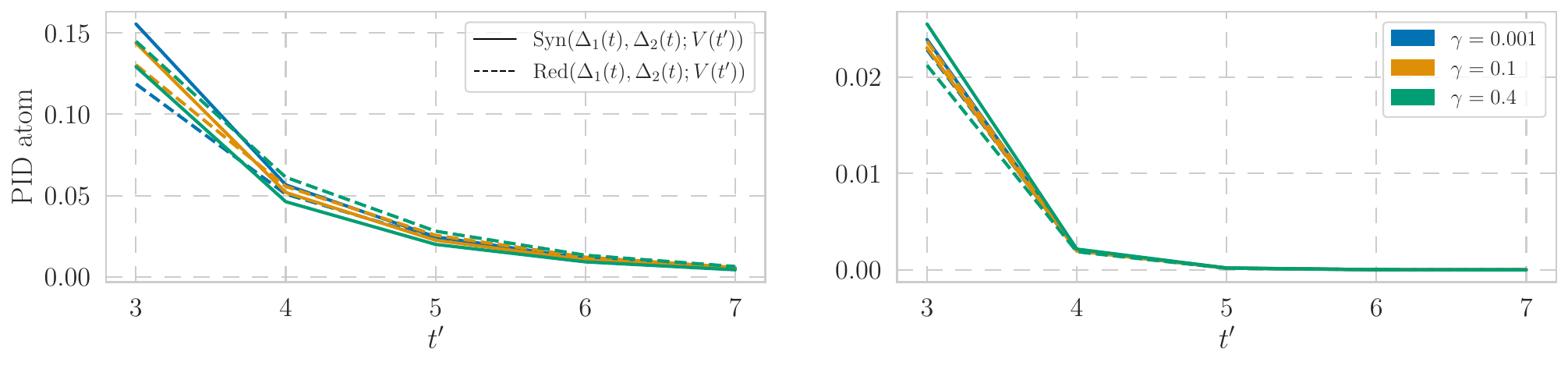}
        \caption{}
        \label{fig:PID_2RWS_theta_1}
    \end{subfigure}
    \vfill
    \begin{subfigure}[]{1\textwidth} 
        \centering
        \includegraphics[width=1\linewidth]{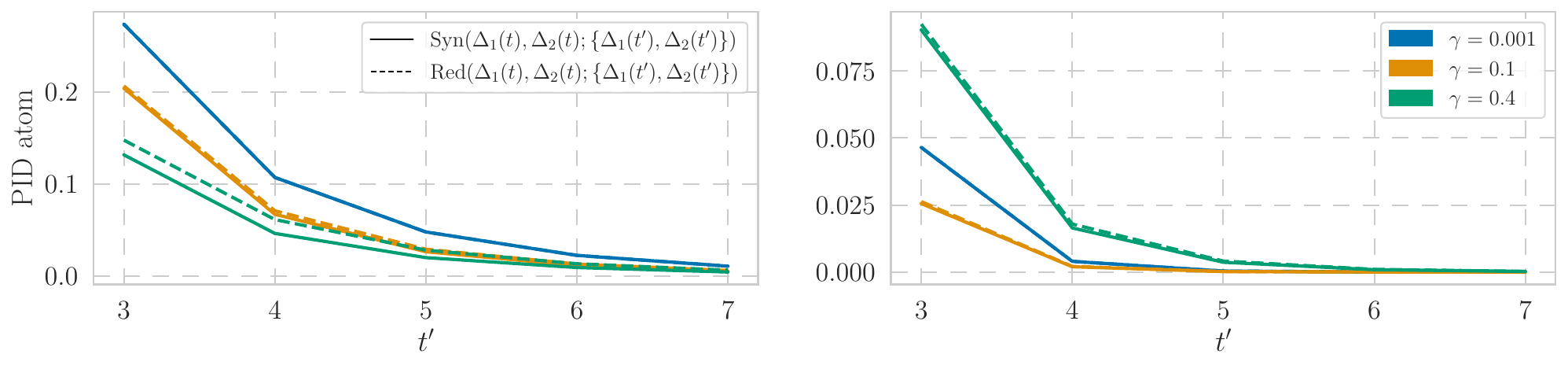}
        \caption{}
        \label{fig:PID_2RWS_theta_2}
    \end{subfigure}
        \caption{\textbf{Synergy and redundancy behave similarly to the non-stationary case for $N=2$ RWs.}
        PID for $N=2$ random walkers with various choices of the parameters $(\theta,\gamma)$. (a) First PID case (1), (b) second and third PID case (2-3). Left and right panels show various $\gamma$ for $\theta=0.3$ and $\theta=0.7$, respectively. Results are shown for $t=2$.}
        \label{fig:PID_2RWs_theta}
\end{figure}

The 2-RW PID results shown in the main article (Fig.~\ref{fig:PID_imgs}) are obtained by the same formulas of the mutual information of App.~\ref{app:N-RWcase_PID}, but using the variances and covariances obtained for the 2 random walkers (App.~\ref{app:2-RWcase_model}). 
Since these calculations are exact, once the determinants of the covariance matrices are calculated there is no need for further mathematical manipulations, and the results can be directly plotted. 
In Fig.~\ref{fig:PID_2RWs_theta} we show how redundancy and synergy behave for different choices of ($\theta$, $\gamma$) in the three PID cases above. 

The behaviours of the PID atoms are qualitatively the same as the ones observed in Fig.~\ref{fig:PID_imgs} in the main text, with synergy decreasing with $\gamma$ in both cases, and redundancy increasing in the first (Fig.~\ref{fig:PID_2RWS_theta_1}), and decreasing in the latter (Fig.~\ref{fig:PID_2RWS_theta_2}). 
As expected from the results on the mutual information, higher $\theta$ suppresses the overall magnitude of both PID atoms. 
Thus, we showed that the findings presented previously do not depend on the non-stationarity of the system.  

\begin{figure*}
    \centering
    \includegraphics[width=\linewidth]{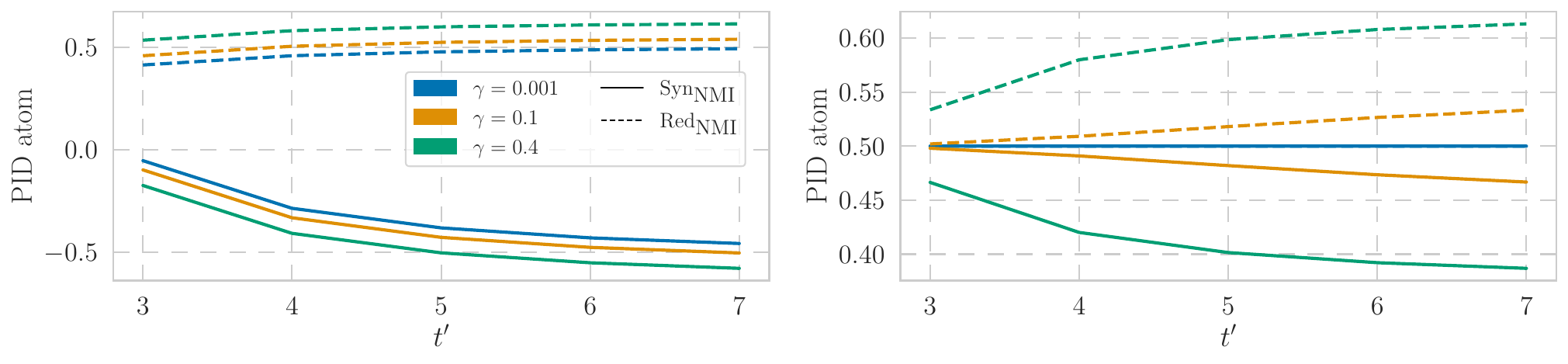}
    \caption{\textbf{Normalised redundancy increases and normalised synergy decreases with interaction strength for $N=2$ RWs.}
    NMI normalised PID for $N=2$ random walkers for $\theta=0.2$ and various $\gamma$. On the left: first PID case (1). On the right: second and third PID case (2-3). Results are shown for $t=2$.}
    \label{fig:PID_2RWS_NMI}
\end{figure*}

Focusing on the normalised quantities, in Fig.~\ref{fig:PID_2RWS_NMI} we show redundancy and synergy calculated above and normalised by the joint mutual information.

In both PID cases, we observe that normalised redundancy increases with the interaction strength, while normalised synergy decreases. This finding differs from the non-normalised result, in which both atoms were decreasing with the coupling of the system. 
More comments about the implications of these behaviours can be found in the main text (Sec.~\ref{sec:results_pid}).

\subsection{Integrated information measures $\Phi_{\mathrm{WMS}}$ and $\Phi_R$} \label{app:2-RWcase_Phi}

In this subsection, we complement the results shown in Fig.~\ref{fig:Phis_imgs} for the integrated information measures.
Recalling Eqs.~\eqref{eq:phi_wms}-\eqref{eq:phi_r_2}, the explicit formula for $\Phi_{\mathrm{WMS}}$ and $\Phi_R$ read:
\begin{align} \label{eq:phi_wms_2}
    \Phi_{\mathrm{WMS}}(\Delta_1,\Delta_2;t,\tp)& = I(\Delta_1(t),\Delta_2(t);\Delta_1(t^{\prime}),\Delta_2(t^{\prime})) -2I(\Delta_1(t);\Delta_1(t^{\prime})) \,,
    \\
    \Phi_{\mathrm{R}}(\Delta_1,\Delta_2;t,\tp) & = \Phi_{\mathrm{WMS}}(\Delta_1,\Delta_2;t,\tp) + I(\Delta_1(t),\Delta_2(t^{\prime})) \,,
\end{align}
where the expressions for the mutual information can be found in Eqs.~\eqref{eq:2_rw_mi_xjxj}-\eqref{eq:2_rw_mi_vx}. 

\begin{figure}[ht]
    \centering
    \begin{subfigure}[]{0.5\textwidth} 
        \hspace*{-5mm}
        \centering
        \includegraphics[width=1\linewidth]{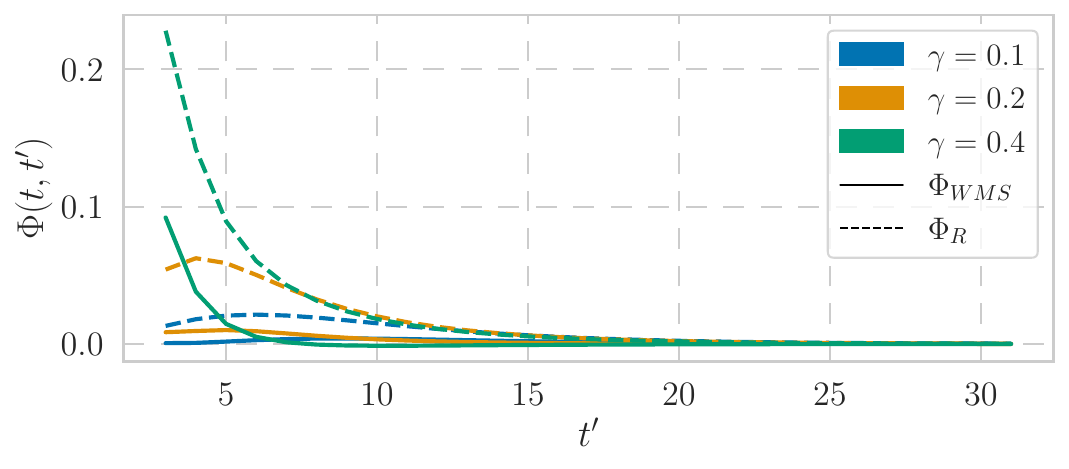}
        \caption{}
        \label{fig:}
    \end{subfigure}
\hspace*{-5mm}
    \begin{subfigure}[]{0.5\textwidth} 
        \centering
        \includegraphics[width=1\linewidth]{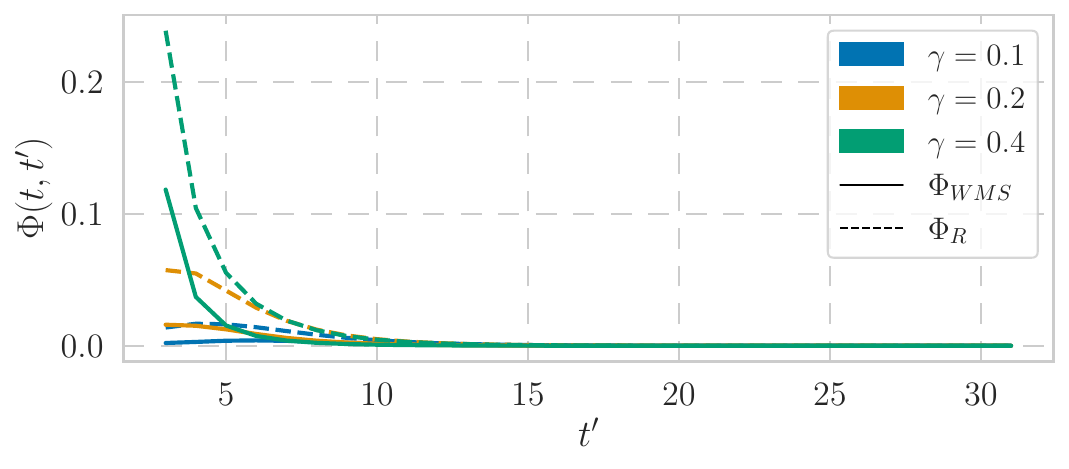}
        \caption{}
        \label{fig:}
    \end{subfigure}
    \vfill
    \begin{subfigure}[]{0.5\textwidth} 
    \hspace*{-5mm}
        \centering
        \includegraphics[width=1\linewidth]{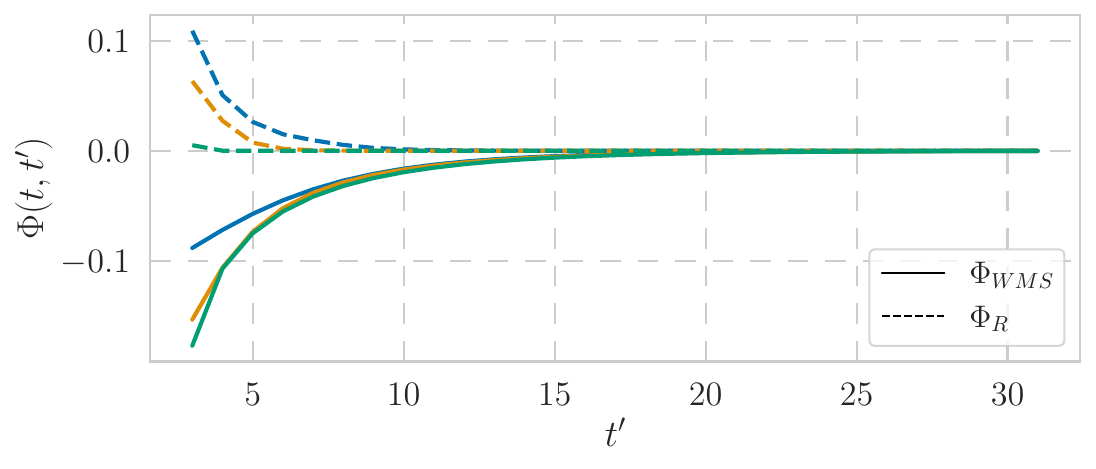}
        \caption{}
        \label{fig:}
    \end{subfigure}
\hspace*{-5mm}
    \begin{subfigure}[]{0.5\textwidth} 
        \centering
        \includegraphics[width=1\linewidth]{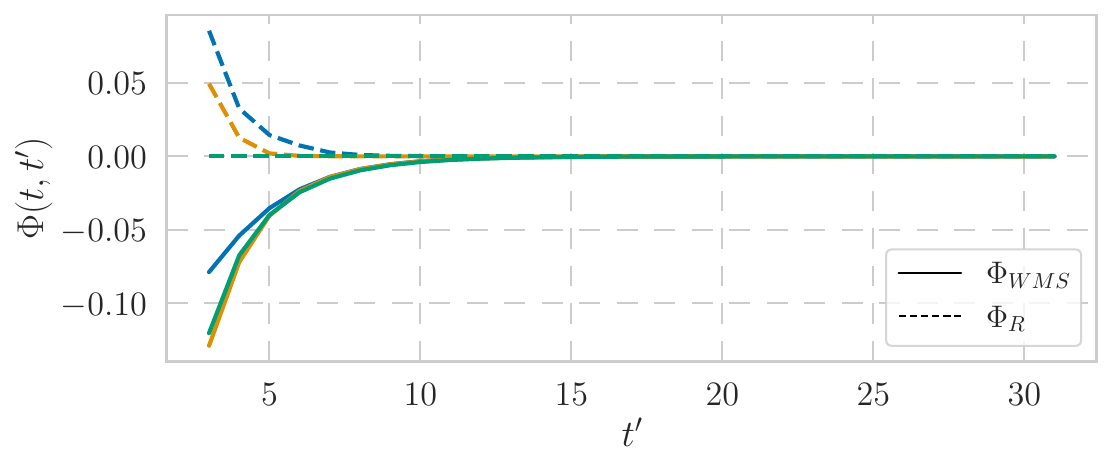}
        \caption{}
        \label{fig:}
    \end{subfigure}
        \caption{\textbf{Integrated information measures do not correlate with the system's interaction strength also in the stationary case. }
        (a) $\Phi_{\mathrm{WMS}}$ and $\Phi_{\mathrm{R}}$ for $\theta=0.1$ and various $\gamma$, (b) same as (a) but with $\theta=0.2$. (c) $\Phi_{\mathrm{WMS}}^V$ and $\Phi_{\mathrm{R}}^V$ for $\theta=0.1$ and various $\gamma$, (d) same as (c) but with $\theta=0.2$. Results are shown for $t=2$.}
        \label{fig:2RWS_phis_OU}
\end{figure}

As for $\Phi^V_{WMS}$ and $\Phi^V_R$, we have:
\begin{align}
    \begin{split} \label{eq:phi_wms_3}
    \Phi_{\mathrm{WMS}}^V(t,\tp) & \equiv\Phi_{\mathrm{WMS}}(\Delta_j,V;t,\tp) \\
    & = I(\Delta_j(t),V(t);\Delta_j(t^{\prime})), V(\tp)) -I(\Delta_j(t);\Delta_j(t^{\prime})) - I(V(t);V(t^{\prime}))\,, 
    \end{split} \\
    \begin{split} \label{eq:phi_r_3}
    \Phi_{\mathrm{R}}^V(t,\tp) & \equiv\Phi_{\mathrm{R}}(\Delta_j,V;t,\tp) \\
    & = \Phi_{\mathrm{WMS}}(\Delta_j,V;t,\tp) + I(\Delta_j(t),\Delta_j(t^{\prime}))\,.
    \end{split}
\end{align}
Using these equations, we can analyse how these measures behave in the case of a OU dynamics, for different choice of the parameters $(\theta,\gamma)$ (Fig.~\ref{fig:2RWS_phis_OU}).

Analogously to the results presented in the main body, $\Phi_{\mathrm{WMS}}$ and $\Phi_{\mathrm{R}}$ behave precisely the same as in the non-OU case. The only difference lies in the quicker decay of integrated information due to the effect of the restoring force.

Furthermore, we can look at how these metrics behave in the infinite time limit, i.e.\ when the random walkers can be explained by a VAR process. Taking the limits $t\to\infty, \tp\to\infty$ and different values of $\Delta t = \tp-t<\infty$, we report the results in Tabs.~\ref{tab:2RWS_phis_wms_deltat}-\ref{tab:2RWS_phis_r_v_deltat}. 

\renewcommand{\arraystretch}{1.2}
\begin{table}[h]
\centering
\begin{minipage}{0.45\textwidth}
\hspace*{15mm}\figuretitle{$\Phi_{\mathrm{WMS}}(t,t+\Delta t)$}
    \centering
    \begin{tabular}%
        {
        >{\centering}p{0.16\columnwidth}|
        >{\centering}p{0.16\columnwidth}
        >{\centering}p{0.16\columnwidth}
        >{\centering\arraybackslash}p{0.16\columnwidth}
        }
    \cline{2-4}
    \backslashbox[14mm]{$\Delta t$}{$\gamma$} & $0.1$ & $0.2$ & $0.4$ \\ \hline
    $1$ & $-0.0890$ & $-0.1375$ & $-0.0678$ \\ %
    $5$ & $-0.0280$ & $-0.0439$ & $-0.0672$ \\ %
    $10$ & $-0.0057$ & $-0.0159$ & $-0.0246$ \\ \hline
    \end{tabular}
    \caption{$\Phi_{\mathrm{WMS}}$ for various $\Delta t$ and $\gamma$ in the limit $t\to\infty, \tp\to\infty$, with $\theta=0.1$.}
    \label{tab:2RWS_phis_wms_deltat}
\end{minipage}
\hspace{1cm}
\begin{minipage}{0.45\textwidth}
\hspace*{15mm}\figuretitle{$\Phi_{\mathrm{R}}(t,t+\Delta t)$}
    \centering
    \begin{tabular}%
        {
        >{\centering}p{0.16\columnwidth}|
        >{\centering}p{0.16\columnwidth}
        >{\centering}p{0.16\columnwidth}
        >{\centering\arraybackslash}p{0.16\columnwidth}
        }
    \cline{2-4}
    \backslashbox[14mm]{$\Delta t$}{$\gamma$} & $0.1$ & $0.2$ & $0.4$ \\ \hline
    $1$ & $0.0332$ & $0.1021$ & $0.3271$ \\ %
    $5$ & $0.0520$ & $0.0779$ & $0.0736$ \\ %
    $10$ & $0.0256$ & $0.0244$ & $0.0201$ \\ \hline
    \end{tabular}
    \caption{$\Phi_{\mathrm{R}}$ for various $\Delta t$ and $\gamma$ in the limit $t\to\infty, \tp\to\infty$, with $\theta=0.1$.}
\end{minipage}
\end{table}

\renewcommand{\arraystretch}{1.2}
\begin{table}[h]
\centering
\begin{minipage}{0.49\textwidth}
\hspace*{15mm}\figuretitle{$\Phi_{\mathrm{WMS}}^V(t,t+\Delta t)$}
    \centering
    \begin{tabular}%
        {
        >{\centering}p{0.16\columnwidth}|
        >{\centering}p{0.16\columnwidth}
        >{\centering}p{0.16\columnwidth}
        >{\centering\arraybackslash}p{0.16\columnwidth}
        }
    \cline{2-4}
    \backslashbox[15mm]{$\Delta t$}{$\gamma$} & $0.1$ & $0.2$ & $0.4$ \\ \hline
    $1$ & $-0.2914$ & $-0.4120$ & $-0.4466$ \\ %
    $5$ & $-0.1140$ & $-0.1289$ & $-0.1408$ \\ %
    $10$ & $-0.0351$ & $-0.0403$ & $-0.0447$ \\ \hline
    \end{tabular}
    \caption{$\Phi_{\mathrm{WMS}}^V$ for various $\Delta t$ and $\gamma$ in the limit $t\to\infty, \tp\to\infty$, with $\theta=0.1$.}
\end{minipage}
\hspace{0cm}
\begin{minipage}{0.49\textwidth}
\hspace*{15mm}\figuretitle{$\Phi_{\mathrm{R}}^V(t,t+\Delta t)$}
    \centering
    \begin{tabular}%
        {
        >{\centering}p{0.16\columnwidth}|
        >{\centering}p{0.16\columnwidth}
        >{\centering}p{0.16\columnwidth}
        >{\centering\arraybackslash}p{0.16\columnwidth}
        }
    \cline{2-4}
    \backslashbox[15mm]{$\Delta t$}{$\gamma$} & $0.1$ & $0.2$ & $0.4$ \\ \hline
    $1$ & $0.1546$ & $0.1076$ & $0.0050$ \\ %
    $5$ & $0.0143$ & $0.0005$ & $0$ \\ %
    $10$ & $0.0004$ & $0$ & $0$ \\ \hline
    \end{tabular}
    \caption{$\Phi_{\mathrm{R}}^V$ for various $\Delta t$ and $\gamma$ in the limit $t\to\infty, \tp\to\infty$, with $\theta=0.1$.}
    \label{tab:2RWS_phis_r_v_deltat}
\end{minipage}
\end{table}

Comparing the different columns in each row, it is easy to see that for $\Phi_{\mathrm{WMS}}$ there is no fixed ordering, and that changing $\Delta t$ can change the hierarchy of integrated information. This is a consequence of the different behaviours of the initial transient and infinite limit of Fig.~\ref{fig:Phis_imgs}. $\Phi_{\mathrm{R}}$ also shows the same erratic behaviour for different $\Delta t$.
On the other hand, $\Phi^V_{WMS}$ and $\Phi^V_{R}$ maintain a constant ordering of higher integrated information for less interactive systems, as already noted above.

\FloatBarrier

\subsection{Causal intervention} \label{app:2-RWcase_intervention}
In this section, we complement the results showed in Sec.~\ref{sec:results_pid} employing the intervention method as a possible approach to investigate the causal interdependencies within a system. 
We achieve this by implementing a perturbative analysis in which the transition from one state to another follows the equation of motion of the original system, but the distribution of $\bv{\Delta}(t)$ at time $t$ of intervention is replaced by a configuration of maximum entropy \cite{pearl1995causal,tononi2001information,  tononi2003measuring, tononi2004information, ay2008information}. 
This way, the information generated in the system is only due to its evolution through different states. 
In the specific scenario of $N=2$ RWs, this consists of investigating the dynamics of the two independent walkers that evolve following Eqs.~\eqref{eq:2RWs_delta1_tt1}-\eqref{eq:2RWs_delta2_tt1}. 
Denoting with $p$ the distributions of the original system and with $q$ the ones undergoing intervention, we have:
\begin{equation} \label{eq:intervention}
    q(\bv{\Delta}(\tp),\bv{\Delta}(t)) = p(\bv{\Delta}(\tp)|\bv{\Delta}(t)) q(\bv{\Delta}(t))\,,
\end{equation}
where ${\bv{\Delta}\equiv\begin{pmatrix} \Delta_1 & \Delta_2 \end{pmatrix}^T}$, ${p(\bv{\Delta}(\tp)|\bv{\Delta}(t))}$ is the conditional distribution of the original system, and ${q(\bv{\Delta}(t))=p(\Delta_1(t))\,p(\Delta_2(t))}$ is the input distribution modelled as a maximum entropy distribution with the marginals equal to the original system. 
Therefore, this analysis allows one to examine the ensemble of states ${q(\bv{\Delta}(\tp))}$ that are reached solely due to the interactions in the evolution of the dynamics.

\begin{figure}
    \centering
    \includegraphics[width=\linewidth]{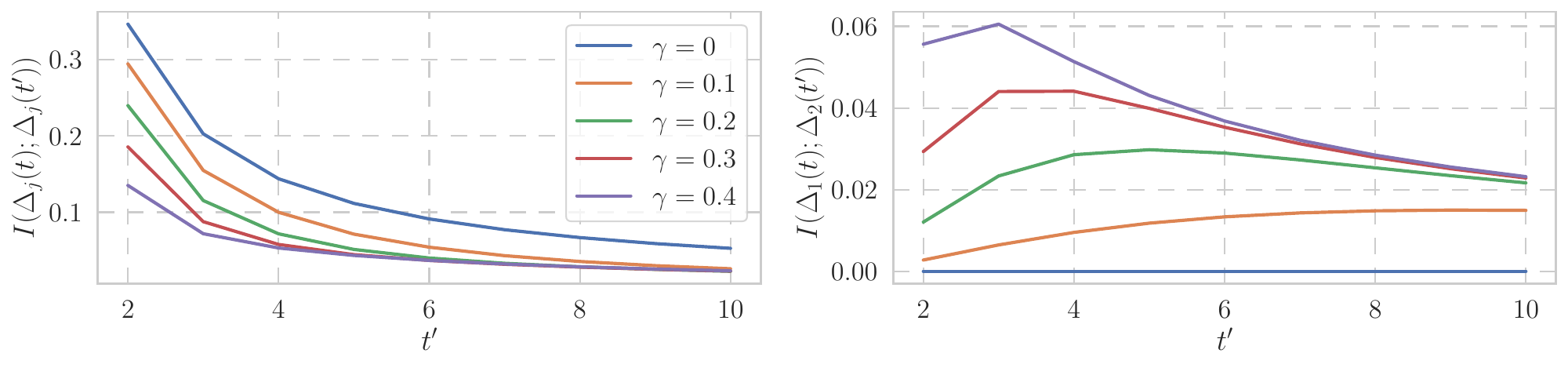}
    \caption{\textbf{Causal intervention changes the mutual information between random walkers.}
    Mutual information for the single random walker ($I(\Djt,\Djtp$) and between random walkers ($I(\Delta_1(t),\Delta_2(\tp)$) for the perturbed system of $N=2$ RWs. Results are shown for $t=1$.}
    \label{fig:intervention_MIs}
\end{figure}

Considering for simplicity the case in which $\theta=0$, the equations of motions between two timepoints $t,\tp$, $\tp>t$ read
\begin{equation}
\displaystyle
    \bv{\Delta}(\tp) = A \bv{\Delta}(t) + V, \quad \text{with }\quad 
    A = \left(\begin{matrix}
        \displaystyle\frac{1+\beta^{\Delta t} }{2} & 
        \displaystyle\frac{1-\beta^{\Delta t}}{2} \vspace{1.5mm} \\
        \displaystyle\frac{1-\beta^{\Delta t}}{2} & 
        \displaystyle\frac{1+\beta^{\Delta t} }{2}
        \end{matrix}\right) ,
    \quad
    V = \left(\begin{matrix}
        \displaystyle\sum_{\tilde{t}=t}^{\tp-1} \Bigl(\eta_+(\tilde{t}) + \beta^{\tp-t-1}\eta_-(\tilde{t})\Bigr) \vspace{1.5mm} \\ 
        \displaystyle\sum_{\tilde{t}=t}^{\tp-1} \Bigl(\eta_+(\tilde{t}) - \beta^{\tp-t-1}\eta_-(\tilde{t})\Bigr)
        \end{matrix}\right) ,
\end{equation}
where we introduced $\Delta t\equiv\tp-t$ and used $\beta=1-2\gamma$. 
With this and Eqs.~\eqref{eq:2RW_var_xjxj}-\eqref{eq:2RW_cov_xjxi} at hand, we start constructing the probability distributions needed in Eq.~\eqref{eq:intervention}. 
We have that
\begin{equation} \label{eq:intervention_marginal}
    \bv{\Delta}(t) \sim \mathcal{N}\left(0\,,\, \Sigma_t \right), \,\quad \text{with} \quad 
    \Sigma_t = 
    \begin{pmatrix}
         \dfrac{\sigma^2}{2}\left(t+\bgeom{t}\right) & 0 \\
         0 & \dfrac{\sigma^2}{2}\left(t+\bgeom{t}\right)
        \end{pmatrix}
\end{equation}
and
\begin{gather} \label{eq:intervention_conditional}
    \bv{\Delta}(\tp)|\bv{\Delta}(t) \sim \mathcal{N}\left(A\bv{\Delta}(t)\,,\, \Sigma_{\tp|t} \right), \quad \text{with} \vspace{3mm} \\
    \Sigma_{\tp|t} \equiv \tilde{\Sigma}_{\tp} - \tilde{\Sigma}_{t;\tp}^T \tilde{\Sigma}_t^{-1} \tilde{\Sigma}_{t;\tp} =
    \dfrac{\sigma^2}{2} \left(\begin{matrix}
         \Delta t + \bgeom{\Delta t}  &  \Delta t - \bgeom{\Delta t} \vspace{2mm} \\
         \Delta t - \bgeom{\Delta t} &  \Delta t + \bgeom{\Delta t}
        \end{matrix}\right) , \notag
\end{gather}
where the $\,\tilde{}\,$ indicates that the covariances belong to the distributions $p$ of the original system. 
Finally, we can calculate the joint distribution by multiplying the Gaussian distributions of Eqs.~\eqref{eq:intervention_marginal}-\eqref{eq:intervention_conditional}, which provides the full covariance of the system:
\begin{gather}
    \bv{\Delta}(\tp),\bv{\Delta}(t) \sim \mathcal{N}\left(0\,,\, \Sigma\right), \quad \text{with} \vspace{2mm} \\
    \Sigma \equiv \begin{pmatrix} \Sigma_{\tp} & \Sigma_{t;\tp}^T \\
                \Sigma_{t;\tp} & \Sigma_{t} \end{pmatrix} = \\
    = \frac{\sigma^2}{2} \left(\begin{smallmatrix}
         \left( \Delta t+\textbgeom{\Delta t} \right) +  \left(t+\textbgeom{t}\right)\frac{1+\beta^{2\Delta t}}{2} \hspace{3mm} & 
         \left(\Delta t - \textbgeom{\Delta t} \right) +  \left(t+\textbgeom{t}\right)\frac{1-\beta^{2\Delta t}}{2} \hspace{3mm} & 
         \left(t+\textbgeom{t}\right) \frac{1+\beta^{\Delta t}}{2} \hspace{3mm} &
         \left(t+\textbgeom{t}\right) \frac{1-\beta^{\Delta t}}{2} \vspace{1mm} \\
         \left(\Delta t - \textbgeom{\Delta t} \right) +  \left(t+\textbgeom{t}\right)\frac{1-\beta^{2\Delta t}}{2} \hspace{3mm} & 
         \left( \Delta t+\textbgeom{\Delta t} \right) +  \left(t+\textbgeom{t}\right)\frac{1+\beta^{2\Delta t}}{2} \hspace{3mm} &
         \left(t+\textbgeom{t}\right) \frac{1-\beta^{\Delta t}}{2} \hspace{3mm} & 
         \left(t+\textbgeom{t}\right) \frac{1+\beta^{\Delta t}}{2} \vspace{1mm} \\
         \left(t+\textbgeom{t}\right) \frac{1+\beta^{\Delta t}}{2} \hspace{3mm} &
         \left(t+\textbgeom{t}\right) \frac{1-\beta^{\Delta t}}{2} \hspace{3mm} &
         \left(t+\textbgeom{t}\right) \hspace{3mm} & 0 \vspace{1mm} \\
         \left(t+\textbgeom{t}\right) \frac{1-\beta^{\Delta t}}{2} \hspace{3mm} & 
         \left(t+\textbgeom{t}\right) \frac{1+\beta^{\Delta t}}{2} \hspace{3mm} & 
         0 & \left(t+\textbgeom{t}\right) 
        \end{smallmatrix}\right). \notag
\end{gather}
Using the terms inside $\Sigma$, we can construct the information-theoretic measures of interest and look at their behaviour. 
Regarding mutual information, we notice that like the unperturbed system (c.f.\ Fig.~\ref{fig:2RWS_MI}) larger coupling strengths result in higher mutual information between the two walkers (Fig.~\ref{fig:intervention_MIs}).

\begin{figure}
    \centering
    \includegraphics[width=0.5\linewidth]{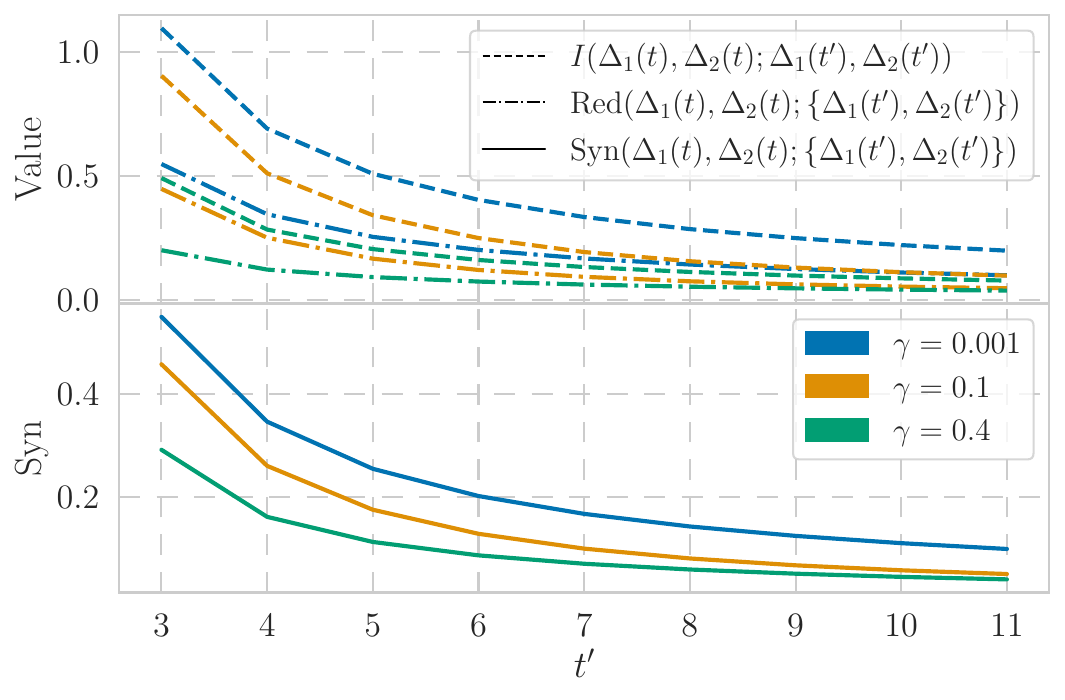}
    \caption{\textbf{Raw PID atoms in the perturbed system behave qualitatively the same as in the original one.}
    Partial Information Decomposition on the perturbed system of $N=2$ random walkers. The dashed lines correspond to the joint mutual information, the dash-dotted lines to redundancy, and the solid to synergy. PID with ${X=\Delta_j(t)}$, ${Y=\Delta_i(t)}$, $Z=\{\Delta_j(t^{\prime}), \Delta_i(t^{\prime})\}$, $i\ne j$. Results are shown for $t=2$.}
    \label{fig:intervention_PID}
\end{figure}

\begin{figure*}[hbt]
\centering
\begin{subfigure}[]{.49\textwidth}
    \centering
    \figuretitle{\hspace{6mm}Original system}
\hspace*{-1.5mm}\includegraphics[width=1\textwidth]{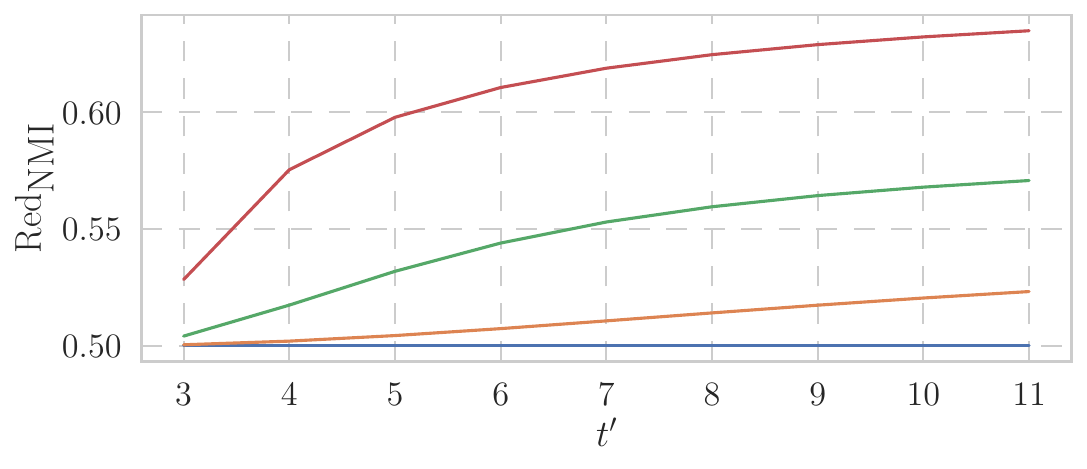}
    \caption{
    }
\end{subfigure}%
\hspace*{-1mm}
\begin{subfigure}[]{.49\textwidth} 
    \centering
    \figuretitle{\hspace{6mm}Perturbed system}
\hspace*{-1.5mm}\includegraphics[width=1\textwidth]{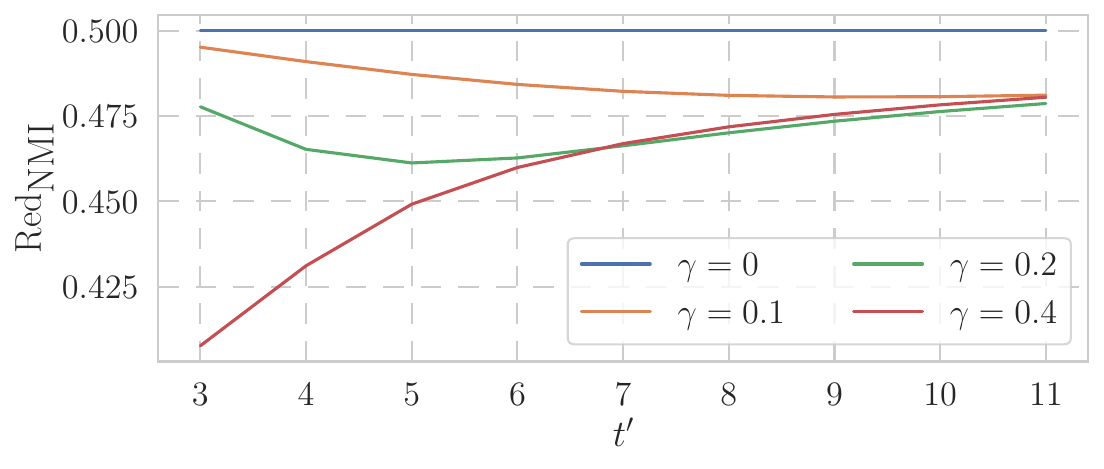}
    \caption{ 
    }
\end{subfigure}%
\caption{\textbf{Normalised redundancy in the perturbed system decreases with the interaction strength.}
NMI normalised redundancy atom for the PID with ${X=\Delta_j(t)}$, ${Y=\Delta_i(t)}$, $Z=\{\Delta_j(t^{\prime}), \Delta_i(t^{\prime})\}$, $i\ne j$ on the system of $N=2$ random walkers. (a) Original system, (b) perturbed system with intervention approach. Results are shown for $t=2$.}
\label{fig:intervention_PID_NMI_redundancy}
\end{figure*}

However, if we now reproduce the PID study of Sec.~\ref{sec:results_pid} in the case (3) -- where the source variables are the past of the RWs and the target is their joint future -- we obtain a profoundly different behaviour: although the raw atoms behave qualitatively similar to the original system (Fig.~\ref{fig:intervention_PID}, c.f.\ Fig.~\ref{fig:PID_imgs}), their normalised version shows that synergy increases with higher $\gamma$ (Fig.~\ref{fig:PID_imgs_NMI_intervention_synergy}) and redundancy decreases (Fig.~\ref{fig:intervention_PID_NMI_redundancy}). 
This finding showcases the information composition related to the system's dynamics, excluding from the picture the effect of the instantaneous statistical distributions of the components.

\FloatBarrier

\section{Numerical tests} \label{app:simul}
In this section, we report the numerical calculations of information-theoretic measures performed on the various systems studied above. 
Since these measures rely on estimating joint and marginal probability distributions, their accuracy is inevitably heavily dependent on numerics and data availability.
Therefore, by comparing them with the theoretical formulas found in previous sections, the objective here is to simultaneously validate our theoretical results and test the robustness of numerical procedures of these metrics. 
Finally, we assess and discuss how numerical approximations and round-up errors can affect the assessment of emergent properties of the system.

\begin{figure}[ht]
    \centering
    \includegraphics[width=1\linewidth]{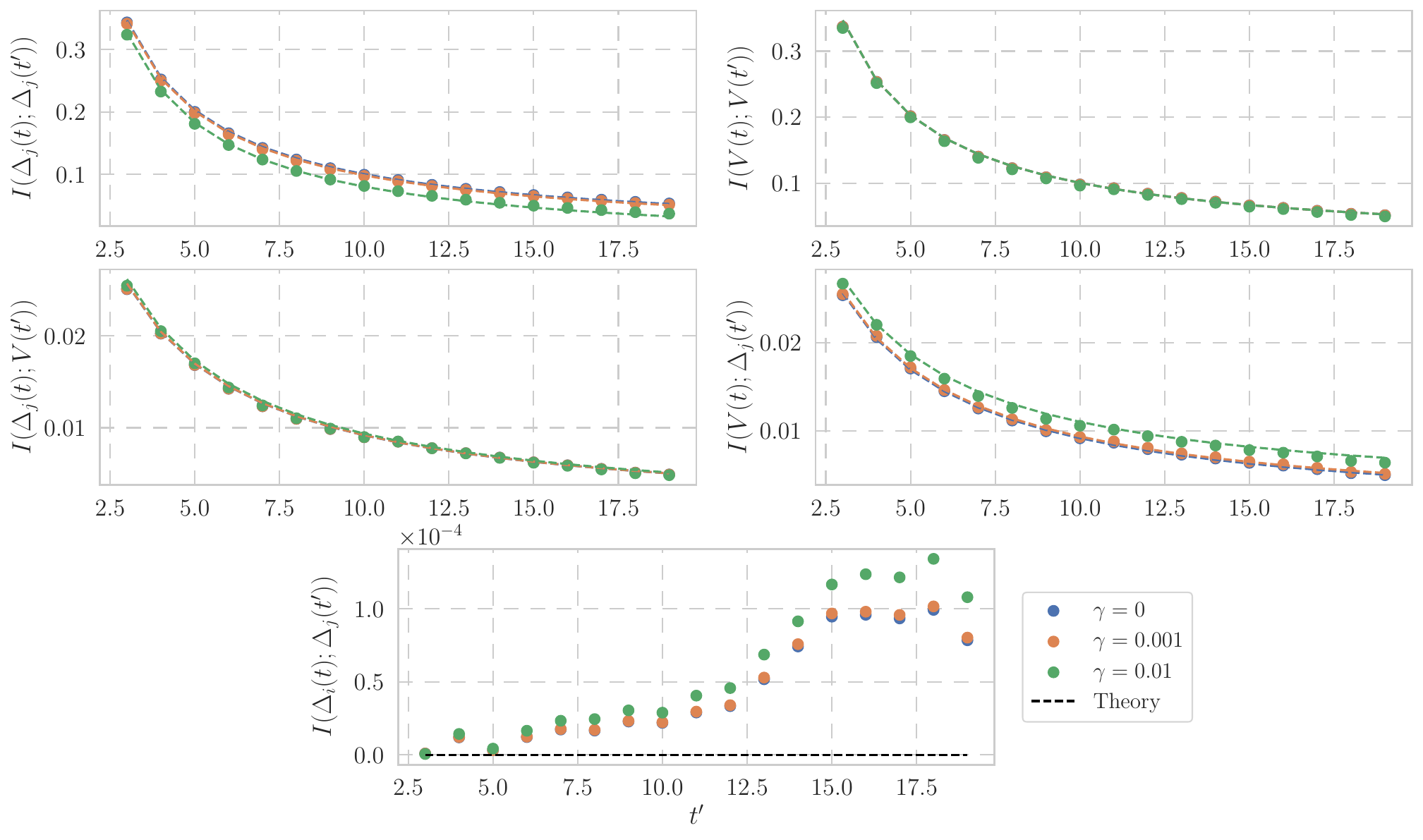}
    \caption{\textbf{Numerical computations of mutual information agree with the analytical findings in the non-stationary case.}
    Comparison of theoretical and numerical results of the mutual information for $\theta=0$ and various $\gamma$ for $N>2$ RWs. Results are shown for $t=2$. For visualisation purposes, the center of mass $V$ in $I(\Djt;V(\tp))$ and $I(V(t);\Djtp)$ was calculated for the 5 central RWs.}
    \label{fig:simul_NRWS_theta_fix}
\end{figure}

\begin{figure}[ht]
    \centering
    \includegraphics[width=1\linewidth]{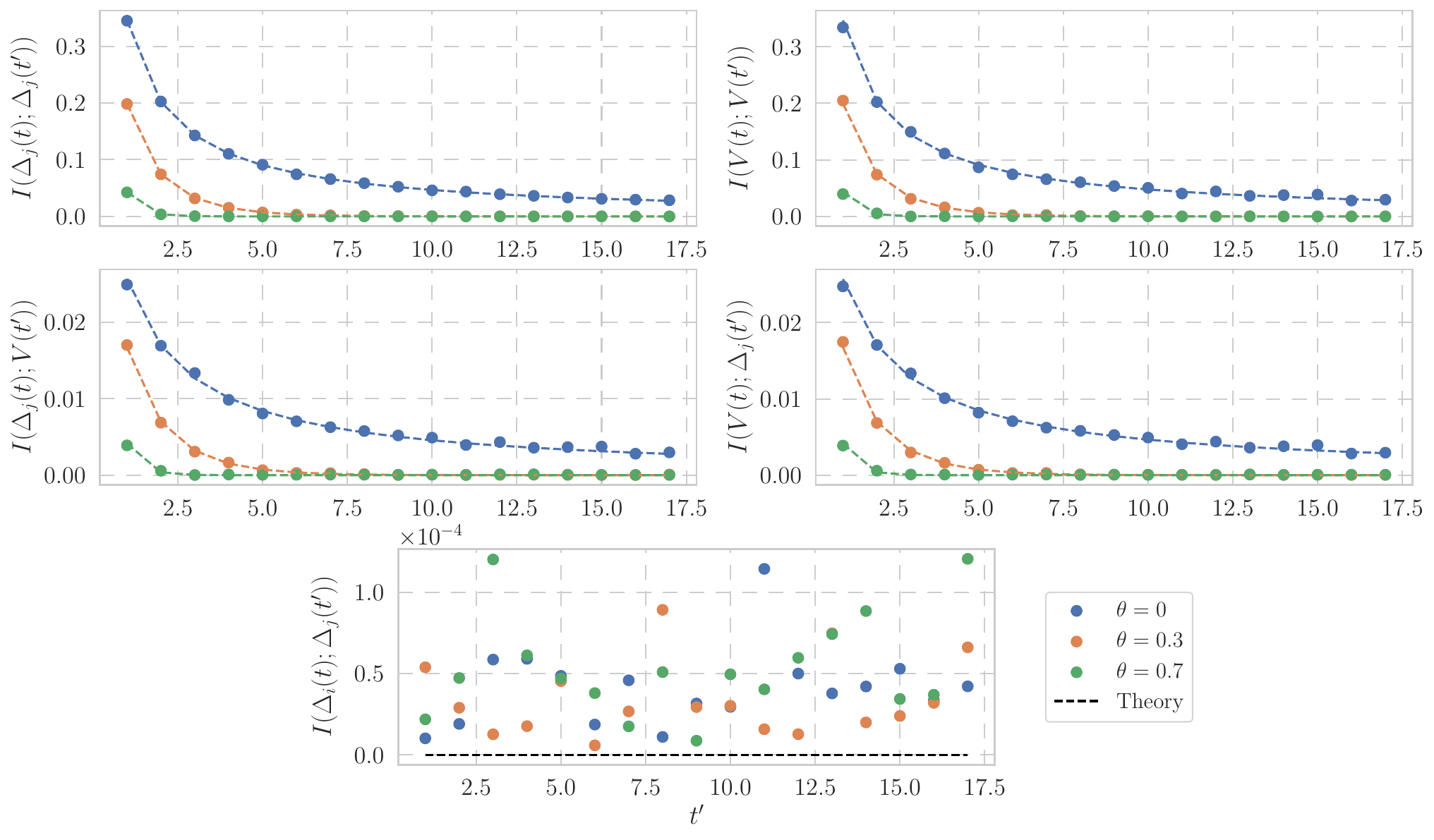}
    \caption{\textbf{Numerical computations of mutual information agree with the analytical findings in the stationary case.}Comparison of theoretical and numerical results of the mutual information for $\gamma=0.001$ and various $\theta$ for $N>2$ RWs. The solid black line in the top left panel marks the zero value. Results are shown for $t=2$. For visualisation purposes, the center of mass $V$ in $I(\Djt;V(\tp))$ and $I(V(t);\Djtp)$ was calculated for the 5 central RWs.}
    \label{fig:simul_NRWS_gamma_fix}
\end{figure}

For brevity, we focus exclusively on the estimation of the mutual information in the case of $N>2$ random walkers (Eqs.~\eqref{eq:NRW_Ixx}-\eqref{eq:NRW_Ivx}). 
For the numerical computations, we simulated the evolution of $N=512$ random walkers with dynamics given by Eq.~\eqref{eq:delta_ev_app} and repeated the procedure for an ensemble of 10000 systems. 
Then, we performed the mutual information calculations with the Java Information Dynamics Toolkit (JIDT) \cite{lizier2014jidt}, which fits a Gaussian distribution on the time series of the data and allows calculations of the MI between any two timepoints. The final estimation is the average across all ensembles. Results are shown in Figs.~\ref{fig:simul_NRWS_theta_fix}-\ref{fig:simul_NRWS_gamma_fix}.

We first note that the numerical values closely resemble the expected theoretical ones for all choices of the parameters $(\theta,\gamma)$. Unsurprisingly, we also notice that the mutual information between different walkers is low, but not exactly zero. 
As mentioned in the main body of the article, we recall the fact that quantities like MI, TE,
are second order quantities in the coupling term. This can  in principle interfere with mechanistic interpretations of the system's properties as higher-order terms and noise can be easily confused.

We conclude by showing how these small effects can come together to significantly change the interpretation of some quantities. We focus on the calculations of $\Psi, \Delta, \Gamma$ metrics, and, as before, we compare the theoretical and numerical results for various parameters $\gamma$. 

\begin{figure}
    \centering
    \includegraphics[width=1\linewidth]{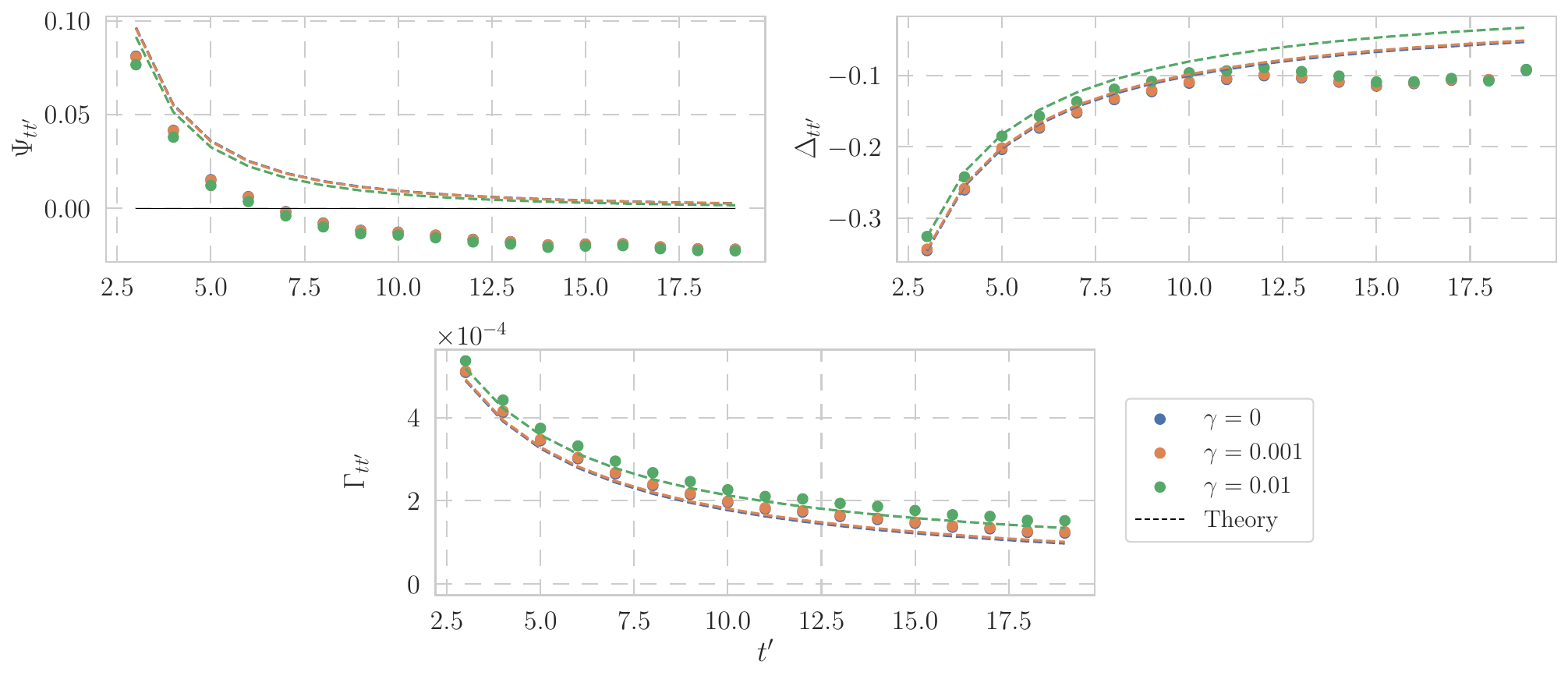}
    \caption{\textbf{Simulated causal emergence quantities show slight discrepancies with analytical results due to numerical errors.}
    Comparison of theoretical and numerical results of $\Psi, \Delta, \Gamma$ emergence measures for $\theta=0$ and various $\gamma$ for $N>2$. Results are shown for $t=2$.}
    \label{fig:simul_NRWS_emergence}
\end{figure}

From Fig.~\ref{fig:simul_NRWS_emergence}, we observe that some deviations appear, especially in the case of $\Psi_{t\tp}$. This behaviour is due to small discrepancies ($\sim10^{-5}$) between theory and computations in $I(\Djt;V(\tp))$ term. 
Although these differences are negligible when taken singularly, once $I(\Djt;V(\tp))$ is multiplied by the size of the system (see Eq.~\eqref{eq:psi_emergence}), these discrepancies become comparable to $I(V(t);V(\tp))$, resulting in an undershooting of $\Psi_{t\tp}$. 
Most importantly, $\Psi_{t\tp}$ now crosses the zero-value line and does not satisfy proposition (3) anymore (Sec.~\ref{sec:results_emergence}). 
We remind here that $\Psi>0$ is only a \textit{sufficient} condition for $V$ to be an emergent quantity, being inconclusive otherwise. 
The purpose of this example is only to underline that a characterisation of the system solely in terms of these quantities can be heavily affected by the propagation of numerical errors, even if small in the first place.

\end{document}